\documentclass[11pt,a4paper]{article}
\pdfoutput=1

\usepackage[colorlinks=true, linkcolor=black!50!blue, urlcolor=blue, citecolor=blue, anchorcolor=blue]{hyperref}
\usepackage[font=small,labelfont=bf,margin=0mm,labelsep=period,tableposition=top]{caption}
\usepackage[a4paper,top=1.5cm,bottom=2cm,left=1.5cm,right=1.5cm,bindingoffset=0mm]{geometry}

\usepackage{placeins,setup,cite}
\usepackage{graphicx}
\usepackage{float}
\usepackage{afterpage}
\usepackage{epsfig}
\usepackage{amssymb}
\usepackage{amsmath}
\usepackage{bm}
\usepackage{multirow}
\usepackage{url}
\usepackage{xcolor}
\usepackage{ulem}
\usepackage{url}
\usepackage{booktabs,multirow,colortbl}
\usepackage[symbol]{footmisc}
\usepackage[colorinlistoftodos]{todonotes}

\usepackage{neuralnetwork}
\usepackage{xstring}
\usepackage{xpatch}
\usepackage{array}

\usepackage{longtable}

\usepackage{caption}
\usepackage{subcaption}

\makeatletter
\def\thickhline{%
             \noalign{\ifnum0 =`}\fi\hrule \@height \thickarrayrulewidth \futurelet
             \reserved@a\@xthickhline}
\def\@xthickhline{\ifx\reserved@a\thickhline
                \vskip\doublerulesep
                \vskip -\thickarrayrulewidth
                \fi
                \ifnum0 =`{\fi}}
\xpatchcmd{\linklayers}{\nn@lastnode}{\lastnode}{}{}
\xpatchcmd{\linklayers}{\nn@thisnode}{\thisnode}{}{}
\makeatother
\newlength{\thickarrayrulewidth}
\usepackage{tikz}
\usetikzlibrary{positioning}
\usetikzlibrary{arrows.meta}
\usepackage{varwidth}
\usepackage{xcolor}
\definecolor{mtplotlib1}{HTML}{1f77b4}
\definecolor{mtplotlib2}{HTML}{ff7f0e}
\definecolor{mtplotlib3}{HTML}{2ca02c}
\definecolor{mtplotlib4}{HTML}{d62728}

\tikzset{%
  >={Latex[width=2mm,length=2mm]},
            base/.style = {rectangle, rounded corners, draw=black,
                           minimum width=4cm, minimum height=1cm,
                           text centered}, 
            mystyle/.style={rectangle, rounded corners, draw=black,
            minimum width=12cm, minimum height=1cm,
            text centered}, 
    col0/.style = {base, fill=white!30},
    col1/.style = {base, fill=mtplotlib1!30},
    col11/.style = {mystyle, fill=mtplotlib1!30},
    col2/.style = {base, fill=mtplotlib2!30},
    col3/.style = {base, fill=mtplotlib3!30},
    col4/.style = {base, minimum width=2.5cm, fill=mtplotlib4!15,}
}

\newcommand{\be}{\begin{equation}}
\newcommand{\ee}{\end{equation}}
\newcommand{\bea}{\begin{eqnarray}}
\newcommand{\eea}{\end{eqnarray}}
\newcommand{\bi}{\begin{itemize}}
\newcommand{\ei}{\end{itemize}}
\newcommand{\ben}{\begin{enumerate}}
\newcommand{\een}{\end{enumerate}}
\newcommand{\la}{\left\langle}
\newcommand{\ra}{\right\rangle}
\newcommand{\lc}{\left[}
\newcommand{\rc}{\right]}
\newcommand{\lp}{\left(}
\newcommand{\rp}{\right)}

\def\frac#1#2{{{#1}\over {#2}}}
\def\gsim{\mathrel{\rlap{\lower4pt\hbox{\hskip1pt$\sim$}}
    \raise1pt\hbox{$>$}}}       
\def\lsim{\mathrel{\rlap{\lower4pt\hbox{\hskip1pt$\sim$}}
    \raise1pt\hbox{$<$}}}

\newcommand{\draft}[1]{}

\def\beq{\begin{equation}}
\def\eeq{\end{equation}}

\numberwithin{equation}{section}
\numberwithin{figure}{section}
\numberwithin{table}{section}

\newcommand*{\tmp}[4]{\ensuremath{%
    {#4%
    \ifx\empty#3\empty\ifx\empty#1\empty\else^{#1}\fi\else^{#1(#3)}\fi%
    \ifx\empty#2\empty\else_{#2}\fi}%
}}
\newcommand*{\ccc}[4][]{\tmp{#2}{#3}{#4}{#1{c}}}
\newcommand{\sss}{\scriptscriptstyle}
\newcommand{\OO}{\ensuremath{\mathcal{O}}}
\newcommand{\Op}[1]{\OO_{\sss #1}}

\def\lra#1{\overset{\text{\scriptsize$\leftrightarrow$}}{#1}}
\newcommand{\fitm}{{\sc\small FitMaker}}
\newcommand{\smefit}{{\sc\small SMEFiT} }
\newcommand{\nnpdf}{\texttt{NNPDF4.0} }
\newcommand{\nnpdfnotop}{\texttt{NNPDF4.0-notop} }
\newcommand{\simunet}{{\texttt{SIMUnet}}}
\DeclareMathOperator*{\argmin}{arg\,min}

\usepackage{tabularx}
\newcolumntype{C}[1]{>{\centering\arraybackslash}p{#1}}

\definecolor{darkblue}{rgb}{0.0,0,0.5}
\definecolor{darkgreen}{rgb}{0.0,0.3,0.0}
\definecolor{redish}{rgb}{0.675,0,0.2}
\definecolor{red}{rgb}{0.8,0,0}
\definecolor{green}{rgb}{0,0.6,0}
\definecolor{bluish}{rgb}{0.2,0.2,0.675}
\definecolor{mygrey}{rgb}{0.6,0.6,0.6}

\begin{document}

\begin{center}
  
{\Large \bf The top quark legacy of the LHC Run II for PDF and SMEFT analyses }\\
  \vspace{1.1cm}
         {\small

           Zahari Kassabov$^1$,
Maeve Madigan$^1$,
Luca Mantani$^1$,
James Moore$^1$,
Manuel Morales Alvarado$^1$,\\[0.2cm]
Juan Rojo$^{2,3}$, and
Maria Ubiali$^1$\\}
  
\vspace{0.7cm}

{\it \small
  ~$^1$DAMTP, University of Cambridge, Wilberforce Road, Cambridge, CB3 0WA, United Kingdom\\[0.1cm]
 ~$^2$Department of Physics and Astronomy, Vrije Universiteit, NL-1081 HV Amsterdam\\[0.1cm]
  ~$^3$Nikhef Theory Group, Science Park 105, 1098 XG Amsterdam, The Netherlands\\[0.1cm]
  
 }

\vspace{1.0cm}

{\bf \large Abstract}

\end{center}

We assess the impact of  top quark
production at the LHC 
on global analyses of parton distributions (PDFs) and of Wilson coefficients
in the SMEFT, both separately and in the framework of a joint interpretation.
We consider the broadest top quark dataset to date
containing all available measurements based on the full Run II luminosity.
First, we determine the constraints that this dataset provides
on the large-$x$ gluon PDF and study its consistency with
other gluon-sensitive measurements.
Second, we carry out a SMEFT interpretation of the same dataset using state-of-the-art
SM and EFT theory calculations, resulting in bounds on 25 Wilson coefficients modifying top quark
interactions.
Subsequently, we integrate the two analyses within the \simunet{} approach
to realise a simultaneous determination of the SMEFT PDFs and the EFT coefficients
and identify regions in the parameter space where their interplay
is most phenomenologically relevant.
We also demonstrate how to separate eventual BSM signals from QCD effects
in the interpretation of top quark measurements at the LHC.

\clearpage

\tableofcontents

\section{Introduction}

The top quark is one of the most remarkable particles within the Standard Model (SM).
Being the heaviest elementary particle known to date, with a mass around 185 times heavier than a proton,
and the only fermion with an $\mathcal{O}(1)$
Yukawa coupling to the Higgs boson, the top quark has long been suspected to play a privileged role
in potential new physics extensions beyond the Standard Model (BSM).
For instance, radiative corrections involving top quarks are responsible
for the so-called hierarchy problem of the SM, and the value of its mass $m_t$ determines
whether the vacuum state of our Universe is stable, metastable,
or unstable~\cite{Isidori:2001bm,Buttazzo:2013uya,DiLuzio:2015iua}.
For these reasons, since its discovery at the Tevatron in 1995~\cite{CDF:1995wbb,D0:1995jca}
the properties of the top quark have been scrutinised with utmost attention and
a large number of BSM searches involving top quarks as final states have been carried out.
The focus on the top quark has further intensified since the start of operations
at the LHC, which has realised an unprecedented top factory producing more than 200 million
top quark pairs so far, for example.

In addition to this excellent potential for BSM studies, top quark production at hadron
colliders also provides unique information on a variety of SM parameters such
as the strong coupling constant $\alpha_s(m_t)$~\cite{CMS:2014rml,Cooper-Sarkar:2020twv}, 
the CKM matrix element $V_{tb}$~\cite{CMS:2020vac}, and the top quark
mass $m_t$~\cite{ATLAS:2022jbw,CMS:2023ebf},
among several others.
Furthermore, top quark production at the LHC constrains the
parton distribution functions (PDFs) of the proton~\cite{Gao:2017yyd,Amoroso:2022eow},
in particular the large-$x$ gluon PDF from inclusive top quark
pair production~\cite{Czakon:2013tha,Czakon:2016olj,Czakon:2019yrx}
and the quark PDF flavour separation
from inclusive single top production~\cite{Nocera:2019wyk,Campbell:2021qgd}.
Indeed, fiducial and differential measurements of top quark pair
production are part of the majority of recent PDF determinations.
Reliably extracting SM parameters, including those parametrising the
subnuclear structure of the proton in the PDFs, from LHC top quark
production data has been made possible thanks to
recent progress in higher order QCD and electroweak calculations
of top quark production.
Inclusive top quark pair production is now known at NNLO in the QCD
expansion both for single-
and double-differential distributions~\cite{Czakon:2016dgf,Czakon:2020qbd}, eventually complemented
with electroweak corrections~\cite{Czakon:2017wor}, threshold
resummation~\cite{Czakon:2018nun}, and matching to parton showers~\cite{Mazzitelli:2021mmm}.
NNLO QCD corrections are also known for single top quark production
at the LHC, both in the
$t$-channel~\cite{Berger:2017zof,Berger:2016oht} and in the $s$-channel~\cite{Liu:2018gxa}.

Even in BSM scenarios where new particles are sufficiently heavy such that direct 
production  lies beyond
the reach of the LHC, current and future measurements can still provide BSM
sensitivity through low-energy signatures.
These are typically revealed in the modification
of  SM particle properties, such as their interactions and coupling strengths.
In this context, a powerful model-agnostic framework to parametrise, identify, and correlate
the low-energy signatures of heavy BSM physics is the
Standard Model Effective Field Theory (SMEFT).
Several groups, both from the theory community and within the experimental collaborations,
have presented interpretations of LHC top quark 
measurements in the SMEFT framework~\cite{Aguilar-Saavedra:2018ksv,
  Buckley:2015lku, Brivio:2019ius, Bissmann:2019gfc, Hartland:2019bjb,
  Durieux:2019rbz, vanBeek:2019evb, Yates:2021udl} to derive bounds
on  higher-dimensional EFT operators that distort the interactions of top quarks.
A key feature of these analyses is that the unprecedented energy reach of the LHC
data  increases the sensitivity to SMEFT operators via energy-growing
effects entering the partonic cross-sections.

Therefore, in the LHC precision era, top quark measurements  are being
interpreted in (at least) two frameworks
with rather different underlying assumptions.
On the one hand, global PDF fits assume the SM and use top data to constrain
the PDFs, producing Standard Model PDFs (denoted ``SM-PDFs'' in the following).
On the other hand, SMEFT analyses assume that  top data does not modify the
SM predictions, and in particular that the proton PDFs
are unchanged; we thus refer to SMEFT fits as ``fixed-PDF'' in the following.
The two assumptions cannot be simultaneously correct, and hence one must answer
two pressing questions concerning the interpretation of LHC top quark measurements.
First, are SM-PDFs contaminated by BSM physics, encapsulated in the SMEFT framework,
which are being reabsorbed into the fitted PDF boundary condition?
Second, are the results of existing SMEFT interpretations  dependent on the choice
of PDFs entering the SM calculations, and is it consistent to use PDF sets that
already include top quark data?
It should be emphasized that for top quark production one cannot 
classify the data in two disjoint ``SM-PDF'' and ``fixed-PDF'' regions, since in both cases
sensitivity arises from the high-energy regime.

These two questions can only be answered by means of the simultaneous determination
of the PDFs and EFT coefficients from a common input dataset resulting in 
so-called ``SMEFT-PDFs''.
A proof of concept of this strategy was presented for deep-inelastic scattering (DIS)
data~\cite{Carrazza:2019sec} and then extended to a joint analysis of DIS and
Drell-Yan (DY) data~\cite{Greljo:2021kvv}
including projections for the HL-LHC; see also~\cite{Liu:2022plj,Gao:2022srd,CMS:2021yzl}
for related work.
The studies of~\cite{Carrazza:2019sec,Greljo:2021kvv} were restricted
to a small number of representative EFT operators,
and extending them to the realistic case of  processes sensitive to
a large number of operators,
such as top or jet production data, required the development of improved techniques.
With this motivation, a new methodology dubbed \simunet{}
was developed~\cite{Iranipour:2022iak}  making possible global SMEFT-PDF interpretations
of LHC data suitable for processes depending on up to several tens of
EFT operators.
A key feature of \simunet{} is that it can be easily projected to both the
SM-PDF case, in which it reduces to the NNPDF fitting methodology~\cite{NNPDF:2021njg,NNPDF:2021uiq},
and to the fixed-PDF case, where it becomes equivalent to global EFT fitting tools such as {\sc\small SMEFiT}~\cite{Ethier:2021bye}.

The aim of this work is to extend the initial explorations
of~\cite{Carrazza:2019sec,Greljo:2021kvv} to a global determination of the SMEFT-PDFs
from top quark production measurements.
To this purpose, we consider the broadest top quark dataset used to date in either
PDF or EFT interpretations, which in particular
contains all available measurements from ATLAS and CMS based on the full Run II luminosity.
By combining this wide dataset with the \simunet{} methodology, 
we derive bounds on  25 independent Wilson coefficients modifying top quark
interactions, identify regions in the parameter space where the interplay between
PDFs and SMEFT signatures
is most phenomenologically relevant, and
demonstrate how to separate eventual BSM signals from QCD effects
in the interpretation of top quark measurements.
As a non-trivial by-product, we also revisit the SM-PDF and fixed-PDF analyses by quantifying the information
that our comprehensive top quark dataset provides. On the one hand, we assess the impact on the large-$x$ gluon (SM-PDF),
and on the other, we study the impact on the EFT coefficients (fixed-PDF),
and compare our findings with related studies in the literature.

The structure of this paper is as follows.
To begin with, in Sect.~\ref{sec:exp} we describe the data
inputs and the theory calculations (both in the SM and in the SMEFT) used in our study,
focusing on top quark sector measurements.
The \simunet{}
methodology deployed for the simultaneous extraction of PDFs and EFT coefficients,
including its application to the fixed-PDF and SM-PDF analyses,
is reviewed in Sect.~\ref{sec:methodology}.
Subsequently, in Sect.~\ref{sec:baseline_sm_fits} we present the results of the SM-PDF fits,
and in particular we quantify the impact on the large-$x$ gluon of recent high-statistics
Run II measurements.
In Sect.~\ref{sec:res_smeft} we consider the fixed-PDF analyses
and present the most extensive SMEFT interpretation of top quark
data from the LHC to date, including comparisons with previous results
in the literature.
The main results of this paper are presented in  Sect.~\ref{sec:joint_pdf_smeft},
namely the simultaneous determinations of the PDFs and EFT coefficients and the comparison
of these  with both the fixed-PDF and SM-PDF cases.
We summarise our results and outline some possible future developments
in Sect.~\ref{sec:summary}.

Technical details of the analysis are collected in the appendices.
App.~\ref{app:recommendations} provides usage recommendations for
interpretations of top quark measurements sensitive both to PDFs and
SMEFT coefficients.
App.~\ref{sec:operators} collects the theory settings for
the SMEFT calculations, mainly concerning input schemes and operator definitions.
App.~\ref{app:benchmark} carries out a benchmark comparison
between \simunet{} (in the fixed-PDF case) and the public
{\sc\small SMEFiT} code, demonstrating the agreement between the two frameworks
at the linear level.
The fit quality to the datasets considered in the analysis
is presented in App.~\ref{app:fit_quality}, and representative data-theory comparisons
are given. 
Finally, in App.~\ref{app:quad}
we discuss the difficulties in extending the simultaneous analysis of
PDFs and EFT coefficients to the case where
terms quadratic in the EFT Wilson coefficients are dominant.

\section{Experimental data and theory calculations}
\label{sec:exp}

We begin by  describing the experimental data and theoretical predictions,
both in the SM and in the SMEFT, used as input for the present analysis.
We start in Sect.~\ref{sec:baseline_data} by 
describing the datasets that we consider, with
emphasis on the top quark production measurements.
Then in Sect.~\ref{sec:dataselection} we use a modified version of the selection criteria
defined in~\cite{NNPDF:2021njg}
to determine a maximally consistent dataset of top quark data to be used
in the subsequent PDF and SMEFT interpretations.
Finally, in Sect.~\ref{sec:theory} we describe the calculation settings
of the SM and SMEFT cross-sections for top quark processes, pointing 
the reader to the appendices for the technical details of their implementation.

\subsection{Experimental data}
\label{sec:baseline_data}

With the exception of the top quark measurements, the dataset used in
this work for fitting the PDFs both in the SM-PDF and SMEFT-PDF cases
overlaps with that of the \nnpdf determination presentend in Ref.~\cite{NNPDF:2021njg}.
In particular,
the no-top variant of the \nnpdf dataset consists of 4535 data points
corresponding to a wide variety of processes in
deep-inelastic lepton-proton scattering~\cite{Arneodo:1996kd,Arneodo:1996qe,Whitlow:1991uw,Benvenuti:1989rh,Onengut:2005kv,Goncharov:2001qe,MasonPhD,Abramowicz:2015mha,H1:2018flt}
and in hadronic proton-proton collisions~\cite{Moreno:1990sf,Webb:2003ps,Towell:2001nh,Aaltonen:2010zza,Abazov:2007jy,Abazov:2013rja,D0:2014kma,Abulencia:2007ez,Aad:2011dm,Aaboud:2016btc,Aad:2014qja,Aad:2013iua,Chatrchyan:2012xt,Chatrchyan:2013mza,Chatrchyan:2013tia,Khachatryan:2016pev, Aaij:2012mda,Aaij:2015gna,Aaij:2015vua,Aaij:2015zlq,Aad:2016zzw,Aaboud:2017ffb,Aad:2019rou,Aaij:2016qqz,Aad:2016naf,Aaij:2016mgv,Aad:2015auj,Khachatryan:2015oaa, Aaboud:2017soa, Sirunyan:2017wgx,Aad:2011fc,Aad:2013lpa,Aad:2014vwa,Chatrchyan:2012bja,Khachatryan:2015luy,Aaboud:2017dvo,Khachatryan:2016mlc,Aad:2016xcr,ATLAS:2017nah}; see~\cite{NNPDF:2021njg}
for more details.

Concerning the LHC top quark measurements considered in the present
analysis,
they partially overlap, but significantly extend, the top datasets included in
global PDF fits such as \nnpdf ~\cite{NNPDF:2021njg} as well as in SMEFT analyses of the top quark sector~\cite{Ellis:2020unq,Ethier:2021bye}.
Here we discuss in turn the different types of measurements to be included: inclusive $t\bar{t}$ cross sections and differential distributions; $t\bar{t}$ production asymmetries; the $W$-helicity fractions;
 associated top pair production with vector bosons and heavy quarks, including $\bar{t}t Z$, $\bar{t}t W$, $\bar{t}t \gamma$, $\bar{t}t\bar{t}t$, $\bar{t}t\bar{b}b$;
 $t-$ and $s-$channel single top production;
 and associated single top and vector boson production.

 \paragraph{Choice of kinematic distribution.}
 Many of these measurements, in particular those targeting
 top quark pair production, are available
  differentially in several kinematic variables,
 as well as either absolute distributions, or distributions normalised
 to the fiducial cross-section.
We must decide which of the
 available kinematic distributions associated to
 a given measurement should be included in the fit,
 and whether it is more advantageous to consider
 absolute or normalised distributions.
 
Regarding the former, we note that correlations between
 kinematic distributions are in general not available, and only one
 distribution at a time can be included without double-counting 
 (one exception is the ATLAS $t\bar{t}$ lepton+jet measurement
 at $\sqrt{s}=8$ TeV~\cite{Aad:2015mbv}
 where the full correlation matrix is provided). Therefore, wherever
 possible we include the top-pair invariant mass $m_{t\bar{t}}$ distributions
 with the rationale that these have enhanced sensitivity to SMEFT
 operators via energy-growing effects; they also provide
 direct information on the large-$x$ PDFs.
 Otherwise, we consider the top or top-pair rapidity distributions,
 $y_{t}$ and $y_{t\bar{t}}$ respectively, which also provide
 the sought-for information on the large-$x$ PDFs;
 furthermore they benefit from moderate
higher-order QCD and electroweak corrections~\cite{Czakon:2016olj}.

Regarding the choice of absolute versus normalised distributions, we elect to use normalised
distributions together with corresponding fiducial cross-sections throughout.
Normalised distributions are typically more precise that their absolute counterparts, since
experimental and theoretical errors partially cancel out when normalising.
In addition, normalisation does not affect the PDF and EFT sensitivity of the measurement,
provided the fiducial cross section measurements used for normalising are also accounted for.
%
%
From the implementation point of view, since in a normalised measurement one bin is dependent on the others,
we choose to exclude the bin with lowest $m_{t\bar{t}}$ value (the production threshold) to avoid
losing sensitivity arising from the high-energy tails.

\paragraph{Inclusive $t\bar{t}$ production.}

\begin{table}[t]
  \begin{center}
{\fontsize{8pt}{8pt}\selectfont
  \centering
   \renewcommand{\arraystretch}{2}
   \setlength{\tabcolsep}{5pt}
   \begin{tabular}{lcccccc|c|c}
     \toprule \textbf{Exp.}   & $\bf{\sqrt{s}}$ \textbf{(TeV)}    &   \textbf{Channel}
    &  \textbf{Observable} & $\mathcal{L}$ (fb${}^{-1}$) & $\mathbf{n_{\rm dat}}$ & \textbf{Ref.}
     &\textbf{New (PDF fits)}
    &  \textbf{New (SMEFT fits)}\\
    \toprule
    \multirow{1}{*}{
      \bf ATLAS}
      & 7
      & dilepton
      & $\sigma(t\bar{t})$
      & $4.6$
      & 1
      & \cite{ATLAS:2014nxi}
      &
      & ($\checkmark$)\\\midrule
      & 8
      & dilepton
      & $\sigma(t\bar{t})$
      & $20.3$
      & 1
      & \cite{ATLAS:2014nxi}
      &
      & ($\checkmark$)\\
      & 
      & 
      & $1/\sigma d\sigma/dm_{t\bar{t}}$
      & $20.2$
      & 5
      & \cite{Aaboud:2016iot}
      & ($y_{t\bar{t}} \rightarrow m_{t\bar{t}}$)
      & (absolute $\rightarrow$ ratio)\\
      & 
      & $\ell+$jets
      & $\sigma(t\bar{t})$
      & $20.2$
      & 1
      & \cite{ATLAS:2017wvi}
      & $\checkmark$
      & ($\checkmark$)\\
      & 
      & 
      & $1/\sigma d\sigma/d|y_{t}|$
      & $20.3$
      & 4
      & \cite{Aad:2015mbv}
      &
      & ($m_{t\bar{t}}, p_t^T \rightarrow |y_t|, |y_{t\bar{t}}|$)\\
      & 
      &
      & $1/\sigma d\sigma/d|y_{t\bar{t}}|$
      & $20.3$
      & 4
      & \cite{Aad:2015mbv}
      &
      & ($m_{t\bar{t}}, p_t^T \rightarrow |y_t|, |y_{t\bar{t}}|$)\\\midrule
      & 13
      & dilepton
      & $\sigma(t \bar{t})$
      & $36.1$
      & 1
      & \cite{ATLAS:2019hau}
      & $\checkmark$
      & $\checkmark$\\
      &
      & hadronic
      & $\sigma(t \bar{t})$
      & $36.1$
      & 1
      & \cite{ATLAS:2020ccu}
      & $\checkmark$
      & $\checkmark$\\
      &
      &
      & $1/\sigma d^2\sigma/d|y_{t\bar{t}}|dm_{t\bar{t}}$
      & $36.1$
      & 10
      & \cite{ATLAS:2020ccu}
      & $\checkmark$
      & $\checkmark$\\
      & 
      & $\ell+$jets
      & $\sigma(t \bar{t})$
      & $139$
      & 1
      & \cite{ATLAS:2020aln}
      & 
      & ($\checkmark$) \\
      &
      & 
      & $1/\sigma d\sigma/dm_{t\bar{t}}$
      & $36$
      & 8
      & \cite{Aad:2019ntk}
      & $\checkmark$
      & (absolute $\rightarrow$ ratio)\\
     \midrule \midrule
       \multirow{1}{*}{\bf CMS}      & 5
      & combination
      & $\sigma(t \bar{t})$
      & 0.027
      & 1
      & \cite{CMS:2017zpm}
      & 
      & $\checkmark$\\ \midrule
      & 7
      & combination
      & $\sigma(t \bar{t})$
      & $5.0$
      & 1
      & \cite{Spannagel:2016cqt}
      & 
      & $\checkmark$\\\midrule
      & 8
      & combination
      & $\sigma(t \bar{t})$
      & $19.7$
      & 1
      & \cite{Spannagel:2016cqt}
      & 
      & $\checkmark$\\     
      &
      & dilepton
      & $1/\sigma d^2\sigma/dy_{t\bar{t}}dm_{t\bar{t}}$
      & $19.7$
      & 16
      & \cite{Sirunyan:2017azo}
      & ($m_{t\bar{t}}, y_t \rightarrow m_{t\bar{t}}, y_{t\bar{t}}$)
      & \\
      &
      & $\ell+$jets
      & $1/\sigma d\sigma/dy_{t\bar{t}}$
      & $19.7$
      & 9
      & \cite{Khachatryan:2015oqa}
      & 
      & \\\midrule
	& 13
	& dilepton
	& $\sigma(t \bar{t})$
	& $43$
	  & 1
	  & \cite{CMS:2015yky}
      & 
      & ($\checkmark$)\\
      &
      &
       & $1/\sigma d\sigma/dm_{t\bar{t}}$
       & $35.9$
      & 5
      & \cite{Sirunyan:2018ucr}
      & 
      & (absolute $\rightarrow$ ratio)\\   
	& 
	& $\ell+$jets
	& $\sigma(t \bar{t})$
	& $137$
        & 1
        & \cite{CMS:2021vhb}
        & $\checkmark$
       & $\checkmark$\\
      &
      & 
      & $1/\sigma  d\sigma/dm_{t\bar{t}}$
      & $137$
      & 14
      & \cite{CMS:2021vhb}
        & $\checkmark$
       & $\checkmark$\\     
\bottomrule
   \end{tabular}
   \vspace{0.3cm}
   \caption{\small
    The inclusive cross-sections and differential distributions
    for top quark pair production from ATLAS
    and CMS that we consider in this analysis.
    For each dataset, we indicate the experiment,
    the centre of mass energy $\sqrt{s}$, the final-state
    channel, the observable(s) used in the fit, the integrated luminosity $\mathcal{L}$ in inverse femtobarns, and the
    number of data points $n_{\rm dat}$, together
    with the corresponding publication reference.
    In the last two columns, we indicate
    with a $\checkmark$ the datasets
    that are included for the first time here in a global PDF fit
    and in a SMEFT interpretation, respectively.
    The sets marked with brackets have already been included in
    previous studies but here we account for their constraints
    in different manner (e.g. by changing spectra or normalisation), as indicated in the table and in the
    text description.
    \label{tab:input_datasets_toppair}
}
}
\end{center}
\end{table}


A summary of the inclusive $t\bar{t}$ fiducial cross sections
and differential distributions considered in this work is provided
in Table~\ref{tab:input_datasets_toppair}.
We indicate in each case
 the centre of mass energy $\sqrt{s}$, the final-state
 channel, the observable(s) used in the fit, the luminosity, and the
 number of data points $n_{\rm dat}$, together
 with the corresponding publication reference.
 In the last two columns, we indicate
 with a $\checkmark$ the datasets
 that are included for the first time here in a global PDF fit (specifically, those which
 are new with respect to \nnpdf)
 and in a SMEFT interpretation (specifically, in comparison
 with the global fits of~\cite{Ellis:2020unq,Ethier:2021bye}).
 The sets marked with brackets have already been included in
 previous studies, but are implemented here in a different manner
 (e.g. by changing spectra or normalisation), as indicated in the
 table; more details are given in each paragraph of the section. 

 The ATLAS dataset comprises six total cross section measurements
 and five differential normalised cross section measurements.
 Concerning the latter, at $8$ TeV we include three distributions
 from the dilepton and $\ell+$jets channels.
 In the $\ell+$jets channel, several kinematic distributions are available
 together with their correlations.
 Following the dataset selection analysis carried out in~\cite{NNPDF:2021njg},
 we select to fit the $y_t$ and $y_{t\bar{t}}$ distributions as done in the
 \nnpdf baseline.
 At $13$ TeV, we include the normalised cross sections
 differential in $m_{t\bar{t}}$ from the $\ell+$jets 
 and  hadronic channels, with both measurements being considered
 for the first time here in the context of a PDF analysis.

 Moving to CMS, in the inclusive $t\bar{t}$ category
 we consider five total cross section  and four normalised differential cross section
 measurements.
 At $\sqrt{s}=8$ TeV we include differential distributions in the 
 $\ell+$jets and dilepton channels, the latter being doubly differential
 in $y_{t\bar{t}}$ and $m_{t\bar{t}}$.
 The double-differential 8 TeV measurement is part of \nnpdf, but there
 the $(y_{t},m_{t\bar{t}})$ distribution was fitted instead.
 At $13$ TeV, we include the $m_{t\bar{t}}$  distributions
 in the dilepton and $\ell+$jets channels.
 In the latter case we  include the single $m_{t\bar{t}}$ distribution
 rather than the double-differential one in $(m_{t\bar{t}},
 y_{t\bar{t}})$, which is also available, since we find that the
 latter cannot be reproduced by the NNLO SM predictions. We present a
 dedicated analysis of the double-differential distribution in Sect.~5.3.
 As mentioned above, we will study the impact of our dataset selection choices
 by presenting variations of the baseline SM-PDF, fixed-PDF, and SMEFT-PDF analyses
 in the following sections.


 \paragraph{$t\bar{t}$ asymmetry measurements.} The $t\bar{t}$ production asymmetry
 at the LHC is defined as:
 \begin{equation}
   \label{eq:ac}
A_C = \frac{N(\Delta |y| > 0) - N(\Delta |y| < 0)}{N(\Delta |y| > 0) + N(\Delta |y| < 0)} \, ,
\end{equation}
with $N(P)$ being the number of events satisfying the kinematical condition $P$, and $\Delta |y| = |y_t| - |y_{\bar{t}}|$ is the difference between the absolute values of the top quark and anti-top quark rapidities.
The asymmetry $A_C$ can be measured either integrating over the fiducial phase space
or differentially, for example binning in the invariant mass $m_{t\bar{t}}$.
Measurements of $A_C$ are particularly important in constraining certain SMEFT directions,
in particular those associated to the two-light-two-heavy operators.
However, they are unlikely to have an impact on PDF fitting due to their large
experimental uncertainties; nevertheless, with the underlying 
motivation of a comprehensive SMEFT-PDF interpretation
of top quark data, we consider here the $A_C$ measurement as part of our baseline dataset,
and hence study whether or not they also provide relevant PDF information. 
A summary of the asymmetry measurements included in this work is given in Table~\ref{tab:input_datasets_topasymmetries}.

\begin{table}[t]
  \begin{center}
{\fontsize{8pt}{8pt}\selectfont
  \centering
   \renewcommand{\arraystretch}{2}
   \setlength{\tabcolsep}{5pt}
   \begin{tabular}{lcccccc|c|c}
     \toprule \textbf{Experiment}     & $\bf{\sqrt{s}}$\textbf{(TeV)}
     & \textbf{Channel}  &  \textbf{Observable} & $\mathcal{L}$ (fb${}^{-1}$) & $\mathbf{n_{\rm dat}}$ & \textbf{Ref.}   &\textbf{New (PDF fits)}
    &  \textbf{New (SMEFT fits)} \\
    \toprule
      \multirow{2}{*}{ATLAS}
      & 8
      & dilepton
      & $A_C$
      & $20.3$
      & 1
      & \cite{Aad:2016ove}
      & $\checkmark$                                                                     
      &  \\
      & 13
      & $\ell+$jets
      & $A_C$
      & $139$
      & 5
      & \cite{ATLAS:2022waa}
      & $\checkmark$                                                                     
      & $\checkmark$   \\
     \midrule
      \multirow{2}{*}{CMS}
      & 8
      & dilepton
      & $A_C$
      & $19.5$
      & 3
      & \cite{Khachatryan:2016ysn}
      & $\checkmark$                                                                     
      &  \\
      & 13
      & $\ell+$jets
      & $A_C$
      & $138$
      & 3
      & \cite{CMS-PAS-TOP-21-014}
      & $\checkmark$                                                                     
      &  \\
\midrule
      ATLAS \& CMS combination
      & 8
      & $\ell+$jets
      & $A_C$
      & $20$
      & 6
      & \cite{Sirunyan:2017lvd}
      & $\checkmark$                                                                     
      &  \\
\bottomrule
   \end{tabular}
   \vspace{0.3cm}
   \caption{\small Same as Table~\ref{tab:input_datasets_toppair} for the $t\bar{t}$
     asymmetry datasets.
   \label{tab:input_datasets_topasymmetries} \label{tab:asymmetries}
}
}
\end{center}
\end{table}


\paragraph{$W$-helicity fractions.}
The $W$-helicity fractions $F_0, F_L$ and $F_R$ are PDF-independent observables
sensitive to SMEFT corrections, and the dependence of the theory predictions
with respect to the Wilson coefficients can be computed
analytically.
Since these $W$-helicity fractions are PDF-independent observables, to include them
in the joint SMEFT-PDF analysis one has to extend the methodology presented in~\cite{Iranipour:2022iak}
to include in the fit datasets that either lack, or have negligible, PDF sensitivity
and depend only on the EFT coefficients.
We describe how this can be achieved within the \simunet{} framework
in Sect.~\ref{sec:methodology}. 

In Table~\ref{tab:input_datasets_helicityfractions} we list
the LHC measurements of the $W$-helicity fractions considered in the current analysis.
At $\sqrt{s}=8$ TeV we include the combined ATLAS and CMS measurement from~\cite{Aad:2020jvx},
while at $13$ TeV we consider the ATLAS measurement of the $W$-helicities from~\cite{ATLAS:2022bdg},
for the first time in a SMEFT fit.

\begin{table}[h!]
  \begin{center}
{\fontsize{7pt}{7pt}\selectfont
  \centering
   \renewcommand{\arraystretch}{2}
   \setlength{\tabcolsep}{5pt}
   \begin{tabular}{lccccc|c}
     \toprule \textbf{Experiment}     & $\bf{\sqrt{s}}$\textbf{(TeV)}
     &   \textbf{Observable} & $\mathcal{L}$ (fb${}^{-1}$) & $\mathbf{n_{\rm dat}}$ & \textbf{Ref.}   
    &  \textbf{New (SMEFT fits)} \\
     \toprule
           ATLAS \& CMS combination
      & 8
      & $F_0, F_L$
      & $20$
      & 2
      & \cite{Aad:2020jvx}
      &  \\
 \midrule
       ATLAS
      & 13
      & $F_0, F_L$
      & $139$
      & 2
      & \cite{ATLAS:2022bdg}
      & $\checkmark$     \\
\bottomrule
   \end{tabular}
   \vspace{0.3cm}
  \caption{\small Same as Table~\ref{tab:input_datasets_toppair} for
    the $W$-helicity fraction measurements.
    These helicity fractions are PDF-independent and hence
    are only relevant in constraining the EFT coefficients.
    \label{tab:whelicities}
   \label{tab:input_datasets_helicityfractions}
}
}
\end{center}
\end{table}


\paragraph{Associated top quark pair production.}
The next class of observables that we discuss is associated $t\bar{t}$ production with a $Z$- or a $W$-boson (Table~\ref{tab:input_datasets2}), a photon $\gamma$ (Table~\ref{tab:input_datasets2b}), or a heavy quark pair ($t\bar{t}b\bar{b}$ or $t\bar{t}t\bar{t}$, Table~\ref{tab:input_datasets2c}).
While measurements of $t\bar{t}V$ have been considered for SMEFT interpretations,
we use them for the first time here in the context of a PDF determination.
The rare processes $t\bar{t}\gamma$, $t\bar{t}b\bar{b}$, and $t\bar{t}t\bar{t}$ exhibit
a very weak PDF sensitivity and hence in the present analysis their theory predictions
are obtained using a fixed PDF, in the same manner as the $W$-helicity fractions
in Table~\ref{tab:input_datasets_helicityfractions}.

\begin{table}[t]
  \begin{center}
{\fontsize{8pt}{8pt}\selectfont
  \centering
   \renewcommand{\arraystretch}{2}
   \setlength{\tabcolsep}{5pt}
   \begin{tabular}{lccccc|c|c}
     \toprule \textbf{Exp.}   & $\bf{\sqrt{s}}$ \textbf{(TeV)}    
    &  \textbf{Observable} & $\mathcal{L}$ (fb${}^{-1}$) & $\mathbf{n_{\rm dat}}$ & \textbf{Ref.}
     &\textbf{New (PDF fits)}
    &  \textbf{New (SMEFT fits)}\\
    \toprule
    {\bf ATLAS}
    & 8
    & $\sigma(t\bar{t}Z)$
    & $20.3$
    & 1
    &  \cite{Aad:2015eua}
      & $\checkmark$                                                                     
      &  \\
    & 
    & $\sigma(t\bar{t}W)$
    & $20.3$
    & 1
    & \cite{Aad:2015eua}
      & $\checkmark$                                                                     
      &  \\\midrule
    & 13
    & $\sigma(t\bar{t}Z)$
    & $36.1$
    & 1
    &\cite{Aaboud:2019njj}
       & $\checkmark$                                                                     
      &  \\
    & 
    & $1/\sigma d\sigma(t\bar{t}Z)/dp_T^Z$
    & $139$
    & 6
    &  \cite{ATLAS:2021fzm}
      & $\checkmark$                                                                     
      & $\checkmark$ \\
    & 
    & $\sigma(t\bar{t}W)$
    & $36.1$
    & 1
    &  \cite{Aaboud:2019njj}
      & $\checkmark$                                                                     
      &  \\
  \midrule
     {\bf  CMS}
    & 8
    & $\sigma(t\bar{t}Z)$
    & $19.5$
    & 1
    & \cite{Khachatryan:2015sha}
       & $\checkmark$                                                                     
      &  \\
    & 
    & $\sigma(t\bar{t}W)$
    & $19.5$
    & 1
    & \cite{Khachatryan:2015sha}
      & $\checkmark$                                                                     
      &  \\ \midrule
    & 13
    & $\sigma(t\bar{t}Z)$
    & $35.9$
    & 1
    & \cite{Sirunyan:2017uzs}
        & $\checkmark$                                                                     
      &  \\
    & 
    & $1/\sigma d\sigma(t\bar{t}Z)/dp_T(Z)$
    & $77.5$
    & 3
    & \cite{CMS:2019too}
      & $\checkmark$                                                                     
      & (absolute $\rightarrow$ ratio) \\
    & 
    & $\sigma(t\bar{t}W)$
    & $35.9$
    & 1
    &  \cite{Sirunyan:2017uzs}
       & $\checkmark$                                                                     
      &  \\
\bottomrule
   \end{tabular}
   \vspace{0.3cm}
   \caption{\small Same as Table~\ref{tab:input_datasets_toppair} for the measurements
     of top quark production
     in association with a vector boson.
     \label{tab:input_datasets2}
   }
}
\end{center}
\end{table}


\begin{table}[t]
{\fontsize{8pt}{8pt}\selectfont
  \centering
   \renewcommand{\arraystretch}{2}
   \setlength{\tabcolsep}{5pt}
   \begin{tabular}{lccccc|c}
     \toprule \textbf{Experiment}     & $\bf{\sqrt{s}}$\textbf{(TeV)}
     &   \textbf{Observable} & $\mathcal{L}$ (fb${}^{-1}$) & $\mathbf{n_{\rm dat}}$ & \textbf{Ref.}   
    &  \textbf{New (SMEFT fits)} \\
    \toprule
    ATLAS
    & 8
    & $\sigma(t\bar{t}\gamma)$
    & $20.2$
    & 1
    &\cite{Aaboud:2017era}
    & \\
  \midrule
      CMS
    & 8
    & $\sigma(t\bar{t}\gamma)$
    & $19.7$
    & 1
    & \cite{Sirunyan:2017iyh}
    & \\
\bottomrule
   \end{tabular}
   \vspace{0.3cm}
  \caption{\small Same as Table~\ref{tab:input_datasets_toppair} for
     $t\bar{t}$ production in association with a photon.
    \label{tab:input_datasets2b}
    Theory predictions for these observables adopt a fixed PDF.
   }
   }
\end{table}


\begin{table}[t]
  \begin{center}
{\fontsize{8pt}{8pt}\selectfont
  \centering
   \renewcommand{\arraystretch}{2.3}
   \setlength{\tabcolsep}{5pt}
   \begin{tabular}{lcccccc|c}
     \toprule \textbf{Experiment}     & $\bf{\sqrt{s}}$\textbf{(TeV)}
     &   \textbf{Channel} & \textbf{Observable} & $\mathcal{L}$ (fb${}^{-1}$) & $\mathbf{n_{\rm dat}}$ & \textbf{Ref.}   
    &  \textbf{New (SMEFT fits)} \\
    \toprule
    ATLAS
    & 13
    & multi-lepton
    & $\sigma_{\text{tot}}(t\bar{t}t\bar{t})$
    & $139$
    & 1
    &\cite{ATLAS:2020hpj}
    & \\
    & 
    & single-lepton
    & $\sigma_{\text{tot}}(t\bar{t}t\bar{t})$
    & $139$
    & 1
    & \cite{ATLAS:2021kqb}
    & $\checkmark$ \\
        & 
    & $\ell+$jets
    & $\sigma_{\text{tot}}(t\bar{t}b\bar{b})$
    & $36.1$
    & 1
    & \cite{ATLAS:2018fwl}
    & \\
  \midrule
      CMS
     & 13
    & multi-lepton
    & $\sigma_{\text{tot}}(t\bar{t}t\bar{t})$
    & $137$
    & 1
    & \cite{CMS:2019rvj}
    &  \\
     & 
    & single-lepton
    & $\sigma_{\text{tot}}(t\bar{t}t\bar{t})$
    & $35.8$
    & 1
    & \cite{CMS:2019jsc}
    &  \\
     & 
    & all-jet
    & $\sigma_{\text{tot}}(t\bar{t}b\bar{b})$
    & $35.9$
    & 1
    & \cite{CMS:2019eih}
    &  \\
     & 
    & dilepton
    & $\sigma_{\text{tot}}(t\bar{t}b\bar{b})$
    & $35.9$
    & 1
    & \cite{CMS:2020grm}
    & \\
     & 
    & $\ell$+jets
    & $\sigma_{\text{tot}}(t\bar{t}b\bar{b})$
    & $35.9$
    & 1
    & \cite{CMS:2020grm}
    &$\checkmark$ \\
\bottomrule
   \end{tabular}
   \vspace{0.3cm}
  \caption{\small  Same as Table~\ref{tab:input_datasets_toppair} for
    the measurements of $t\bar{t}$ production in association with a heavy quark
    pair.
    Theory predictions for these observables adopt a fixed PDF.
     \label{tab:input_datasets2c}
   }
}
\end{center}
\end{table}


Concerning the $t\bar{t}Z$ and $t\bar{t}W$ data,
from both ATLAS and CMS we use four fiducial cross section measurements at 8 TeV
and 13 TeV,
and one distribution differential in $p_T^Z$ at 13 TeV.
These measurements are particularly interesting to probe SMEFT coefficients that
modify the interactions between the top quark and the electroweak sector.
For top-quark production associated with a photon, we include the fiducial
cross-section measurements from ATLAS and CMS at 8 TeV; also available is a
differential distribution at 13 TeV from ATLAS binned in the photon transverse
momentum $p_T^\gamma$~\cite{Aad:2020axn}, but we exclude this from our analysis because of the difficulty in
producing SMEFT predictions in the fiducial phase space (in the \fitm analysis, its inclusion
is only approximate, and in \smefit this distribution is neglected entirely).
Finally, we include fiducial measurements of
$t\bar{t}b\bar{b}$ and $t\bar{t}t\bar{t}$ production at 13 TeV considering
the data with highest luminosity for each available final state.
%

\paragraph{Inclusive single-top pair production.}
The inclusive single-top production data considered here
and summarised in Table~\ref{tab:input_datasets_3}
comprises measurements of
single-top production in the $t$-channel, which have previously been included
in PDF fits~\cite{Nocera:2019wyk,NNPDF:2021njg}, as well as measurements of single-top production in the $s$-channel, which in the context of PDF studies have been implemented for the first time in this study.
For $t$-channel production, we consider the ATLAS and CMS top and anti-top fiducial cross sections
$\sqrt{s}=7,8,$ and 13 TeV, as well as normalised $y_t$ and $y_{\bar{t}}$ distributions
at 7 and 8 TeV (ATLAS) and at 13 TeV (CMS).
For  $s$-channel production, no differential measurements are available and hence
we consider fiducial cross-sections at 8 and 13 TeV from ATLAS and CMS.

\begin{table}[t]
  \begin{center}
{\fontsize{8pt}{8pt}\selectfont
  \centering
   \renewcommand{\arraystretch}{2}
   \setlength{\tabcolsep}{5pt}
   \begin{tabular}{lcccccc|c|c}
     \toprule \textbf{Exp.}   & $\bf{\sqrt{s}}$ \textbf{(TeV)}    &   \textbf{Channel}
    &  \textbf{Observable} & $\mathcal{L}$ (fb${}^{-1}$) & $\mathbf{n_{\rm dat}}$ & \textbf{Ref.}
     &\textbf{New (PDF fits)}
    &  \textbf{New (SMEFT fits)}\\
    \toprule
    {\bf ATLAS}
    & 7
    & $t$-channel
    & $\sigma_\text{tot}(t)$ 
    & $4.59$
    & 1
    & \cite{ATLAS:2014sxe}
      & ($\checkmark$)                                                                    
      &   $\checkmark$     \\
    & 
    & 
    & $\sigma_\text{tot}(\bar{t})$ 
    & $4.59$
    & 1
    & \cite{ATLAS:2014sxe}
          & ($\checkmark$)                                                                    
      &   $\checkmark$     \\
    & 
    &
    & $1/\sigma d\sigma(tq)/dy_t$ 
    & $4.59$
    & 3
    & \cite{ATLAS:2014sxe}
     &                                                                  
      & $\checkmark$       \\
    & 
    &
    & $1/\sigma d\sigma(\bar{t}q)/dy_{\bar{t}}$ 
    & $4.59$
    & 3
    & \cite{ATLAS:2014sxe}
     &                                                                
    &  $\checkmark$      \\
    \midrule
    & 8
    & $t$-channel
    & $\sigma_{\text{tot}}(t)$
    & $20.2$
    & 1
    & \cite{Aaboud:2017pdi}
      & ($\checkmark$)                                                                    
      &  $\checkmark$      \\
    & 
    & 
    & $\sigma_{\text{tot}}(\bar{t})$
    & $20.2$
    & 1
    & \cite{Aaboud:2017pdi}
          & ($\checkmark$)                                                                    
      &   $\checkmark$     \\
    & 
    &
    & $1/\sigma d\sigma(tq)/dy_t$ 
    & $20.2$
    & 3
    & \cite{Aaboud:2017pdi}
     &
       & ($\checkmark$)\\
    & 
    &
    & $1/\sigma d\sigma(\bar{t}q)/dy_{\bar{t}}$ 
    & $20.2$
    & 3
    & \cite{Aaboud:2017pdi}
     &
       & ($\checkmark$)\\
    & 
    & $s$-channel
    & $\sigma_{\text{tot}}(t + \bar{t})$
    & $20.3$
    & 1
    & \cite{Aad:2015upn} 
     & $\checkmark$
    & \\
    \midrule
    & 13
    & $t$-channel
    & $\sigma_{\text{tot}}(t)$
    & $3.2$
    & 1
    & \cite{Aaboud:2016ymp}  
     & ($\checkmark$)
     & \\
    & 
    &
    & $\sigma_{\text{tot}}(\bar{t})$
    & $3.2$
    & 1
    &  \cite{Aaboud:2016ymp}  
     & ($\checkmark$)
     & \\
    & 
    & $s$-channel
    & $\sigma_{\text{tot}}(t + \bar{t})$ 
    & $139$
    & 1
    & \cite{ATLAS:2022wfk}
     & $\checkmark$
     & $\checkmark$\\
    \midrule
    \midrule
      {\bf CMS}
     & 7
    & $t$-channel
    & $\sigma_{\text{tot}}(t) + \sigma_{\text{tot}}(\bar{t})$  
    & $1.17, 1.56$
    & 1
    & \cite{CMS:2012xhh}
     &
      & $\checkmark$\\
      \midrule
    & 8
    & $t$-channel
    & $\sigma_{\text{tot}}(t)$
    & $19.7$
    & 1
    & \cite{Khachatryan:2014iya}
     & ($\checkmark$)
     & \\
    & 
    &
    & $\sigma_{\text{tot}}(\bar{t})$
    & $19.7$
    & 1
    & \cite{Khachatryan:2014iya}
     & ($\checkmark$)
     & \\
    & 
    & $s$-channel
    & $\sigma_{\text{tot}}(t + \bar{t})$
    & $19.7$
    & 1
    & \cite{Khachatryan:2016ewo}  
     & $\checkmark$
      & \\
      \midrule
    & 13
    & $t$-channel
    & $\sigma_{\text{tot}}(t)$
    & $2.2$
    & 1
    & \cite{Sirunyan:2016cdg} 
        & ($\checkmark$)
     & \\
    &
    & 
    & $\sigma_{\text{tot}}(\bar{t})$
    & $2.2$
    & 1
    & \cite{Sirunyan:2016cdg} 
      & ($\checkmark$)
     & \\
    &
    & 
    & $1/\sigma d\sigma/d|y^{(t)}|$
    & $35.9$
    & 4
    & \cite{Sirunyan:2019hqb}
     & $\checkmark$
     & \\
\bottomrule
   \end{tabular}
   \vspace{0.3cm}
  \caption{\small Same as Table~\ref{tab:input_datasets_toppair} for
    the inclusive single-top production datasets.
     \label{tab:input_datasets_3}
   }
}
\end{center}
\end{table}
\paragraph{Associated single top-quark production with weak bosons.}
Finally, Table~\ref{tab:input_datasets_4}
lists the measurements of associated single-top production with vector bosons
included in  our analysis.
We consider fiducial cross-sections for $tW$ production at 8 and 13 TeV
from ATLAS and CMS in the dilepton and single-lepton final states,
as well as the $tZj$ fiducial cross-section at 13 TeV from ATLAS and CMS in
the dilepton final state.
In addition, kinematical distributions in $tZj$ production from CMS
at 13 TeV are considered
 for the first time here in an EFT fit.
 For these differential distributions, the measurement is presented binned in either $p_T^Z$ or $p_T^t$;
 here, we take the former as default for consistency with the corresponding $t\bar{t}Z$ analysis.

\begin{table}[t]
  \begin{center}
{\fontsize{8pt}{8pt}\selectfont
  \centering
   \renewcommand{\arraystretch}{2}
   \setlength{\tabcolsep}{5pt}
   \begin{tabular}{lcccccc|c}
     \toprule \textbf{Experiment}     & $\bf{\sqrt{s}}$\textbf{(TeV)}
     &   \textbf{Channel} & \textbf{Observable} & $\mathcal{L}$ (fb${}^{-1}$) & $\mathbf{n_{\rm dat}}$ & \textbf{Ref.}   
     &  \textbf{New (SMEFT fits)} \\
    \toprule
      {\bf ATLAS}
      & 8
      & dilepton
      & $\sigma_{\text{tot}}(tW)$
      & $20.3$
      & 1
      & \cite{Aad:2015eto}
      & \\
      & 
      & single-lepton
      & $\sigma_{\text{tot}}(tW)$
      & $20.2$
      & 1
      & \cite{Aad:2020zhd}
      & \\
      \midrule
      & 13
      & dilepton
      & $\sigma_{\text{tot}}(tW)$
      & $3.2$
      & 1
      & \cite{Aaboud:2016lpj}
     &  \\
      & 
      & dilepton
      & $\sigma_{\text{fid}}(tZj)$
      & $139$
      & 1
      & \cite{Aad:2020wog}
      & \\
     \midrule
      {\bf CMS}
      & 8
      & dilepton
      & $\sigma_{\text{tot}}(tW)$
      & $12.2$
      & 1
      & \cite{Chatrchyan:2014tua}
      &   \\
      \midrule
      & 13
      & dilepton
      & $\sigma_{\text{tot}}(tW)$
      & $35.9$
      & 1
      & \cite{Sirunyan:2018lcp}
      & \\
      & 
      & dilepton
      & $\sigma_{\text{fid}}(tZj)$
      & $77.4$
      & 1
      & \cite{Sirunyan:2018zgs}
     & \\
      &
      & dilepton
      & $d\sigma_{\text{fid}}(tZj)/dp_T^t$
      & $138$
      & 3
      & \cite{CMS:2021rvz}
      & $\checkmark$\\
      & 
      & single-lepton
      & $\sigma_{\text{tot}}(tW)$
      & $36$
      & 1
      & \cite{CMS:2021vqm}
      & $\checkmark$\\
\bottomrule
   \end{tabular}
   \vspace{0.3cm}
  \caption{\small Same as Table~\ref{tab:input_datasets_toppair} for
    single-top production in association with an electroweak bosons.
}
   \label{tab:input_datasets_4}
}
\end{center}
\end{table}


\clearpage

\subsection{Dataset selection}
\label{sec:dataselection}

The top quark production measurements listed in
Tables~\ref{tab:input_datasets_toppair}-\ref{tab:input_datasets_4} 
summarise all datasets that have been considered for the present analysis.
In principle, however, some of these may need to be excluded from the baseline 
fit dataset to ensure that the baseline dataset is maximally consistent.
Following the dataset selection procedure adopted
in~\cite{NNPDF:2021njg}, here our baseline dataset is chosen to exclude datasets that may be either internally
inconsistent or inconsistent with other measurements of the same process type.
These inconsistencies can be of experimental origin, for instance due to
unaccounted (or underestimated) systematic errors,
or numerically unstable correlation models, as well as originating in theory,
for example whenever a given process is affected
by large missing higher-order perturbative uncertainties.
Given that the ultimate goal of a global SMEFT analysis, such as the present one,
is to unveil deviations from the SM, one should strive to deploy
objective dataset selection criteria that exclude datasets affected by such
inconsistencies, which are unrelated to BSM physics.

The first step is to run a global SM-PDF fit including all the datasets
summarised in Tables~\ref{tab:input_datasets_toppair}-\ref{tab:input_datasets_4}
(and additionally a fit with the data summarised therein, but with the CMS measurement of 
the differential $t\bar{t}$ cross-section at $13$ TeV in the $\ell+$jets channel replaced with 
the double-differential measurement) and monitor in each case the following two statistical estimators:

\begin{itemize}
\item The total $\chi^2$ per data point and the number of standard deviations $n_\sigma$ by which the value of the
  $\chi^2$ per data point differs from the median of the $\chi^2$ distribution
  for a perfectly consistent dataset,
  \begin{equation}\label{eq:nsigma}
    n_\sigma\equiv
    \frac{|\chi^2-1|}{\sigma_{\chi^2}}=\frac{|\chi^2-1|}{\sqrt{2/n_{\rm dat}}} \, ,
  \end{equation}
  where the $\chi^2$ in this case (and in the rest of the paper unless
  specified) is the experimental $\chi^2$ per
  data point, which is defined as
  \begin{equation}\label{eq:chi2exp}
\chi^2 \equiv \chi^2_{\rm exp}/n_{\rm dat} \,=  \,\frac{1}{n_{\rm
    dat}}\,\sum_{i,j=1}^{n_{\rm dat}} (D_i-T_i^0)\,\,\left({\rm cov}_{\rm exp}^{-1}\right)_{ij}  \,(D_j-T_j^0),
  \end{equation}
  where $T_i^0$ are the theoretical predictions computed with the
  central PDF replica, which is the average over the PDF replicas, and
  the experimental covariance matrix is the one defined for example in
  Eq.~(3.1) of Ref.~\cite{Ball:2022uon}.
  
  Specifically, we single out for further examination datasets for which $n_\sigma\ge  3$
  and $\chi^2 \ge 2$ per data point, where the poor description of the data is unlikely
  to be caused by a statistical fluctuation (note that these conditions relax those given 
  in~\cite{NNPDF:2021njg}, which we hope gives the opportunity for the EFT to account for
  poor quality fits to data, rather than immediately attributing poor fits to inconsistencies).
  The question is then to ascertain whether this poor $\chi^2$ can be explained
  by non-zero EFT coefficients (and in such case it should be retained
  for the fit) or if instead there one can find other explanations,
  such as the ones mentioned above, that justify removing it from the baseline
  dataset.

\item The metric $Z$ defined in Ref.~\cite{Kassabov:2022pps} which
quantifies the stability of the $\chi^2$ with respect
to potential inaccuracies affecting the modelling of the experimental correlations.
The calculation of $Z$ relies  exclusively on the experimental covariance matrix
and is independent of the theory predictions.
A large value of the stability metric $Z$ corresponds to
datasets with an unstable covariance matrix, in the sense that small changes
in the values of the correlations between data points lead to large
increases in the corresponding $\chi^2$.
Here we single out for further inspection  datasets with $Z\ge 4$.

As also described in~\cite{Kassabov:2022pps}, it is possible to regularise
covariance matrices in a minimal manner to assess the impact of these numerical
instabilities at the PDF or SMEFT fit level, and determine how
they affect the resulting pre- and post-fit $\chi^2$.
To quantify whether datasets with large $Z$ distort the fit results in a sizable
manner, one can run fit variants applying this decorrelation procedure such that all datasets
exhibit a value of the $Z$-metric below the threshold. We do not find it necessary
to run such fits in this work.

\end{itemize}

\noindent
In  Tables~\ref{tab:datasets_selection_atlas} and~\ref{tab:datasets_selection_cms}
we list the outcome of such a global SM-PDF fit, where entries that lie above the corresponding threshold values for $\chi^2$, $n_{\sigma}$, or $Z$ are highlighted in boldface.
In the last column, we indicate whether the dataset is flagged.
For the flagged datasets, we carry out the following tests to ascertain whether
it should be retained in the fit:

\begin{itemize}

\item For datasets with  $n_\sigma> 3$ and $Z>4$,  we run a fit variant in which the covariance matrix is regularised.
  If, upon regularisation of the covariance matrix, the PDFs are
  stable and both the $\chi^2$ per data point and the $|n_\sigma|$ decrease to a value below the
  respective thresholds of 2.0 and 3.0, we retain the dataset, else we exclude it.

\item For datasets with $\chi^2>2.0$ and $n_\sigma>3$ we carry out a fit variant
  where this dataset is given a very high weight.
  If in this high-weight fit variant the $\chi^2$ and $n_\sigma$ estimators
  improve to the point that their values lie below the thresholds  without
  deteriorating the description of any of the other datasets included
  the dataset is kept, then the specific measurement is not
  inconsistent, it just does not have enough weight compared to the
  other datasets. See Ref.~\cite{NNPDF:2021njg} for a detailed
  discussion on the size of the weight depending on the size of the
  dataset. 
\end{itemize}

\begin{table}[htbp]
  \begin{center}
{\fontsize{8pt}{8pt}\selectfont
  \centering
   \renewcommand{\arraystretch}{2}
   \setlength{\tabcolsep}{5pt}
   \begin{tabularx}{\textwidth}{ccXccccl}
  \toprule
  \textbf{Experiment}
  &\textbf{$\sqrt{s}$ (TeV)}
  &\textbf{Observable, Channel}
  & \textbf{$n_{\rm dat}$}
  & $\chi^2_{\rm exp}/n_{\rm dat}$
  & \textbf{$n_\sigma$}
  & $Z$
  & \textbf{flag}\\
  \midrule
  {\bf ATLAS }
  & 7
  & $\sigma_{t\bar{t}}^{\rm tot}$, dilepton
  & 1
  & {\bf 4.63}  
  & 2.57
  & 1.00
  & no  \\
  & 
  & $\sigma^{\rm tot}(t)$, $t$-channel
  & 1
  &  0.76 
  & -0.17
  & 1.00
  & no  \\
  & 
  & $\sigma^{\rm tot}(\bar{t})$, $t$-channel
  & 1
  &  0.29
  & -0.50
  & 1.00
  & no  \\
  & 
  & $1/\sigma d(tq)/dy_t$, $t$-channel
  & 3
  & 0.97 
  & -0.04
  & 1.28
  & no  \\
  & 
  & $1/\sigma d(\bar{t}q)/dy_{\bar{t}}$, $t$-channel
  & 3
  & 0.06 
  & -1.15
  & 1.39
  & no  \\ \midrule
  & 8
    & $\sigma_{t\bar{t}}^{\rm tot}$, dilepton
  & 1
  & 0.03
  & -0.69
  & 1.00
  & no \\
    &
    & $1/\sigma d\sigma/dm_{t\bar{t}}$, dilepton
  & 5
  & 0.29
  & -1.12
  & 1.61
  & no \\
    &
    & $\sigma_{t\bar{t}}^{\rm tot}$, $\ell+$jets 
  & 1
  & 0.28 
  & -0.51
  & 1.00 
  & no \\
     &
     & $1/\sigma d\sigma/d|y_t|$, $\ell+$jets
  & 4
  & {\bf 2.86} 
  & 2.63 
  & 1.65
  & no \\
  &
     & $1/\sigma d\sigma/d|y_{t\bar{t}}|$, $\ell+$jets 
  & 4
  & {\bf 3.37}
  & {\bf 3.35}
  & 2.19 
  & {\bf yes (kept)} \\
  &
  & $A_C$, dilepton
  & 1
  & 0.67
  & -0.23
  & 1.00
  & no\\
  & 
  & $\sigma(t\bar{t}Z)$
  & 1
  & 0.23
  & -0.54
  & 1.00
  & no \\
  & 
  & $\sigma(t\bar{t}W)$
  & 1
  & {\bf 2.44}
  & 1.01
  & 1.00
  &  no \\
  & 
  & $\sigma^{\rm tot}(t + \bar{t})$, $s$-channel
  & 1
  & 0.21 
  & -0.56
  & 1.00
  & no  \\
  & 
  & $\sigma^{\rm tot}(t W)$, dilepton
  & 1
  & 0.54 
  & -0.33
  & 1.00
  & no  \\
  & 
  & $\sigma^{\rm tot}(t W)$, single-lepton
  & 1
  & 0.71  
  & -0.21
  & 1.00
  & no  \\
  \midrule
 &13
  & $\sigma_{t\bar{t}}^{\rm tot}$, dilepton
  & 1
  & 1.41
  & 0.29
  & 1.00
  & no \\
     &
     & $\sigma_{t\bar{t}}^{\rm tot}$, hadronic
  & 1
  & 0.23 
  & -0.54
  & 1.000
  & no \\
     &
     & $1/\sigma d^2\sigma/d|y_{t\bar{t}}|dm_{t\bar{t}}$, hadronic 
  & 10
  & 1.95 
  & 2.12
  & 2.33
  & no\\ 
     &
     & $\sigma_{t\bar{t}}^{\rm tot}$, $\ell+$jets 
  & 1
  & 0.50
  & -0.35
  & 1.00 
  & no\\ 
     &
     & $1/\sigma d\sigma/dm_{t\bar{t}}$, $\ell+$jets
  & 8
  & 1.83
  & 1.66
  & {\bf 7.61}
  & no\\
  &
  & $A_C$, $\ell+$jets
  & 5
  & 0.99
  & -0.02
  & 1.41
  & no\\
  & 
  & $\sigma(t\bar{t}Z)$
  & 1
  & 0.75
  & -0.18
  & 1.00
  &  no \\
  &
  & $1/\sigma d\sigma(t\bar{t}Z)/dp_T(Z)$
  & 5
  & 1.93
  & 1.47
  & 2.27 
  & no\\
  &
  & $\sigma(t\bar{t}W)$
  & 1
  & 1.43
  & 0.30
  & 1.00 
  & no\\
  & 
  & $\sigma^{\rm tot}(t)$, $t$-channel
  & 1
  & 0.72  
  & -0.20
  & 1.00
  & no  \\
  & 
  & $\sigma^{\rm tot}(\bar{t})$, $t$-channel
  & 1
  & 0.39 
  & -0.43
  & 1.00
  & no  \\
  & 
  & $\sigma^{\rm tot}(t + \bar{t})$, $s$-channel
  & 1
  & 0.70
  & -0.21
  & 1.00
  & no  \\
  & 
  & $\sigma^{\rm tot}(t W)$, dilepton
  & 1
  & 1.15
  & 0.36
  & 1.00
  & no  \\
  \bottomrule
   \end{tabularx}
   \vspace{0.3cm}
   \caption{\small For the ATLAS measurements that we consider in this work,
     we list the outcome of a global SM-PDF fit with all
     measurements listed in
     Tables~\ref{tab:input_datasets_toppair}-\ref{tab:input_datasets_4}
     included.
     We display for each dataset the  number of
data points, the $\chi^2$ per data point (Eq.~\eqref{eq:chi2exp}, the number of
standard deviations $n_\sigma$ (Eq.~\eqref{eq:nsigma}), and the
stability metric $Z$  defined in~\cite{Kassabov:2022pps}.
The entries that lie above the corresponding threshold values
are  highlighted in boldface,
In the
last column, we indicate whether the dataset is flagged and
is either kept or removed.
See text for more details.
   \label{tab:datasets_selection_atlas}
}
}
\end{center}
\end{table}

\begin{table}[htbp]
{\fontsize{8pt}{8pt}\selectfont
  \centering
   \renewcommand{\arraystretch}{2}
   \setlength{\tabcolsep}{5pt}
   \begin{tabularx}{\textwidth}{ccXccccl}
  \toprule
  \textbf{Experiment}
  &\textbf{$\sqrt{s}$ (TeV)}
  &\textbf{Observable}
  & \textbf{$n_{\rm dat}$}
  & $\chi^2_{\rm exp}/n_{\rm dat}$
  & \textbf{$n_\sigma$}
  & $Z$
  & \textbf{flag} \\
  \midrule
  CMS
  & 5
  & $\sigma_{t\bar{t}}^{\rm tot}$, combination
  & 1
  & 0.56
  & -0.31
  & 1.00
  & no \\
  & 7
  & $\sigma_{t\bar{t}}^{\rm tot}$, combination
  & 1
  & 1.08
  & 0.06
  & 1.00
  & no \\
  & 
  & $\sigma^{\rm tot}(t) + \sigma^{\rm tot}(\bar{t})$, $t$-channel
  & 1
  & 0.72
  & -0.20
  & 1.00
  & no \\
  & 8
  & $\sigma_{t\bar{t}}^{\rm tot}$, combination
  & 1
  & 0.27 
  & -0.52
  & 1.00
  & no \\
   & 
  & $1/\sigma d^2\sigma/dy_{t\bar{t}}dm_{t\bar{t}}$, dilepton 
  & 16
  & 0.98
  & -0.06
  & 2.33
  & no \\
   & 
  & $1/\sigma d\sigma/dy_{t\bar{t}}$, $\ell+$jets 
  & 9
  & 1.15
  & 0.31
  & 1.63
  & no \\
  & 
  & $A_C$, dilepton
  & 3
  & 0.05
  & -1.16
  & 1.16
  & no \\
  & 
  & $\sigma(t\bar{t}Z)$
  & 1
  & 0.47
  & -0.37
  & 1.00
  & no\\
  & 
  & $\sigma(t\bar{t}W)$
  & 1
  & {\bf 2.27}
  & 0.90
  & 1.00
  &  no\\
  & 
  & $\sigma^{\rm tot}(t)$, $t$-channel
  & 1
  & 0.01
  & -0.70
  & 1.00
  & no \\
  & 
  & $\sigma^{\rm tot}(\bar{t})$, $t$-channel
  & 1
  & 0.09
  & -0.64
  & 1.00
  & no \\
  & 
  & $\sigma^{\rm tot}(t + \bar{t})$, $s$-channel
  & 1
  & 1.11 
  & 0.08
  & 1.00
  & no \\
  & 
  & $\sigma^{\rm tot}(t W)$, dilepton
  & 1
  & 0.38
  & -0.44
  & 1.00
  & no \\
   & 13
  & $\sigma_{t\bar{t}}^{\rm tot}$, dilepton  
  & 1
  & 0.06 
  & -0.66
  & 1.00
  & no \\
   & 
  & $1/\sigma d\sigma/dm_{t\bar{t}}$, dilepton  
  & 5
  & {\bf 2.49}
  & 2.36 
  & 1.61
  & no\\
   & 
  & $\sigma_{t\bar{t}}^{\rm tot}$, $\ell+$jets channel
  & 1
  & 0.22
  & -0.55
  & 1.00
  & no \\
   & 
  & $1/\sigma d\sigma/dm_{t\bar{t}}$, $\ell+$jets
  & 14
  & 1.41 
  & 1.08
  & {\bf 4.57}
  & no \\
   & 
  & $1/\sigma d\sigma/dm_{t\bar{t}}dy_{t}$, $\ell+$jets
  & 34
  & {\bf 6.43} 
  & 22.4
  & 3.88
  & {\bf yes (excl)} \\
  & 
  & $A_C$, $\ell+$jets
  & 3
  & 0.29
  & -0.87
  & 1.00
  & no \\
  & 
  & $\sigma(t\bar{t}Z)$
  & 1
  & 1.24
  & 0.17
  & 1.00
  &  no \\
  & 
  & $1/\sigma d\sigma(t\bar{t}Z)/dp_T(Z)$
  & 3
  & 0.59
  & -0.50
  & 1.28
  &  no\\
  & 
  & $\sigma(t\bar{t}W)$
  & 1
  & 0.66
  & -0.24
  & 1.00
  &  no \\
  & 
  & $\sigma^{\rm tot}(t)$, $t$-channel
  & 1
  & 0.88
  & -0.08
  & 1.00
  &  no \\
  & 
  & $\sigma^{\rm tot}(\bar{t})$, $t$-channel
  & 1
  & 0.13 
  & -0.62
  & 1.00
  &  no \\
  & 
  & $1/\sigma d\sigma/d|y^{(t)}|$, $t$-channel
  & 4
  & 0.38
  & -0.88
  & 1.70
  &  no \\
  & 
  & $\sigma^{\rm tot}(tW)$, dilepton
  & 1
  & 0.43  
  & -0.40
  & 1.00
  &  no \\
  & 
  & $\sigma^{\rm tot}(tW)$, single-lepton
  & 1
  & {\bf 2.84} 
  & 1.30
  & 1.00
  &  no \\
  \midrule
  ATLAS-CMS combination
  & 8
  & $A_C$, $\ell$+jets
  & 6
  & 0.602
  & -0.69
  & 1.65
  & no \\
  \bottomrule
   \end{tabularx}
   \vspace{0.3cm}
  \caption{\small Same as Table \ref{tab:datasets_selection_atlas} for the
    CMS and combined ATLAS-CMS datasets. \textit{Note carefully:} the row
    corresponding to the CMS doubly-differential distribution at 13 TeV in the $\ell+$jets
    channel comes from a separate fit, where the corresponding 1D distribution is replaced
    by this dataset.
   \label{tab:datasets_selection_cms}
}
}
\end{table}


    From the analysis of Tables~\ref{tab:datasets_selection_atlas} and~\ref{tab:datasets_selection_cms},
    one finds that only two datasets in the inclusive
    top quark pair production (lepton+jets final state) category are flagged as potentially
    problematic: the ATLAS $|y_{t\bar{t}}|$ distribution at 8 TeV
    and the CMS double-differential distributions in $m_{t\bar{t}}$ and $y_{t}$
    at 13 TeV.
    The first of these was already discussed in the \nnpdf
    analysis~\cite{NNPDF:2021njg}. It was observed that each of the
    four distributions measured by ATLAS and presented in
    Ref.~\cite{Aad:2015mbv} behave somewhat differently upon being
    given large weight. The $\chi^2$ of all distributions significantly improves when given
large weight. However, while for the top transverse momentum and top pair invariant mass distributions this improvement is accompanied by a rather significant deterioration of the global fit quality, in the case of the top and
top pair rapidity distributions the global fit quality is very similar and only the description of jets
deteriorates moderately. The rapidity distributions thus remain
largely compatible with the rest of the dataset, hence they are kept.

    %
    Also shown in one row of Table~\ref{tab:datasets_selection_cms} is the 
    fit-quality information for the CMS double-differential distribution at 13 TeV in the $\ell+$jets 
    channel, from a separate fit wherein the CMS single differential distribution at 13 TeV in the 
    $\ell+$jets channel is replaced by this dataset. We find that the
    2D set is described very poorly, with a $\chi^2=6.43$,
    corresponding to a $22\sigma$ deviation from the median of the $\chi^2$ distribution
  for a perfectly consistent dataset. To investigate this further, we performed a weighted fit; however, we find that the $\chi^2$
    improves only moderately (from $\chi^2$=
    6.43 to $\chi^2$ = 4.56) and moreover the $\chi^2$-statistic of the other datasets deteriorates
    significantly (with total $\chi^2$ jumping from 1.20 to 1.28). The test
    indicates that the double-differential distribution is both incompatible
    with the rest of the data and also internally inconsistent given the
    standard PDF fit. Hence we exclude this dataset from our baseline and
    include instead the single-differential distribution in $m_{t\bar{t}}$,
    which is presented in the same publication~\cite{CMS:2021vhb} and is
    perfectly described in the baseline fit.  To check whether the
    incompatibility we observe in the double-differential distribution can be
    cured by the inclusion of SMEFT corrections, we will run a devoted analysis
    presented in Sect.~5.3.

\subsection{Theoretical predictions}
\label{sec:theory}

In this section we describe the calculation settings adopted
for the SM and SMEFT cross-sections used in the present analysis.

\paragraph{SM cross-sections.}
Theoretical predictions for SM cross-sections are evaluated at NNLO in perturbative QCD, whenever
available, and at NLO otherwise.
Predictions accurate to NLO QCD are obtained in terms of fast interpolation grids from {\sc\small MadGraph5\_aMC@NLO}~\cite{Frederix:2018nkq,Alwall:2014hca}, 
interfaced to {\sc\small APPLgrid}~\cite{Carli:2010rw} or {\sc\small FastNLO}~\cite{Kluge:2006xs,Wobisch:2011ij,Britzger:2012bs} together 
with {\sc\small aMCfast}~\cite{Bertone:2014zva} and {\sc\small APFELcomb}~\cite{Bertone:2016lga}.
Wherever available, NNLO QCD corrections to matrix elements are  implemented  by multiplying the NLO predictions by bin-by-bin $K$-factors, see Sect.~2.3 in~\cite{Ball:2014uwa}.
The top mass is set to $m_t = 172.5\ \text{GeV}$ for all processes considered.

In the case of inclusive $t\bar{t}$ cross sections and charge asymmetries,
a dynamical scale choice of $\mu_R = \mu_F = H_T/4$
is adopted, where $H_T$ denotes the sum of the transverse masses of the top and anti-top,
following the recommendations of Ref.~\cite{Czakon:2016dgf}.
This scale choice ensures that
the ratio of fixed order NNLO predictions to the NNLO+NNLL ones is minimised,
allowing us to neglect theory uncertainties associated to missing higher orders beyond NNLO. 
To obtain the corresponding
NNLO $K$-factors, we use the {\sc\small HighTEA} public software~\cite{hightea}, 
an event database for distributing and analysing the results of fixed order NNLO
calculations for LHC processes.
The NNLO PDF set used in the computation of these $K$-factors is either 
{\tt NNPDF3.1} or \nnpdf, depending on whether a given dataset was already included
in the \nnpdf global fit or not, respectively.

For associated $t\bar{t}$ and $W$ or $Z$ production, dedicated fast NLO grids  have been generated.
Factorisation and renormalisation scales are fixed to $\mu_F = \mu_R = m_t + \frac{1}{2}m_V$, where $m_V = m_W, m_Z$ is the mass of the associated weak boson,
as appropriate.
This scale choice follows the recommendation of Ref.~\cite{Kulesza:2018tqz} and minimises the ratio of the NLO+NLL over the fixed-order NLO prediction.
We supplement the predictions for the total cross section for associated $W$ and $Z$-production at 13 TeV with NLO+NNLL QCD $K$-factors taken from Table 1 of~\cite{Kulesza:2018tqz}.
On the other hand, the $t\bar{t}\gamma$, $t\bar{t}t\bar{t}$ and $t\bar{t}b\bar{b}$ data are implemented as PDF independent observables, and the corresponding theory predictions are taken directly from the relevant experimental papers in each case.

The evaluation of theoretical predictions for single top production follows~\cite{Nocera:2019wyk}.
Fast NLO interpolation grids are generated for both  $s$- and $t$-channel
single top-quark and top-antiquark datasets in the 5-flavour scheme,  with fixed
factorisation and renormalisation scales set to $m_t$.
Furthermore, 
for the $t$-channel production we include the NNLO QCD corrections to
both total and differential cross sections~\cite{Berger:2017zof}.
When the top decay is calculated, it is done in the narrow-width approximation, under which the QCD corrections to the top-(anti)quark production
and the decay are factorisable and the full QCD corrections are approximated by the vertex corrections.

\paragraph{SMEFT cross-sections.}
SMEFT corrections to SM processes are computed both at LO and at NLO in QCD, 
and both at the linear and the quadratic level in the EFT expansion.
Flavour assumptions follow the LHC TOP WG prescription of~\cite{Aguilar-Saavedra:2018ksv}
which were also used in the recent {\sc\small SMEFiT} analysis~\cite{Ethier:2021bye}.
The flavour symmetry group is
given by $U(3)_l \times U(3)_e \times U(3)_d \times U(2)_u \times U(2)_q$, i.e. we single out operators that contain
top quarks (right-handed $t$ and $SU(2)$ doublet $Q$).
This also means that one works in a five-flavour scheme in which the
only massive fermion in the theory is the top.
As far as the electroweak input scheme is concerned, we work in the $m_W$-scheme, meaning that the $4$
electroweak inputs are
$\{m_W, G_F, m_h, m_Z\}$.

At dimension-six, SMEFT operators modify the SM Lagrangian as:
\begin{equation}
    \label{eq:lagrangian}
    \mathcal{L}_{\rm SMEFT} = \mathcal{L}_{\rm SM} + \sum_{n=1}^{N} \frac{c_n}{\Lambda^2} \mathcal{O}_n \, ,
\end{equation}
where $\Lambda$ is the UV-cutoff energy scale, $\{\mathcal{O}_{n} \}$ are dimension-six operators, and
$\{ c_n \}$ are Wilson coefficients.
The $25$ operators considered for this study 
are listed in Table~\ref{tab:ops} in the Warsaw basis~\cite{Grzadkowski:2010es}.
In this work we neglect renormalisation group effects on the Wilson coefficients~\cite{Aoude:2022aro}. 
For hadronic data, i.e. for proton-proton collisions, which are the only data affected by the SMEFT in this study,
 the linear effect of the $n$-th SMEFT operator on a theoretical prediction can be quantified by:
\begin{equation}
  \label{eq:mult_k_fac_app1}
   R_{\rm SMEFT}^{(n)} \equiv \displaystyle \left( {\cal L}_{ ij}^{\rm NNLO} \otimes d\widehat{\sigma}_{ij,{\rm SMEFT}}^{(n)}\right)
 \big/ \left( {\cal L}_{ ij}^{\rm NNLO} \otimes d\widehat{\sigma}_{ij,{\rm SM}} \right) \, , \quad
 n=1\,\ldots, N \, ,
\end{equation}
where $i, j$ are parton indices, ${\cal L}_{ ij}^{\rm NNLO}$ is the NNLO partonic luminosity defined as 
\begin{equation}
\label{eq:lumidef}
\mathcal{L}_{ij}(\tau,M_X) = \int_{\tau}^1 \frac{d x}{x}~f_i (x,M_X) f_j (\tau/x,M_X) ~, \quad \tau=M_X^2/s,
\end{equation}
$d\widehat{\sigma}_{ij,{\rm SM}}$ the bin-by-bin partonic SM cross section, and $d\widehat{\sigma}_{ij,{\rm SMEFT}}^{(n)}$
the corresponding partonic cross section associated to the interference between 
$\mathcal{O}_n$ and the SM amplitude $\mathcal{A}_{\rm SM}$ when setting $c_n = 1$. This value of $c_n$ 
is only used to initialize the potential contributions of the
SMEFT operator; the effective values of the Wilson coefficient 
are found after the fit is performed. 
Quadratic effects of the interference between the $n$-th and $m$-th
SMEFT operators can be evaluated as
\begin{equation}
  \label{eq:mult_k_fac_app2}
  R_{\rm SMEFT}^{(n,m)} \equiv \displaystyle \left( {\cal L}_{ ij}^{\rm NNLO} \otimes d\widehat{\sigma}_{ij,{\rm SMEFT}}^{(n,m)}\right)
  \big/ \left( {\cal L}_{ ij}^{\rm NNLO} \otimes d\widehat{\sigma}_{ij,{\rm SM}} \right) \, , \quad
 n,m=1\,\ldots, N \, ,
\end{equation}
with the bin-by-bin partonic cross section
$d\widehat{\sigma}_{ij,{\rm SMEFT}}^{(n,m)}$ now being evaluated from the squared amplitude $\mathcal{A}_n\mathcal{A}_m$
associated to the operators $\mathcal{O}_n$ and $\mathcal{O}_m$ when $c_n = c_m = 1$.

The computation of the SMEFT contributions is
performed numerically with the FeynRules~\cite{Alloul:2013bka} model SMEFTatNLO~\cite{Degrande:2020evl}, which allows one to include NLO QCD corrections to the observables. The obtained cross sections are then combined in so-called \textit{BSM factors} by taking the ratio with the respective SM cross sections, in order to 
produce $R_{\rm SMEFT}^{(n)}$ and $R_{\rm SMEFT}^{(n,m)}$, respectively the linear and quadratic corrections.

With these considerations, we can account for SMEFT effects in our theoretical predictions
by mapping the SM prediction $T^{\rm SM}$ to 
\begin{equation}
\label{eq:theory_k_fac_app3}
 T = T^{\rm SM} \times  K(\{c_n\})\, , 
\end{equation}
with 
\begin{equation}
\label{eq:theory_k_fac}
    K(\{c_n\}) = 1+\sum_{n=1}^{N} c_n R_{\rm SMEFT}^{(n)}
    +\sum_{1 \leq n \leq m \leq N} c_{nm} R_{\rm SMEFT}^{(n,m)} \, ,
\end{equation}
with $c_{nm} = c_n c_m$. Eq. (\ref{eq:theory_k_fac_app3}) is at the centre of the \simunet{} methodology, which we discuss in Sect.~\ref{sec:methodology}.

\section{Fitting methodology}
\label{sec:methodology}

In this work, the joint determination of the PDFs and the EFT coefficients is
carried out using the \simunet{} methodology, first presented
in~\cite{Iranipour:2022iak}, which is substantially extended in this work.
The core idea of  \simunet{}  is to incorporate the Wilson coefficients into the optimisation problem
that enters the PDF determination, by accounting explicitly for their dependence
in the theoretical predictions used to fit the PDFs.
Specifically, the neural network model used in the SM-PDF fits of \nnpdf is
augmented with an additional layer, which encodes the dependence of the theory
predictions entering the fit on the Wilson coefficients.

In this section, first we provide an overview of the \simunet{}
methodology, highlighting the new features that have been implemented
for the present study.
%

\subsection{\simunet{} overview}
\label{sec:simunet}

The  \simunet{}~\cite{Iranipour:2022iak}
methodology extends the  \nnpdf framework~\cite{NNPDF:2021njg, NNPDF:2021uiq} to
account for the EFT dependence (or, in principle, any parametric dependence) 
of the theory cross-sections entering the
PDF determination.
This is achieved by adding an extra layer to the \nnpdf neural network to
encapsulate the dependence of the theory predictions on the EFT coefficients,
including the free parameters in the general optimisation procedure. This
results in a simultaneous fit of the PDF as well as EFT coefficients to the
input data.
As in the NNPDF methodology, the error uncertainty estimation makes use of the Monte
Carlo replica method, which yields an uncertainty estimate on both PDF and EFT
parameters.  We discuss the limitations of this method in App.~\ref{app:quad}.

The SM theoretical observables are encoded using interpolation grids,
known as {\tt FK}-tables~\cite{Ball:2010de,Ball:2012cx,Bertone:2016lga}, which 
encode the contribution of both the DGLAP evolution
and the hard-scattering matrix elements and interface it with the
initial-scale PDFs in a fast and efficient way. 
%

\begin{figure}[t]
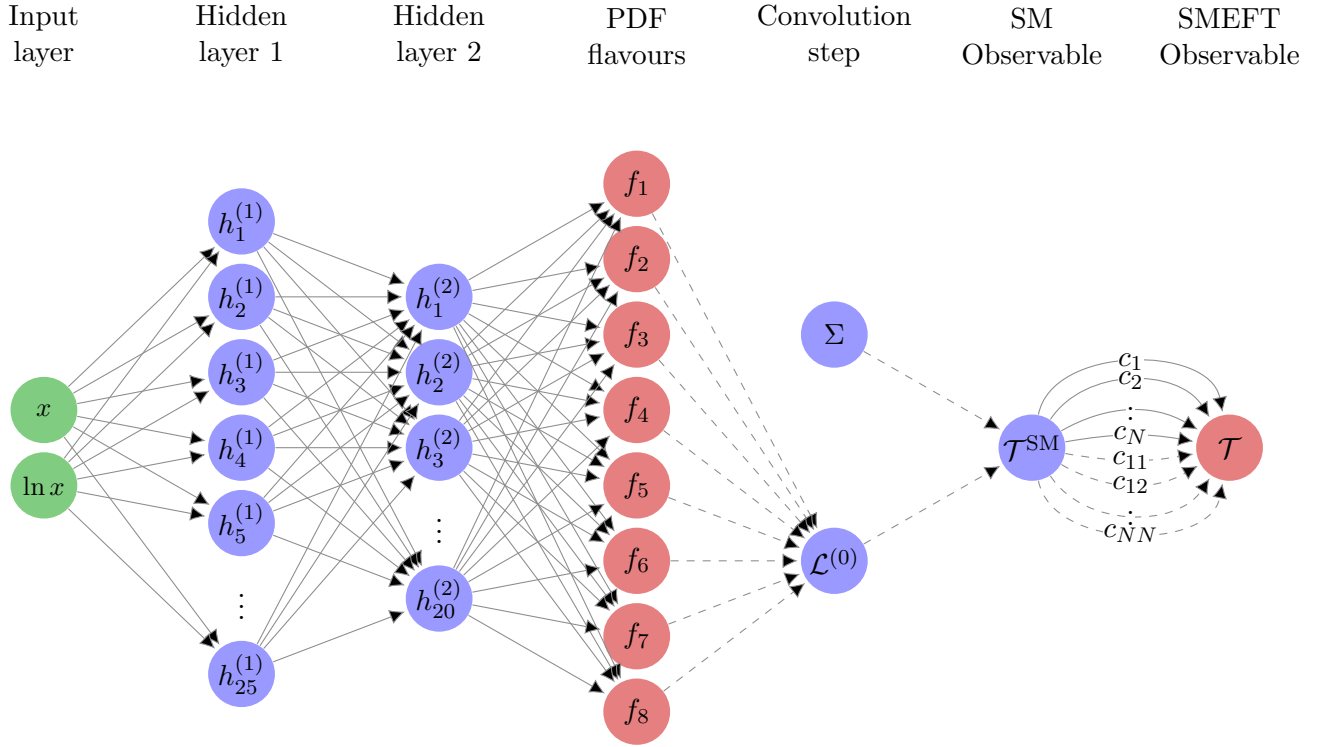

  \centering
  \begin{neuralnetwork}[height=10, layerspacing=26mm, nodesize=25pt]
    \newcommand{\x}[2]{\IfEqCase{#2}{{1}{\small $x$}{2}{\small $\ln{x}$}}}
    \newcommand{\y}[2]{$f_#2$}
    \newcommand{\basis}[2]{\ifnum #2=4 \(\mathcal{L}^{(0)}\) \else $\Sigma$ \fi}
    \newcommand{\hfirst}[2]{\ifnum #2=7 $h^{(1)}_{25}$ \else $h^{(1)}_#2$ \fi}
    \newcommand{\hsecond}[2]{\ifnum #2=5 $h^{(2)}_{20}$ \else $h^{(2)}_#2$ \fi}
    \newcommand{\standardmodel}[2] {$\mathcal{T}^\text{SM}$}
    \newcommand{\eft}[2] {$\mathcal{T}$}
    \newcommand{\co}[4]{$c_1$}
    \newcommand{\ct}[4]{$c_2$}
    \newcommand{\cnn}[4]{$c_{N-1}$}
    \newcommand{\cn}[4]{$c_{N}$}
    \newcommand{\vd}[4]{$\vdots$}
    \newcommand{\FK}[2]{$\sigma$}
    \newcommand{\convolve}[4]{$\otimes$}
    \newcommand{\ci}[4]{$c_1$}
    \newcommand{\cj}[4]{$c_2$}
    \newcommand{\ck}[4]{$c_N$}
    \newcommand{\calpha}[4]{$c_{11}$}
    \newcommand{\cbeta}[4]{$c_{12}$}
    \newcommand{\cgamma}[4]{$c_{NN}$}
    \inputlayer[count=2, bias=false, title=Input\\layer, text=\x]
    \hiddenlayer[count=7, bias=false, title=Hidden\\layer 1, text=\hfirst, exclude={6}]\linklayers[not to={6}]
    \hiddenlayer[count=5, bias=false, title=Hidden\\layer 2, text=\hsecond, exclude={4}]\linklayers[not from={6}, not to={4}]
    \outputlayer[count=8, title=PDF\\flavours, text=\y] \linklayers[not from={4}]
    \hiddenlayer[count=4, bias=false, text=\basis, title=Convolution\\step, exclude={2,3}]\linklayers[not to={1,2,3}, style={dashed}]
    \hiddenlayer[count=1, bias=false, title=SM\\Observable, text=\standardmodel]\linklayers[not from={2,3}, style={dashed}]
    \outputlayer[count=1, bias=false, text=\eft, title=SMEFT \\ Observable]
    \link[from layer = 5, to layer = 6, from node = 1, to node = 1, style={bend left=79}, label={\ci}]
    \link[from layer = 5, to layer = 6, from node = 1, to node = 1, style={bend left=57}, label={\cj}]
    \link[from layer = 5, to layer = 6, from node = 1, to node = 1, style={bend left=30}, label={\vd}]
    \link[from layer = 5, to layer = 6, from node = 1, to node = 1, style={bend left=10}, label={\ck}]
    \link[from layer = 5, to layer = 6, from node = 1, to node = 1, style={dashed, bend right=10}, label={\calpha}]
    \link[from layer = 5, to layer = 6, from node = 1, to node = 1, style={dashed, bend right=30}, label={\cbeta}]
    \link[from layer = 5, to layer = 6, from node = 1, to node = 1, style={dashed, bend right=57}, label={\vd}]
    \link[from layer = 5, to layer = 6, from node = 1, to node = 1, style={dashed, bend right=79}, label={\cgamma}]

    \path (L1-5) -- node{$\vdots$} (L1-7);
    \path (L2-3) -- node{$\vdots$} (L2-5);
  \end{neuralnetwork}
  \caption{Schematic representation of the \simunet{} architecture for a
    general observable.
    Trainable weights are represented by solid arrows, and
  non-trainable weights by dashed arrows.  Through a
  forward pass across the network, the inputs ($x$-Bjorken and its logarithm, in green) proceed through
  2 hidden layers (in blue) to output the PDFs $f_{1}, \cdots, f{_8}$ (in red) at
  the initial parametrisation scale $Q_0$.
  For each of the experimental observables entering the fit, these
  PDFs are combined into a partonic luminosity $\mathcal{L}^{(0)}$ at $Q_0$,
  which is then convolved with precomputed {\tt FK}-tables $\Sigma$ to obtain the SM
  theoretical prediction $\mathcal{T}^\text{SM}$.
  Subsequently, the effects of the EFT coefficients $c_i$
  are accounted for by means of an extra layer. In linear EFT fits this layer simplifies 
  to just $N$ trainable weights to account for each coefficient, and in quadratic EFT fits a set of 
  $N(N+1)/2$ non-trainable weights has to be added to account for the EFT-EFT interference. The forward-pass 
  of this layer results in the final prediction for the observable $\mathcal{T}$ entering the SMEFT-PDF fit.
  By setting the weights in the EFT layer to zero, one recovers the SM-PDF case.
  By freezing the PDF-related weights in the network architecture, one can carry out a fixed-PDF EFT
  determination or include in the joint  SMEFT-PDF fit observables whose PDF dependence can be neglected.
  \label{fig: architecture}}
\end{figure}

The simultaneous fit is represented as a neural network using the
\texttt{Tensorflow}~\cite{tensorflow2015:whitepaper} and
\texttt{Keras}~\cite{chollet2015keras} libraries. The architecture is
schematically represented in Fig.~\ref{fig: architecture}.
Trainable weights are represented by solid arrows, and
non-trainable weights by dashed arrows.  Through a
forward pass across the network, the inputs ($x$-Bjorken and its logarithm) proceed through
 hidden layers  to output the eight fitted PDFs at
the initial parametrisation scale $Q_0$.
For each of the experimental observables entering the fit, these
PDFs are then combined into a partonic luminosity $\mathcal{L}^{(0)}$ at $Q_0$,
which is convolved with the precomputed {\tt FK}-tables $\Sigma$ to obtain the SM
theoretical prediction $\mathcal{T}^\text{SM}$.
Subsequently, the effects of the $N$ EFT coefficients $\boldsymbol{c}=(c_1,\ldots,c_N)$,
associated to the operator basis considered,
are accounted for by means of an extra layer, resulting in the
final prediction for the observable $\mathcal{T}$ entering the SMEFT-PDF fit.
The \simunet{} code allows for both linear and quadratic dependence on the EFT
coefficients. In linear EFT fits, the last layer consists of $N$ trainable
weights to account for each Wilson coefficient. In quadratic EFT fits, in
addition to the $N$ trainable weights, a set of $N(N+1)/2$ non-trainable
parameters, which are functions of the trainable weights, is included to account
for all diagonal and non-diagonal contributions of EFT-EFT interference to the
cross-sections. The results obtained with the quadratic functionality
of \simunet{} are, however, not displayed in this work, for the reasons explained in
App.~\ref{app:quad}. 
The PDF parameters $\boldsymbol{\theta}$
and the EFT coefficients $\boldsymbol{c}$ entering the evaluation
of the SMEFT observable in Fig.~\ref{fig: architecture} are then determined
simultaneously from the minimisation of the fit figure of merit (also
known as loss function).

%
The \simunet{} architecture can be minimally modified
to deal with the fixed-PDF case, in which only the EFT coefficients
are treated as free parameters in the optimisation process.
This can be achieved by freezing the PDF-related weights in the network
architecture to the values obtained in some previous fit, for example a SM-PDF
determination based on \nnpdf.
In this manner, \simunet{} can also be used to carry out traditional EFT fits where the
PDF dependence of the theory predictions is neglected.
Furthermore,
for PDF-independent observables, computing an {\tt FK}-table $\Sigma$ is not required
and the SM cross-section $\mathcal{T}^\text{SM}$ can be evaluated separately
and stored to be used in the fit.
%


As illustrated in Fig.~\ref{fig: architecture}, within
the \simunet{} framework a single neural network
encapsulates both the PDF and the EFT dependence of physical observables,
with the corresponding parameters being simultaneously constrained from the experimental
data included in the fit.
Specifically, we denote the prediction of the neural network as:
\be
\mathcal{T} = \mathcal{T}(\boldsymbol{\hat\theta})=\lp T_1(\boldsymbol{\hat\theta}),\ldots, T_n(\boldsymbol{\hat\theta})\rp \, ,
\ee
with $n=n_{\rm dat}$ and 
$\boldsymbol{\hat{\theta}} = (\boldsymbol{\theta}, \boldsymbol{c})$, where $\boldsymbol{\theta}$ and $\boldsymbol{c}=(c_1, \ldots, c_N)$ represent
the weights associated to the PDF nodes
of the network, and to the $N$ Wilson coefficients from
the operator basis, respectively.
The uncertainty estimation uses the Monte Carlo replica method, where a large number $N_{\rm rep}$
of replicas $D^{(k)}=\lp D_1^{(k)}, \ldots, D_n^{(k)}\rp$
of the experimental measurements $D=\lp D_1, \ldots, D_n\rp$ are sampled from the
distribution of experimental uncertainties with $k=1,\ldots,N_{\rm rep}$.
The optimal values for the fit parameters $\boldsymbol{\hat{\theta}}^{(k)}$ associated
to each replica are obtained by means of a Stochastic Gradient Descent (SGD) algorithm
that minimises the corresponding figure of merit:
\begin{equation}
  \label{eq:simunet_loss}
  E_{\rm tot}^{(k)}\lp \boldsymbol{\hat{\theta}}\rp  = \frac{1}{n_{\rm dat}}\sum_{i,j=1}^{n_{\rm dat}} \lp D_i^{(k)} - T_i(\boldsymbol{\hat\theta}) \rp \lp {\rm cov}_{t^0}^{-1} \rp_{ij}
  \lp D_j^{(k)} - T_j(\boldsymbol{\hat\theta}) \rp \, ,
\end{equation}
where the covariance matrix in Eq.~(\ref{eq:simunet_loss})
is the $t_0$ covariance matrix, which is constructed from all sources of statistical and
systematic uncertainties that are made available by the experiments
with correlated multiplicative uncertainties treated via the ‘t0’ prescription~\cite{Ball:2009qv} 
in the fit to avoid fitting bias associated with multiplicative uncertainties. 

Once Eq.~(\ref{eq:simunet_loss}) is minimised for each replica, subject
to the usual cross-validation stopping, one ends up with a sample of
best-fit values for both the EFT
coefficients and the PDF parameters:
\begin{equation}
  \label{eq:minimiser}
  \boldsymbol{\hat{\theta}}^{(k)}
  = \lp \boldsymbol{{\theta}}^{(k)}, \boldsymbol{c}^{(k)}  \rp
  = \argmin_{\boldsymbol{\hat{\theta}}}  E_{\rm tot}^{(k)}\lp \boldsymbol{\hat{\theta}}\rp  \, ,\qquad k=1,\ldots, N_{\rm rep} \, ,
\end{equation}
from which one can evaluate statistical properties such as averages,
variances, higher moments, or confidence level intervals.
For example, the preferred value of the EFT coefficients
could be evaluated over the mean over the replica sample,
\begin{equation}
  c_\ell^* = \left\langle c_\ell^{(k)} \right\rangle_{\rm rep} =
  \frac{1}{N_{\rm rep}}\sum_{k=1}^{N_{\rm rep}}  c_\ell^{(k)}
  \ ,
\end{equation}
though one could also define the preferred value as the median or mode
of the distribution. Note that, in this methodology, the Monte Carlo
error propagation automatically propagates the PDF uncertainty to the
distribution of the best-fit values of the EFT
coefficients. Hence the variance on the EFT coefficients reflects not
only the experimental uncertainty of the data included in the fit, but
also the functional uncertainty associated with the PDFs.

As we discuss below, the current implementation of the 
\simunet{} methodology also allows performing fixed-PDF fits, where
only the Wilson coefficients are optimised. This is done by freezing the weights
 of the PDF part of the neural network during the minimisation of the loss function
 (\ref{eq:simunet_loss}) from some other previous fit, $\boldsymbol{{\theta}}^{(k)}=
 \boldsymbol{\widetilde{\theta}}^{(k)}$, such that Eq.~(\ref{eq:minimiser})
 reduces to
 \begin{equation}
  \label{eq:minimiser2}
  \boldsymbol{\hat{\theta}}^{(k)}
  = \lp \boldsymbol{\widetilde{\theta}}^{(k)}, \boldsymbol{c}^{(k)}  \rp
  = \argmin_{\boldsymbol{c}}  E_{\rm tot}^{(k)}\lp \boldsymbol{\widetilde{\theta}}^{(k)},
    \boldsymbol{c} \rp  \, ,\qquad k=1,\ldots, N_{\rm rep} \, .
 \end{equation}
In this limit, \simunet{} reduces to a fixed-PDF EFT fit such as the
 MCfit variant of {\sc\small SMEFiT}~\cite{Giani:2023gfq}. 
 Likewise, by setting to zero the EFT coefficients,
 \begin{equation}
  \label{eq:minimiser3}
  \boldsymbol{\hat{\theta}}^{(k)}
  = \lp \boldsymbol{{\theta}}^{(k)}, \boldsymbol{c}^{(k)} =\boldsymbol{0} \rp
  = \argmin_{\boldsymbol{{\theta}}}  E_{\rm tot}^{(k)}\lp \boldsymbol{{\theta}}^{(k)},
    \boldsymbol{c}^{(k)}=\boldsymbol{0} \rp  \, ,\qquad k=1,\ldots, N_{\rm rep} \, ,
 \end{equation}
 one recovers the same PDF weights $\boldsymbol{{\theta}}^{(k)}$
 as in \nnpdf, or those of the SM-PDF fit being used as baseline in
 the analysis. 

 An important {\it caveat} here is that, while in the \simunet{}
 methodology the PDF uncertainty is propagated to the posterior
 distribution of the EFT coefficients via the Monte Carlo replica
 method, in the MCfit variant of the \smefit methodology the fit of
 the EFT only considers the central PDF member (which in the \nnpdf
 case corresponds to the average of the PDF replicas) for all $N_{\rm rep}$
 replicas, and the PDF uncertainty is propagated to the EFT
 coefficients by utilising an additional covariance matrix (both in the fit of the EFT
 coefficients and in the generation of the Monte Carlo replicas of the
 experimental data) that is added to $t_0$ covariance matrix. Namely, 
 \begin{equation} \label{eq:expthcov}
{\rm cov}_{\rm exp+th} = {\rm cov}_{t^0} + {\rm cov}_{\rm th},
\end{equation}
where ${\rm cov}_{\rm th}$ 
includes the PDF contribution~\cite{Ethier:2021bye,Hartland:2019bjb}, computed as
\begin{equation}
\left( {\rm cov}_{\rm th}\right)_{ij} = \langle T_i^{(k)}T_j^{(k)}\rangle_k
\,-\, \langle T_i^{(k)}\rangle_k \, \langle T_j^{(k)}\rangle_k,
  \end{equation}
in which the average is taken over PDF replicas. The two ways of
propagating PDF uncertainties to the distribution of the EFT
coefficients are equivalent assuming that PDF uncertainties are Gaussian and uncorrelated. 
 
 \simunet{} adopts the same optimisation settings as those set in the
 \nnpdf analysis for the PDF-dependent
 part of the network. On the other hand it adjusts only those hyperparameters
 associated to the EFT-dependent layer.
 Within the joint SMEFT-PDF fit, several of the fit settings
 such as the prior ranges for the EFT parameters and the learning rates are improved
 in an iterative way until convergence is achieved.
 In doing so, we also iterate the $t_0$ covariance matrix and the preprocessing
 exponents as customary in the NNPDF procedure.
 In the fixed-PDF EFT fit, the user can decide
 both the ranges and the prior distributions to be used in the initial
 sampling of EFT coefficients as determined e.g. from a previous
 fit or from one-parameter scans.
 %
 %
 
\subsection{New features}
\label{sec:new_simunet}

We now discuss some of the new features that have been implemented in \simunet{}, in comparison
with~\cite{Iranipour:2022iak}, which are motivated by the needs of the SMEFT-PDF fits
to LHC top quark data presented in this work.
We consider in turn the following new features: the implementation of the quadratic contributions
to the EFT cross-sections in the joint fits; fitting observables whose PDF dependence
is negligible or non-existent; initialising the PDF weights of the neural network with the results
of a previous fit; and finally, the improved initialisation of the EFT
coefficients. 

\paragraph{Quadratic EFT contributions.}
The  version of \simunet{} used in~\cite{Iranipour:2022iak} for the SMEFT-PDF fits
of high-mass Drell-Yan data allowed the inclusion of quadratic contributions
to the EFT cross-sections only under the approximation in which the cross-terms proportional to $c_ic_j$
with $i\ne j$ in Eq.~(\ref{eq:theory_k_fac}) were neglected.
In the current implementation,  \simunet{} can instead account for the full quadratic
contributions to the EFT cross-sections, including the non-diagonal cross-terms.
This feature can be especially important for the interpretation of top quark
measurements at the LHC, given that for many observables quadratic corrections, including
cross-terms relating different operators, can be sizeable specially in the high-energy region.
%

The implementation consists of explicitly accounting for the cross terms, as
parameters which depend on the Wilson coefficients and can be differentiated as
a function of them during the training procedure. 

\paragraph{PDF-independent observables.}
In the original version of \simunet{}, only physical observables with explicit
dependence on both the PDFs (via the {\tt FK}-tables interface) and the EFT coefficients
  could be included in a simultaneous fit.
  We have now extended the \simunet{} framework
  to describe observables that are independent of the PDF parameters $\boldsymbol{\theta}$,
  namely the weights and thresholds of the network depicted in Fig.~\ref{fig: architecture}
  that output the SM partonic luminosity $\mathcal{L}^{(0)}$.
  For these PDF-independent observables, the SM predictions $T^{\rm SM}$ are evaluated separately
  and stored in theory tables which can be used to evaluate the SMEFT cross-sections
  after applying the rescaling of Eq.~(\ref{eq:theory_k_fac});
  hence, these observables only depend on
the Wilson coefficients $c_n$.

In the current analysis the four-heavy cross-sections $\sigma_{\rm tot}(t\bar{t}b\bar{b})$
and $\sigma_{\rm tot}(t\bar{t}t\bar{t})$, the $W$-helicity measurements,
and the associated top quark production cross-sections $tZ$, $tW$ and
$t\bar{t}\gamma$ are treated as PDF-independent observables, as
 for those cross-sections the PDF dependence can be neglected
in comparison with other sources of theoretical and experimental
uncertainty. 
%
%
%
The possibility to include PDF-independent observables makes \simunet{} a global SMEFT analysis tool 
on the same footing as \smefit~\cite{Hartland:2019bjb,Giani:2023gfq}, \fitm~\cite{Ellis:2020unq},
{\sc\small HepFit}~\cite{DeBlas:2019ehy}, {\sc\small EFTfitter}~\cite{Castro:2016jjv},
and {\sc\small Sfitter}~\cite{Brivio:2019ius} among others. This is
demonstrated in App.~\ref{app:benchmark}, where it is shown that the
results of a linear fixed-PDF SMEFT analysis
performed with \simunet{} coincide with those obtained with
\smefit~\cite{Giani:2023gfq} once the same
experimental data and theory calculations are used. 
Moreover, the new feature will allow us to include in future analyses any non-hadronic
observables, such as electroweak precision observables (EWPO)~\cite{Han:2004az}. 

\paragraph{Fixed-PDF weight initialisation.}
Within the current  \simunet{} implementation, one can also choose to initialise
the PDF-dependent weights of the network in Fig.~\ref{fig: architecture}
using the results of a previous Monte Carlo fit of PDFs, for example
an existing SM-PDF analysis obtained with the \nnpdf methodology.
The weights of the latter are written to file and then read by  \simunet{} for the network
initialisation.

This feature has a two-fold application.
First, instead of initialising at random the network weights in a
simultaneous SMEFT-PDF fit,
one can set them to the results of a previous SM-PDF fit, thus
speeding up the convergence of the simultaneous fit, with the rationale that
EFT corrections are expected to represent a perturbation of the SM predictions.
Second, we can use this feature to compute EFT observables in the fixed-PDF case
described above using the {\tt FK}-table convolution with this previous PDF set as input,
as opposed to having to rely on an independent calculation of the SM cross-section.
Therefore, this PDF weight-initialisation feature helps realise
\simunet{} both as a fixed-PDF EFT analysis framework, and to assess the stability of the
joint SMEFT-PDF fits upon a different choice of initial state of the network
in the minimisation.

\paragraph{Improved initialisation of the EFT coefficients.}
In the original implementation of \simunet{} it was only possible
to initialise the EFT coefficients at specific values, selected beforehand
by the user.
In this work, we have developed more flexible initialisation schemes for the Wilson
coefficients, in the sense that they can now be sampled from a prior probability
distributed defined by the user.
Specifically, each Wilson coefficient $c_i$ can be
sampled from either a uniform $\mathcal{U} [a_i, b_i]$ or a normal $\mathcal{N}(\mu_i,
\sigma_i)$ distribution.
The ranges  $(a_i, b_i)$  of the uniform distribution $\mathcal{U}$
and the mean and standard deviation $(\mu_i, \sigma_i)$ of the Gaussian
distribution $\mathcal{N}$ are now user-defined parameters, which can
be assigned independently to each Wilson coefficients that enter the fit.
This feature enhances the effectiveness and flexibility of the minimisation procedure by
starting from the regions of the parameter space with the best sensitivity
to the corresponding Wilson coefficient.

Another option related to the improved initialisation of EFT coefficients
is the possibility to adjust the overall normalisation of each coefficient
by means of a user-defined scale factor.
The motivation to introduce such a coefficient-dependent scale factor
is to end up with (rescaled) EFT coefficients entering the fit which all have
a similar expected range of variation.
This feature is advantageous, since the resulting gradients entering the SGD algorithm
will all be of the same order, and hence use a unique learning rate which is appropriate for the fit at hand.

\section{Impact of the top quark Run II dataset on the SM-PDFs}
\label{sec:baseline_sm_fits}

Here we present the results of a global SM-PDF determination which accounts for the constraints
of the most extensive top quark dataset considered to date
in such analyses and described in Sect.~\ref{sec:exp}.
The fitting methodology adopted follows closely the settings of the \nnpdf
study~\cite{NNPDF:2021njg}, see also~\cite{Ball:2022uon} for a rebuttal of some critical arguments.
This dataset includes not only the most up-to-date measurements of top quark
pair production from Run II, but it also includes all available single top
production cross-sections and distributions 
and for the first time new processes not considered in PDF studies before,
such as the $A_C$ asymmetry in $t\bar{t}$ production and the $t\bar{t}V$ and
$tV$ associated production (with $V = Z,W$).

We begin by summarising the methodological settings used for these
fits in Sect.~\ref{sec:setting}.
Then in Sect.~\ref{sec:individual} we assess the impact of adding to
a no-top baseline PDF fit various subsets of the top quark data
considered in this study. 
In particular, we assess the impact of including updated Run II $t\bar{t}$
and single-top measurements in comparison
with the subset used in the \nnpdf analysis, see the second-to-last
column of Tables~\ref{tab:input_datasets_toppair}--
\ref{tab:input_datasets_3}.
Furthermore, we quantify for the first time the impact of
associated vector boson and single-top ($tV$) as well as associated vector boson and
top-quark pair production ($t \bar{t} V$) data in a PDF fit.
Finally in Sect.~\ref{sec:alltop} we combine these results and present a
\nnpdf fit variant including all data described in Sect.~\ref{sec:exp},
which is compared to both the \nnpdfnotop baseline and to the original
\nnpdf set.

\subsection{Fit settings}
\label{sec:setting}

An overview of the SM-PDF fits that are discussed in this section is
presented in Table~\ref{tab:fit_list}. 
First of all, we produce a baseline fit, denoted
by \nnpdfnotop, which is based on the same dataset as \nnpdf
but with all top quark measurements excluded.
Then we produce fit variants A to G, which quantify the impact of
including in this baseline various subsets of
top data (indicated by a check mark in the table).
Finally, fit variant H is the main result of this section, namely the fit
in which the full set of top quark measurements described in
Sect.~\ref{sec:exp} is added to the no-top baseline.

In these fits, methodological settings such as network architecture,
learning rates, and other hyperparameters are kept the same as in
\nnpdf, unless otherwise specified.
One difference is the training fraction $f_{\rm tr}$ defining
the training/validation split used for the cross-validation stopping criterion.
 In \nnpdf, we used $f_{\rm tr}=0.75$  for all datasets.
 Here instead we adopt $f_{\rm tr}=0.75$ only for the no-top datasets
 and $f_{\rm tr}=1.0$ instead for the top datasets.
 The rationale of this choice is to ensure that the fixed-PDF 
 SMEFT analysis, where overfitting is not possible~\cite{Ethier:2021bye},
 exploits all the information contained in the top quark data considered
 in this study,  and then for consistency to maintain the same settings in
 both the SM-PDF fits (this section) and
 in the joint SMEFT-PDF fits (to be discussed in Sect.~\ref{sec:joint_pdf_smeft}).
 Nevertheless, we have verified that the resulting SM-PDF fits are statistically equivalent to the fits obtained by setting the training fraction to be $0.75$ for all data, including for the top quark observables.

 Fits A--G in Table~\ref{tab:fit_list}  are composed by $N_{\rm rep}=100$ Monte Carlo replicas after post-fit selection criteria, while the \nnpdfnotop baseline fit and fit H are instead composed by $N_{\rm rep}=1000$  replicas.
 As customary, all fits presented in this section are iterated with respect
 to the $t_0$ PDF set and the pre-processing exponents.

\begin{table}[tb]
  \begin{center}
  \centering
  \footnotesize
   \renewcommand{\arraystretch}{2.1}
   \setlength{\tabcolsep}{5pt}
  \begin{tabular}{l|C{1.9cm}C{1.6cm}C{1.9cm}C{1.6cm}C{1.6cm}C{1.6cm}C{1.6cm}}
  \multirow{2}{*}{\textbf{Fit ID}}    &  \multicolumn{6}{c}{\textbf{Datasets included in fit}}  \\
   & \textbf{No-top baseline, Sect.~\ref{sec:baseline_data}} & \textbf{Incl. $t\bar{t}$, Table~\ref{tab:input_datasets_toppair}} & \textbf{Asymm., Table~\ref{tab:input_datasets_topasymmetries}} & \textbf{Assoc. $t\bar{t}$, Table~\ref{tab:input_datasets2}} & \textbf{Single-$t$, Table~\ref{tab:input_datasets3}} & \textbf{Assoc. single$-t$, Table~\ref{tab:input_datasets_4}}  \\
    \midrule
    \nnpdfnotop (Baseline) & \checkmark  \\
    A (inclusive $t\bar{t}$) & \checkmark & \checkmark \\
    B (inclusive $t\bar{t}$ and charge asymmetry) & \checkmark & \checkmark & \checkmark \\
    C (single top) & \checkmark & & & & \checkmark \\
    D (all $t\bar{t}$ and single top) & \checkmark & \checkmark & \checkmark & & \checkmark \\
    E (associated $t\bar{t}$) & \checkmark & & & \checkmark & & \\
    F (associated single top) & \checkmark & & & & & \checkmark \\
    G (all associated top) & \checkmark & & & \checkmark & & \checkmark \\
    H (all top data) & \checkmark & \checkmark & \checkmark & \checkmark & \checkmark & \checkmark \\
    \bottomrule
  \end{tabular}
  \vspace{0.3cm}
  \caption{Overview of the SM-PDF fits discussed in this section.
    The baseline fit, \nnpdfnotop, is based on the same dataset as \nnpdf
    with all top quark measurements excluded.
    The fit variants A to G consider the impact of including in this baseline
    various subsets of top data, while in fit H the full set of top quark measurements described in Sect.~\ref{sec:exp} is added to the baseline.
     \label{tab:fit_list}
  }
  \end{center}
\end{table}


\subsection{Impact of individual top quark datasets}
\label{sec:individual}

First we assess the impact of specific subsets of LHC top quark data when added
to the \nnpdfnotop baseline, fits A--G in Table~\ref{tab:fit_list}. 
In the next section we discuss the outcome of fit H, which contains the full
top quark dataset considered in this work.

\begin{figure}[t]
\centering
	\begin{subfigure}[b]{0.49\textwidth}
	\includegraphics[scale=0.5]{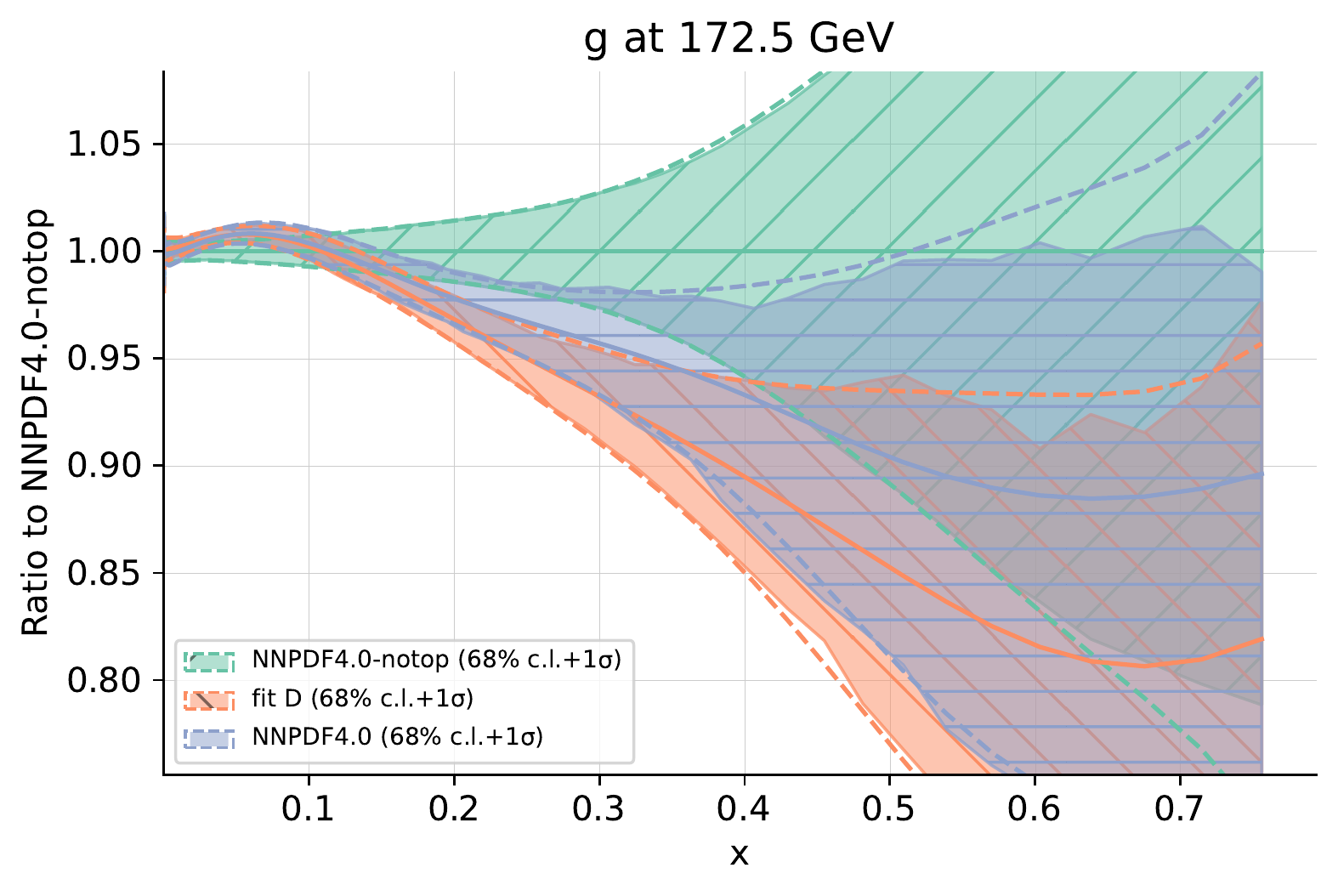}
	\end{subfigure}
	\begin{subfigure}[b]{0.49\textwidth}
	\includegraphics[scale=0.5]{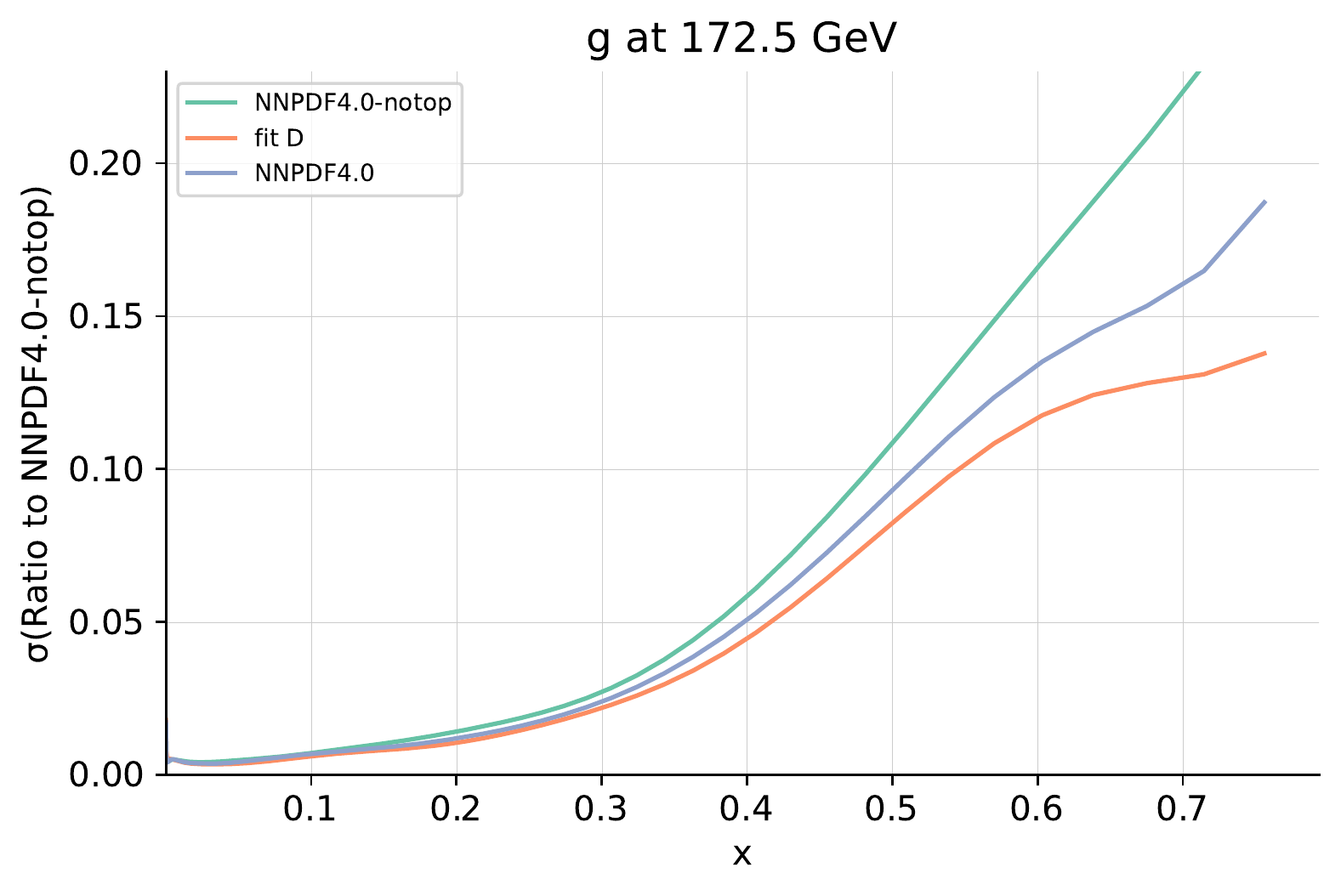}
	\end{subfigure}
        \caption{Left: comparison between the gluon PDF at $Q=m_t=172.5$ GeV
          obtained in the \nnpdf and \nnpdfnotop fits
          against fit D in Table~\ref{tab:fit_list}, which includes all
          top-quark
          pair (also the charge asymmetry $A_C$)
          and single-top quark production data considered in this analysis.
          Results are normalised to the central value of the \nnpdfnotop set.
          Right: same comparison now for the PDF uncertainties (all
          normalised to the central value of the \nnpdfnotop set).
}
\label{fig:fitD_gluon}
\end{figure}

Fig.~\ref{fig:fitD_gluon}
displays the comparison between the gluon PDF at $Q=m_t=172.5$ GeV obtained
in the \nnpdf and \nnpdfnotop fits against fit D, which includes all top-quark
pair (also the charge asymmetry $A_C$) and all single-top quark production
data considered in this analysis.
Results are normalised to the central value of the \nnpdfnotop fit,
and in the right panel we show the corresponding PDF uncertainties, all 
normalised to the central value of the \nnpdfnotop baseline. 
%
%
From Fig.~\ref{fig:fitD_gluon} one finds that the main impact of the
additional LHC Run II $t\bar{t}$ and sinle-top data included in fit D as compared to that
already present in \nnpdf is a further depletion of the large-$x$ gluon
PDF as compared to the \nnpdfnotop baseline, together with a reduction of
the PDF uncertainties in the same kinematic region.
While fit D and \nnpdf agree within uncertainties in the whole range of $x$,
fit D and \nnpdfnotop agree only at the $2\sigma$ level
in the region $x\approx [0.2,0.4]$.
These findings imply that the effect on the SM-PDFs of the new Run II top
data is consistent with, and strengthens, that of the data already part of
\nnpdf, and suggests a possible tension
between top quark data and other measurements in the global PDF sensitive
to the large-$x$ gluon, in particular inclusive jet and di-jet production.
The reduction of the gluon PDF uncertainties from the new measurements
can be as large as about 20\% at $x\approx 0.4$.
Differences are much reduced for the quark PDFs, and restricted
to a 5\% to 10\% uncertainty reduction in the region around $x\sim 0.2$
with central values essentially unchanged.

\begin{figure}[t]
\centering
	\begin{subfigure}[b]{0.49\textwidth}
	\includegraphics[scale=0.55]{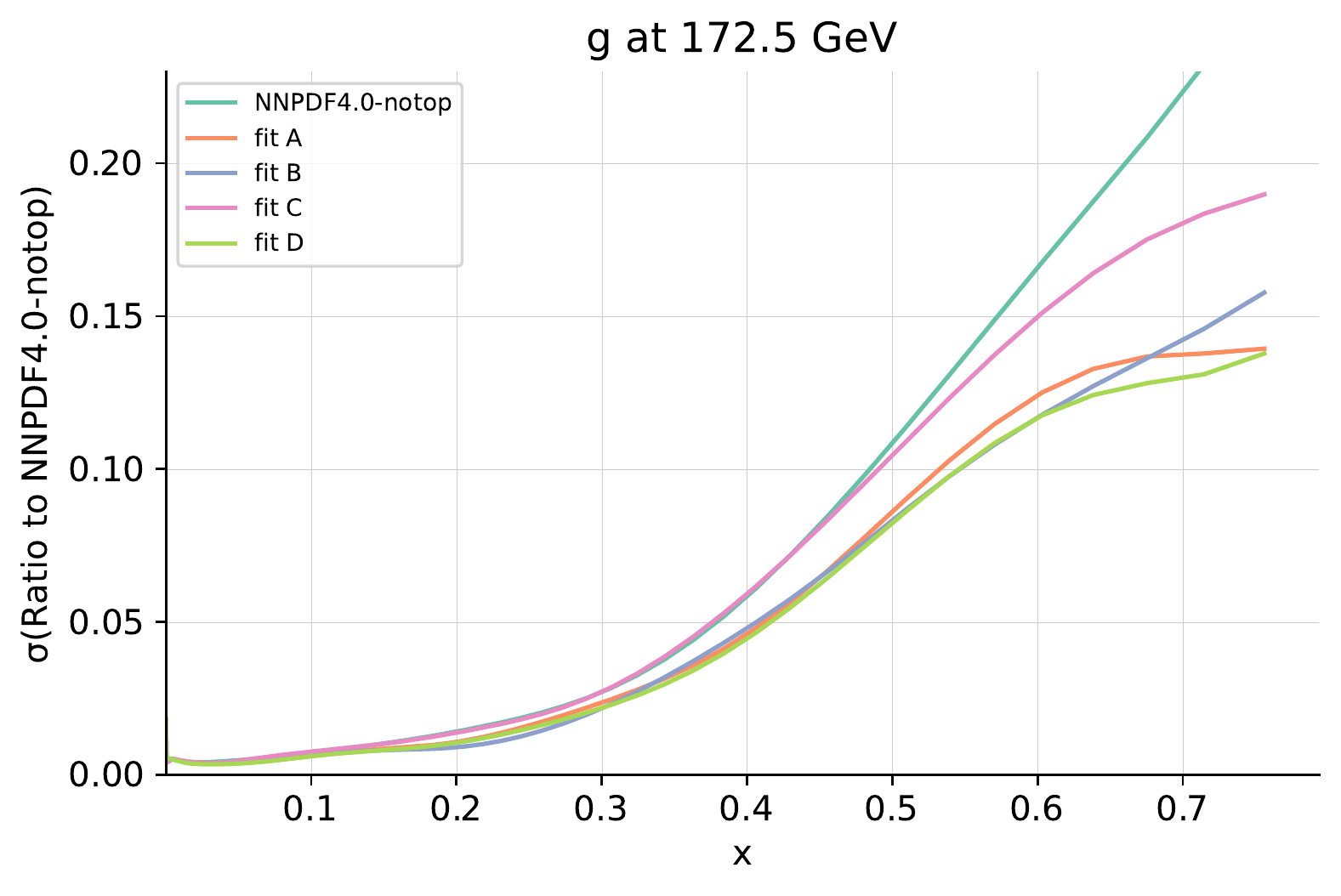}
	\end{subfigure}
	\begin{subfigure}[b]{0.49\textwidth}
	\includegraphics[scale=0.54]{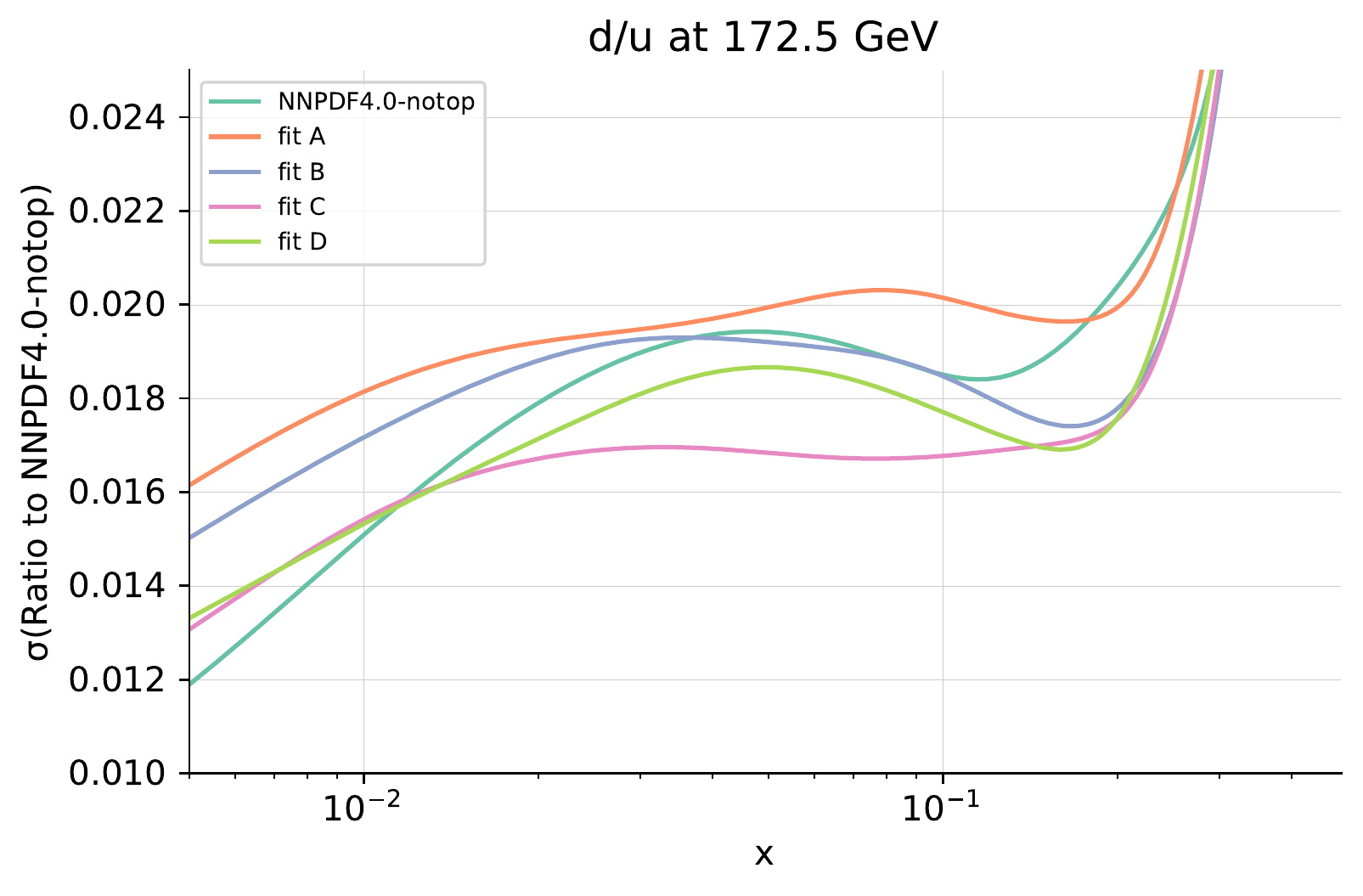}
	\end{subfigure}
        \caption{The ratio between the PDF 1$\sigma$ uncertainty and the
          central value of the \nnpdfnotop baseline in the case of the gluon
          (left panel) and in the case of the $d/u$ ratio (right panel).
          The uncertainty of  the  baseline fit at $Q=m_t=172.5$ GeV
          is compared with the uncertainty associated with fits A, B, C, and
          D in Table~\ref{tab:fit_list}.
          These fit variants include the following top datasets:  inclusive $t\bar{t}$ (A),
          inclusive $t\bar{t}$ + $A_C$ (B), single top (C), and their sum (D).}
\label{fig:fitABCD}
\end{figure}

To disentangle which of the processes
considered dominates the observed effects on the gluon and the light
quarks PDFs, Fig.~\ref{fig:fitABCD} compares the relative PDF uncertainty
on the gluon and on the $d/u$ quark ratio in the \nnpdfnotop 
baseline fit at $Q=m_t=172.5$ GeV with the results from fits A, B, C, and D.
As indicated in Table~\ref{tab:fit_list}, these fit variants include the following top datasets:  inclusive $t\bar{t}$ (A),
inclusive $t\bar{t}$ + $A_C$ (B), single top (C), and their sum (D).
The inclusion of the top charge asymmetry $A_C$ data does not have any
impact on the PDFs; indeed fits A and B are statistically equivalent.
This is not surprising, given that in  Eq.~\eqref{eq:ac} the dependence
on PDFs cancels out.
Concerning the inclusion of single top data (fit C), it does not affect
the gluon PDF but instead leads to a moderate reduction on the PDF
uncertainties on the light quark PDFs in the intermediate-$x$ region,
$x\approx[0.01,0.1]$, as shown in the right panel
displaying the relative uncertainty reduction for the $d/u$ ratio.
This observation agrees with what was pointed out by a previous
study~\cite{Nocera:2019wyk}, and the impact of  LHC single-top
measurements is more marked now
as expected since the number of data points considered here is larger.
We conclude that the inclusive $t\bar{t}$ measurements dominate
the impact on the large-$x$ gluon observed in Fig.~\ref{fig:fitD_gluon}, with
single top data moderately helping to constrain the light quark PDFs
in the intermediate-$x$ region. 

We now consider the effect of the inclusion of data that were not included before in any PDF fit, namely either $t\bar{t}$ or single-top production in association with a weak vector boson.
Although current data exhibits large experimental and theoretical uncertainties, it is interesting to
study whether  they impact PDF fits at all, in view of their increased precision expected in future measurements; in particular, it is useful to know which parton flavours are most affected.

\begin{figure}[t]
\centering
	\begin{subfigure}[b]{0.49\textwidth}
	\includegraphics[scale=0.5]{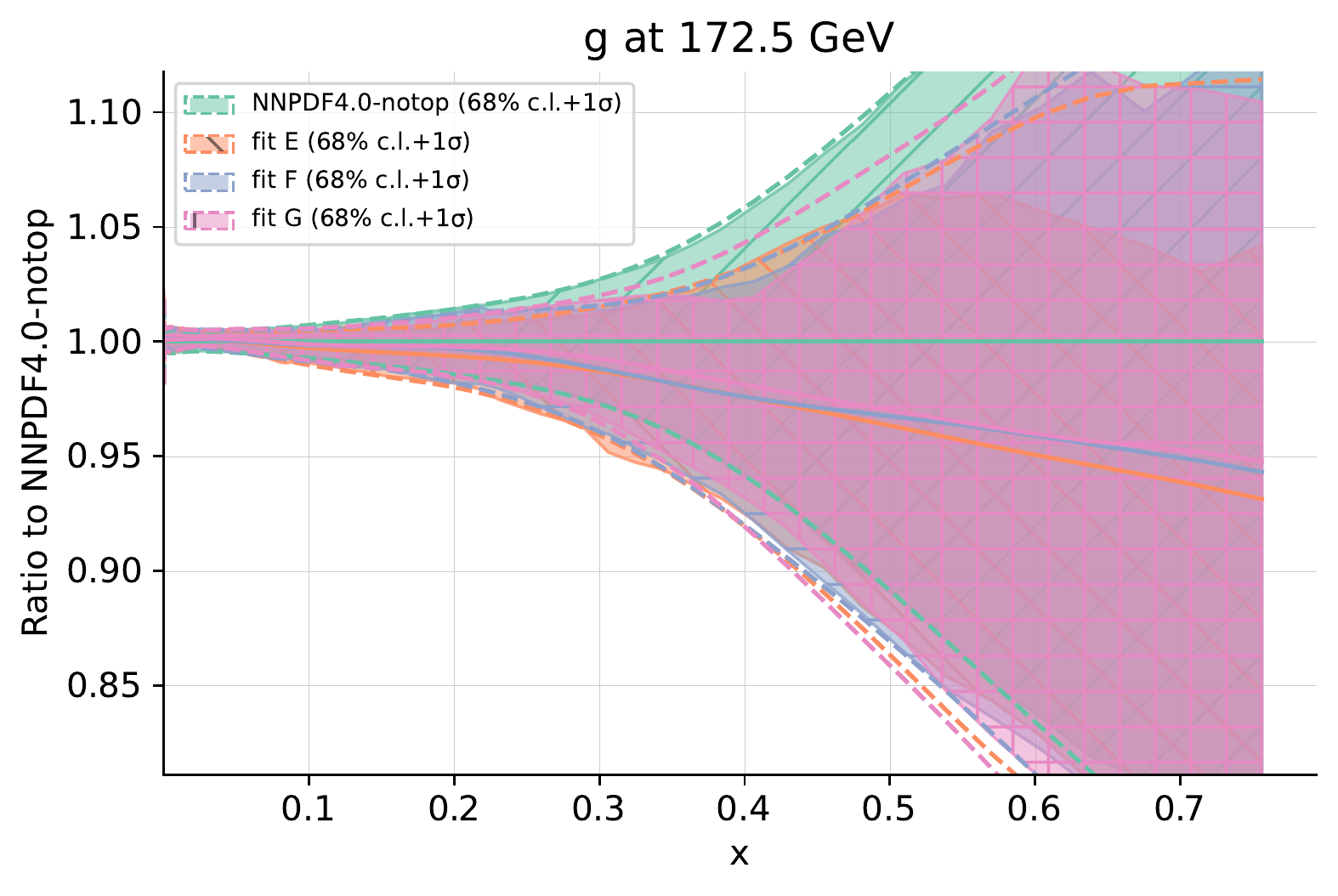}
	\end{subfigure}
	\begin{subfigure}[b]{0.49\textwidth}
	\includegraphics[scale=0.5]{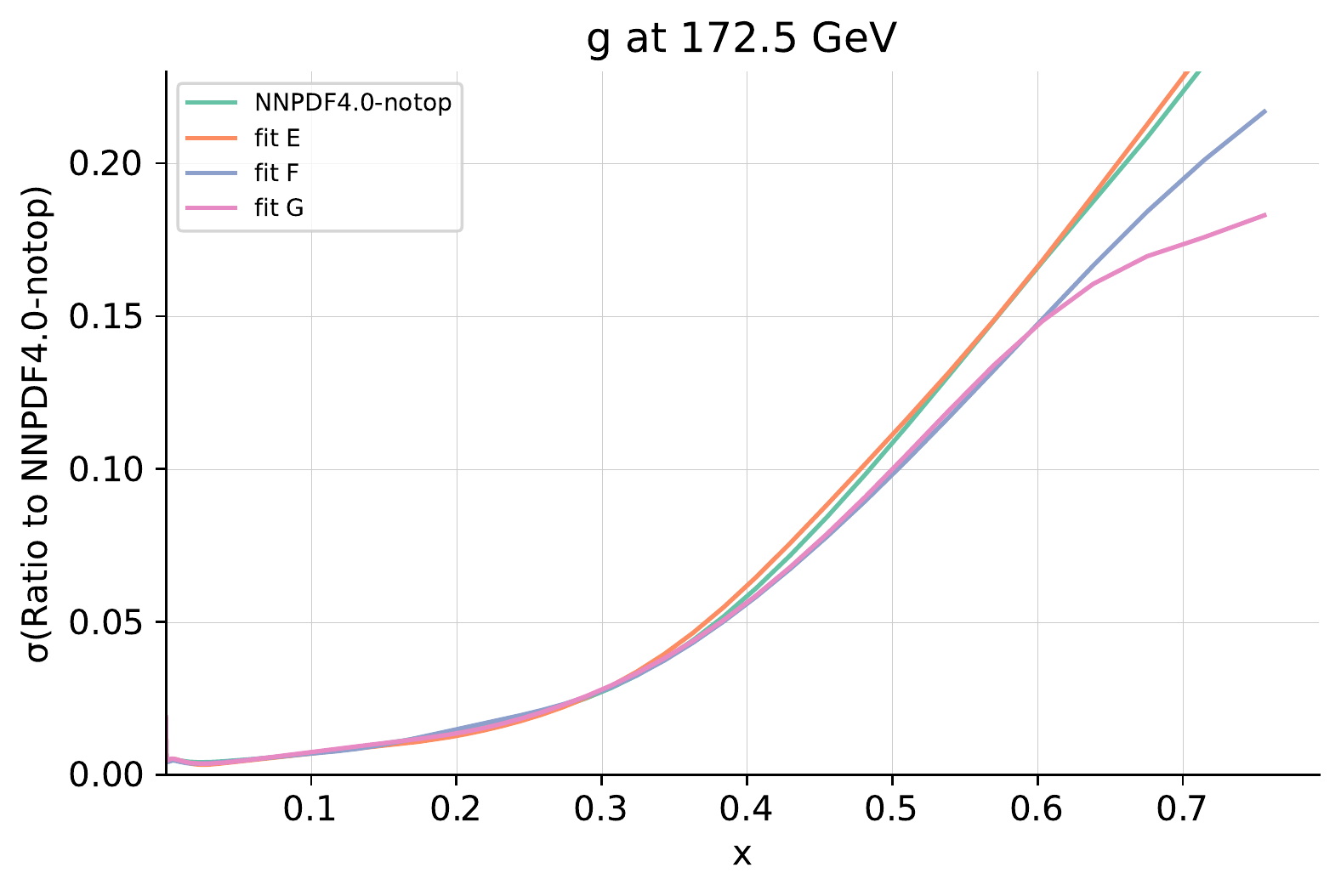}
	\end{subfigure}
        \caption{Same as Fig.~\ref{fig:fitD_gluon} comparing the \nnpdfnotop
          baseline fit with variants E, F, and G from Table~\ref{tab:fit_list}.
          These variants include associated $t\bar{t}$ and vector boson
          production data (E), associated single top and vector boson
          production data (F) and their sum (G).
        }
\label{fig:fitEFG}
\end{figure}

Fig.~\ref{fig:fitEFG} displays the same comparison as in
Fig.~\ref{fig:fitD_gluon} now for the \nnpdfnotop baseline
and the variants E, F, and G from Table~\ref{tab:fit_list}, which 
include the $t\bar{t}V$ (E) and $tV$ (F) data
as well as their sum (G).
The pull of $t\bar{t}V$ is very small, while the pull of the $tV$
measurements is in general small, but consistent
with those of the inclusive  $t\bar{t}$ measurements, namely preferring a
depletion of the large-$x$ gluon.
This result indicates that $t\bar{t}V$  and $tV$  data may be
helpful in constraining PDF once both future experimental data 
and theoretical predictions become more precise, although
the corresponding inclusive measurements are still expected to provide
the dominant constraints. 

\subsection{Combined effect of the full top quark dataset}
\label{sec:alltop}

\begin{figure}[t]
\centering
	\begin{subfigure}[b]{0.49\textwidth}
	\includegraphics[scale=0.5]{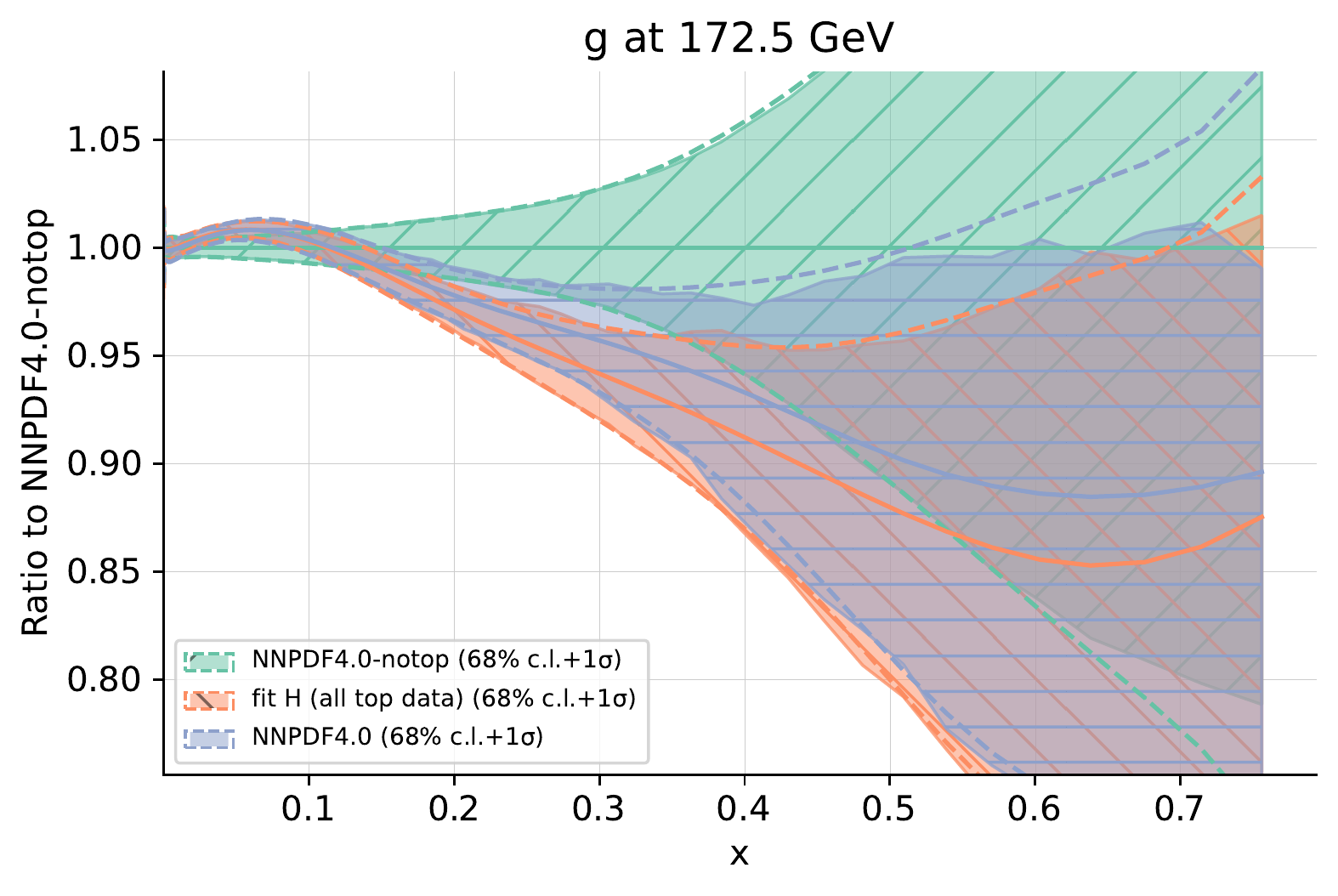}
	\end{subfigure}
	\begin{subfigure}[b]{0.49\textwidth}
	\includegraphics[scale=0.5]{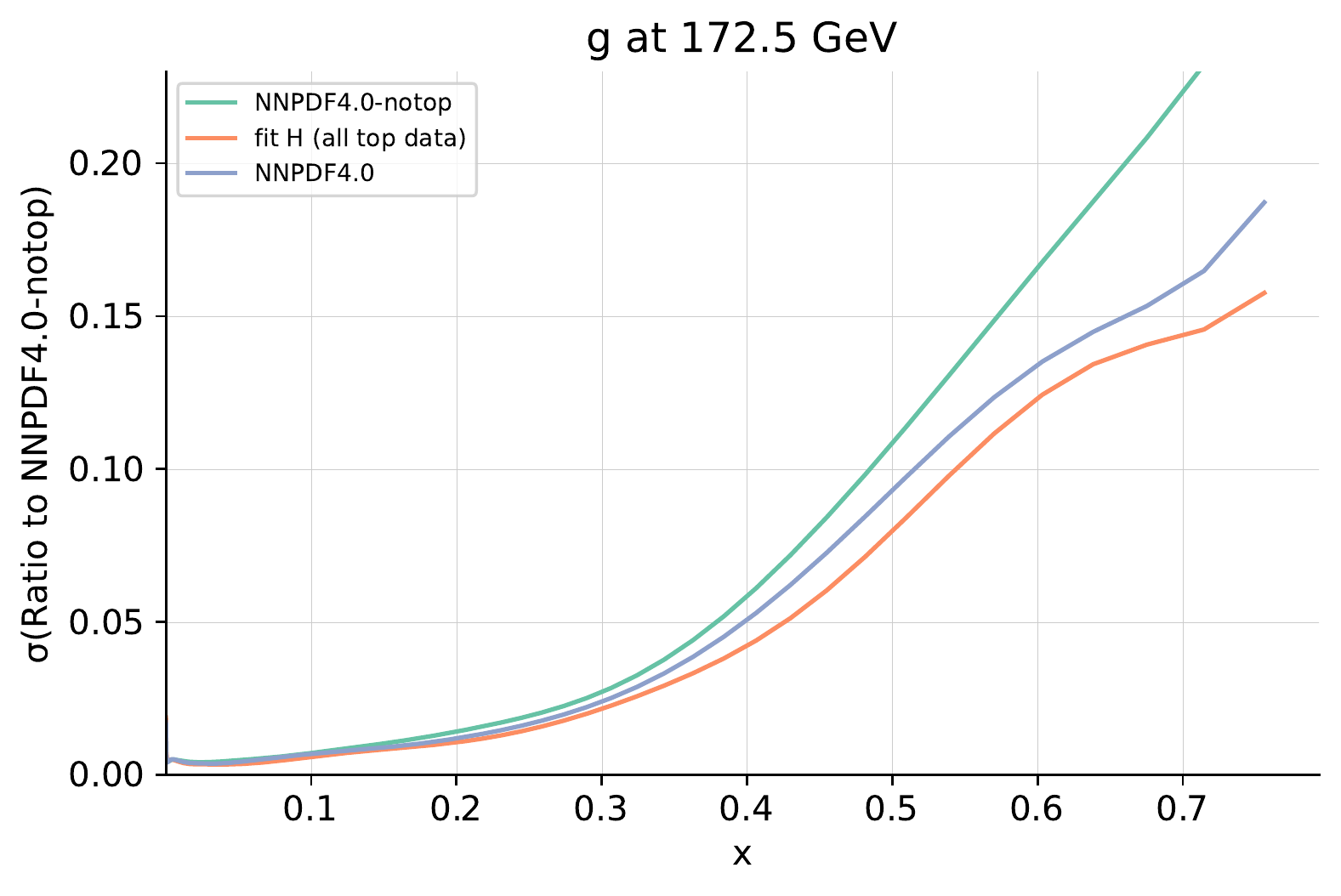}
	\end{subfigure}
        \caption{ Same as Fig.~\ref{fig:fitD_gluon} comparing NNPDF4.0
          and NNPDF4.0 (no top) with fit variant H in Table~\ref{tab:fit_list},
          which includes the full set of top quark measurements considered in this analysis.}
\label{fig:fitH_gluon}
\end{figure}

The main result of this section is displayed in Fig.~\ref{fig:fitH_gluon},
which compares the \nnpdf and the \nnpdfnotop fits with variant H in
Table~\ref{tab:fit_list},
namely with the fit where the full set of top quark measurements considered
in this analysis has been added to the no-top baseline.
As in the case of Fig.~\ref{fig:fitD_gluon}, we show the large-$x$ gluon
normalised to the central value of \nnpdfnotop and the associated
1$\sigma$ PDF uncertainties (all normalised to the central value of the
baseline).
The results of fit H are similar to those of fit D, although slightly
more marked. This is expected, 
since as shown above the associated production datasets $t\bar{t}V$
and $tV$ carry little weight in the fit.

From Fig.~\ref{fig:fitH_gluon} one observes
how the gluon PDF of fit H deviates from the \nnpdfnotop baseline
markedly in the data region $x \in [0.1, 0.5]$.
The shift in the gluon PDF can be up to the $2\sigma$ level, and
in particular the two PDF uncertainty bands do not overlap
in the region $x\in \lc 0.2, 0.35\rc $.
As before, we  observe that the inclusion of
the latest Run II top quark measurements enhances the effect of the top
data already
included in \nnpdf, by further depleting the gluon in the large-$x$ region
 and  by reducing its uncertainty by a factor up to 25\%.
 Hence, one finds again that the new Run II top quark production
 measurements lead to
 a strong pull on the large-$x$ gluon, qualitatively
 consistent but stronger as compared
 with the pulls associated from the datasets already included in \nnpdf.

\begin{figure}[t]
\centering
\includegraphics[width=0.49\textwidth]{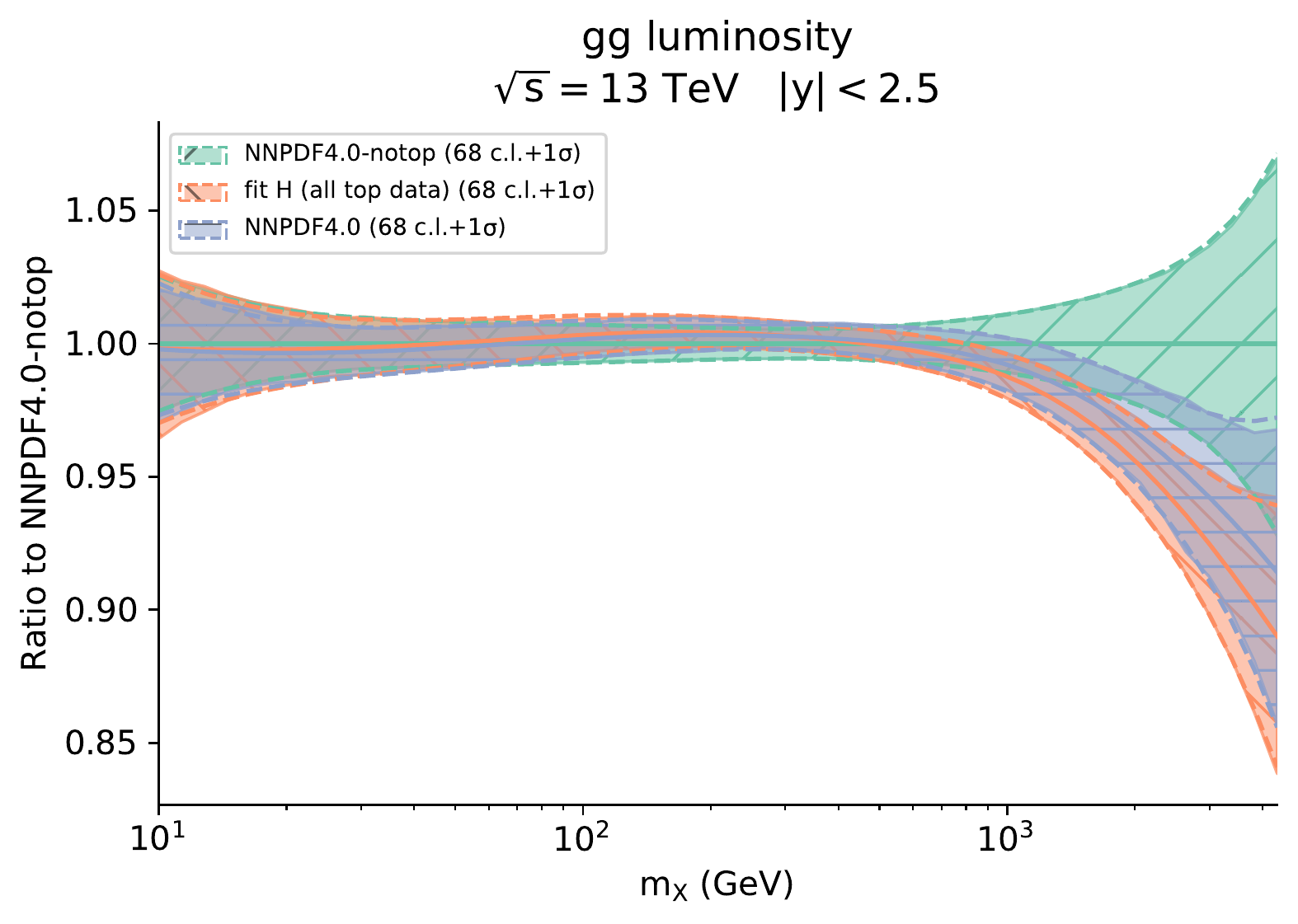}
\includegraphics[width=0.49\textwidth]{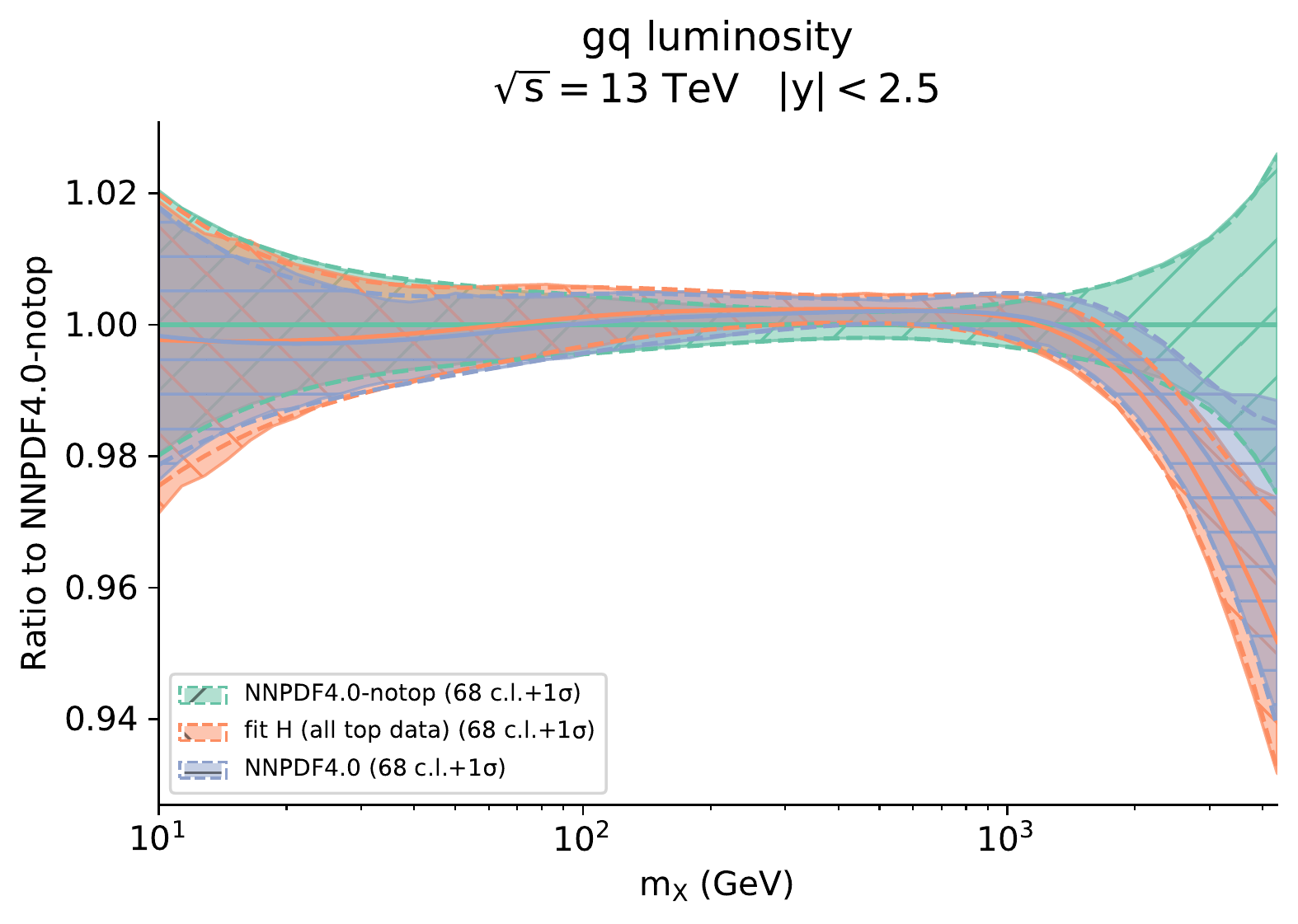}
\caption{The gluon-gluon (left) and quark-gluon (right panel)
  partonic luminosities at $\sqrt{s}=13$ TeV restricted to
  the central acceptance region with $|y|\le 2.5$.
  We compare the \nnpdf and \nnpdfnotop fits with the
  predictions based on fit H, which includes the full top quark dataset
  considered here.
  Results are presented as the ratio to the \nnpdfnotop baseline fit.
}
\label{fig:fith_luminosities}
\end{figure}

 To assess the phenomenological impact of our analysis at the level
 of LHC processes, Fig.~\ref{fig:fith_luminosities}
 compares the gluon-gluon and quark-gluon 
  partonic luminosities at $\sqrt{s}=13$ TeV (restricted to
  the central acceptance region with $|y|\le 2.5$)
between  \nnpdf, \nnpdfnotop, and fit H including the full top quark dataset
  considered here and Fig.~\ref{fig:fith_luminosities} compares their
  uncertainties. 
  Results are presented as the ratio to the no-top baseline fit.
  The $qq$ and $q\bar{q}$ luminosities of fit H are essentially
  identical to those of the no-top baseline, as expected given the negligible changes in the quark PDFs observed   in fit H, and hence are not discussed further. 

\begin{figure}[t]
\centering
\includegraphics[width=0.49\textwidth]{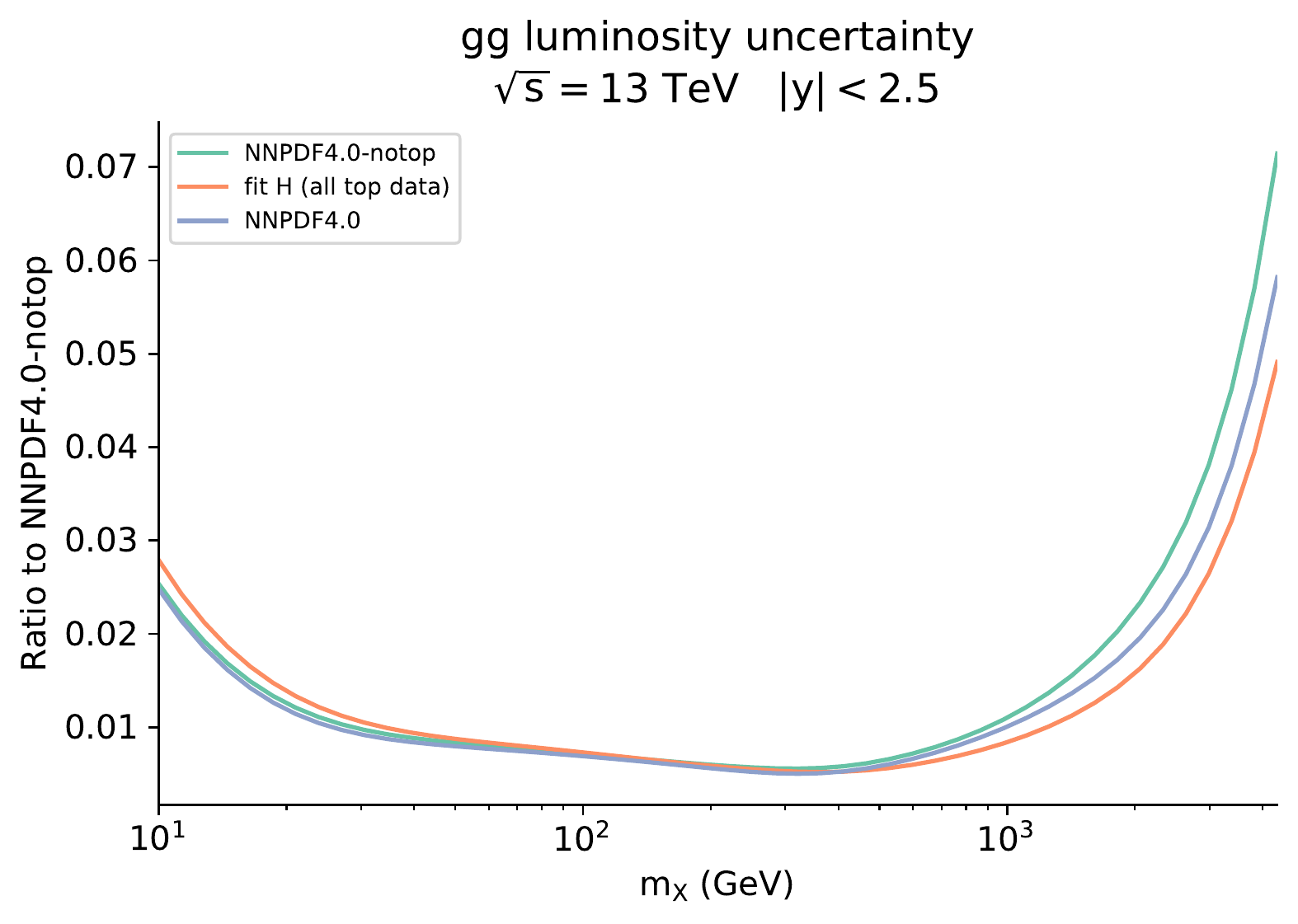}
\includegraphics[width=0.49\textwidth]{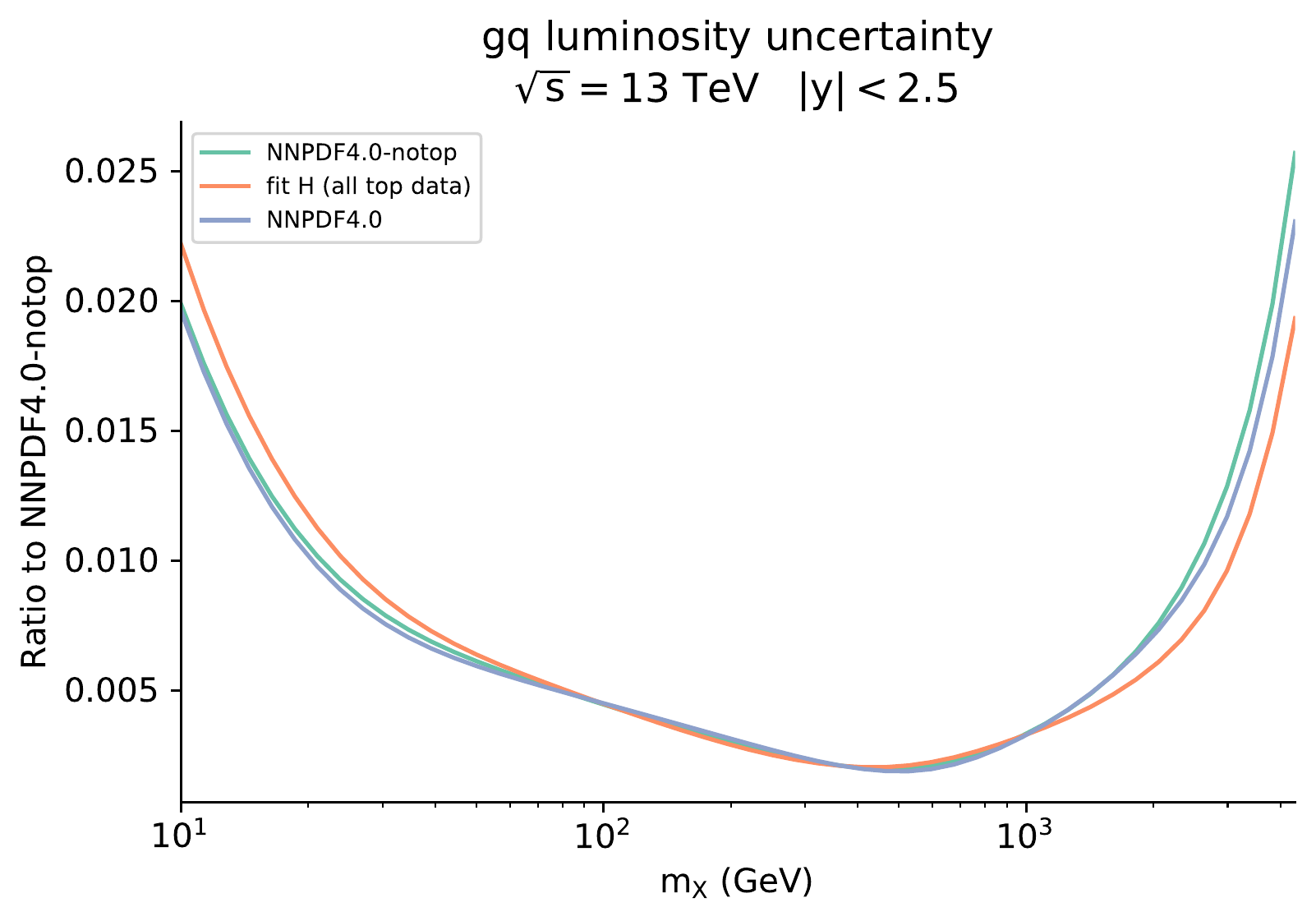}
\caption{Same as Fig.~\ref{fig:fith_luminosities}, now for the
  relative luminosity uncertainties, all normalised to the
  \nnpdfnotop baseline fit.
}
\label{fig:fith_luminosities_unc}
\end{figure}

From Figs.~\ref{fig:fith_luminosities}-\ref{fig:fith_luminosities_unc} one
observes that both
for the quark-gluon and gluon-gluon luminosity the impact
of the LHC top quark data is concentrated on the region above $m_X \gsim 1$ TeV.
As already reported for the case of the gluon PDF, also for the luminosities
the net effect of the new LHC Run II top quark data is to further reduce
the luminosities for invariant masses in the TeV range, with a qualitatively
similar but stronger pull as compared to that obtained in \nnpdf.
While \nnpdf and its no-top variant agree at the $1\sigma$ level in the full
kinematical range considered, this is not true for fit H,  whose error bands
do not overlap with those of \nnpdfnotop for invariant masses $m_X$ between
2 and 4 TeV.
On the other hand, \nnpdf and fit H are fully consistent across the full
$m_X$ range, and hence we conclude that predictions for LHC observables
based on \nnpdf will not be significantly affected by the inclusion
of the latest LHC top quark data considered in this work.

Finally, concerning the fit quality of fit H is essentially stable, actually 
better relative to the \nnpdfnotop baseline. The experimental $\chi^2$ per data point on
their respective datasets is $1.156$ for the no-top baseline, whilst for
fit H is reduced to $1.144$. A complete summary of the $\chi^2$ information
for all of the fits in this section is given in App.~\ref{app:fit_quality}.
It is interesting to observe that all new top data included in fit H are
already described well by using the \nnpdf set, although clearly the $\chi^2$
per data point improves (from $1.139$ to $1.102$) once all data are included
in the fit. This confirms the overall consistency of the analysis.

\section{Impact of the top quark Run II dataset on the SMEFT}
\label{sec:res_smeft}

We now quantify the impact of the LHC Run II legacy measurements, described in Sect.~\ref{sec:exp},
on the top quark sector of the SMEFT.
As compared to previous investigations of SMEFT operators modifying top-quark
interactions~\cite{Ellis:2020unq,Ethier:2021bye}~\cite{Aguilar-Saavedra:2018ksv,
  Buckley:2015lku, Brivio:2019ius, Bissmann:2019gfc, Hartland:2019bjb,
  Durieux:2019rbz, vanBeek:2019evb, Yates:2021udl}, the current analysis
considers a wider dataset, in particular extended
to various measurements based on the full LHC Run II luminosity.
In the last column of Tables~\ref{tab:input_datasets_toppair}--\ref{tab:input_datasets_4} we indicated
which of the datasets included here were considered
for the first time within a SMEFT interpretation.
Here we assess the constraints that the available LHC top quark measurements
provide on the SMEFT parameter space, and in particular study the
impact of the new measurements as compared to those used in~\cite{Ellis:2020unq,Ethier:2021bye}.
In this section we restrict ourselves to fixed-PDF EFT fits, where the input PDFs used in the calculation
of the SM cross-sections are kept fixed.
Subsequently, in Sect.~\ref{sec:joint_pdf_smeft},
we generalise to the case in which PDFs are extracted simultaneously
together with the EFT coefficients.

The structure of this section is as follows. We begin in Sect.~\ref{subsec:smeftfitsettings} 
by describing the methodologies used
to constrain the EFT parameter space both at linear and quadratic
order in the EFT expansion.
%
We also present results for the Fisher information matrix, which indicates which datasets
provide constraints on which operators.
In Sect.~\ref{subsec:smeftresults}, we proceed to give the results of the fixed-PDF EFT 
fits at both linear and quadratic order,
highlighting the impact of the new Run II top quark data by comparison with 
previous global SMEFT analyses.
In Sect.~\ref{subsec:cms1dvs2d}, we assess the impact of replacing the CMS 13 TeV differential
measurement of $t\bar{t}$ in the $\ell+$jets channel, binned with respect to invariant top quark pair
mass, by the corresponding double-differential measurement binned with respect to both invariant
top quark pair mass and top quark pair rapidity. In the dataset selection performed in Sect~\ref{sec:dataselection}
we rejected the double-differential distribution due to its poor $\chi^2$-statistic in the SM, which could
not be improved by a weighted fit of PDFs; in the present section, it is interesting to see whether the
SMEFT can help account for the poor fit of this dataset.
Finally, in Sect.~\ref{subsec:pdfeftcorr} we evaluate the correlation between PDFs and EFT coefficients
to identify the kinematic region and EFT operators which are potentially
sensitive to the SMEFT-PDF interplay to be studied in Sect.~\ref{sec:joint_pdf_smeft}.

\subsection{Fit settings}
\label{subsec:smeftfitsettings}

Throughout this section, we will allow only the SMEFT coefficients to vary in the fit,
keeping the PDFs fixed to the SM-PDFs baseline obtained in the
\nnpdfnotop 
fit discussed in Sect.~\ref{sec:baseline_sm_fits}; with this choice, one removes the overlap
between the datasets entering the PDF fit and the EFT coefficients determination.
Our analysis is sensitive to the $N=25$ Wilson coefficients defined
in Table~\ref{tab:ops}, except at the linear level where the four-heavy
coefficients $c_{Qt}^{8}$, $c_{QQ}^{1}$, $c_{QQ}^{8}$, $c_{Qt}^{1}$ and $c_{tt}^{1}$ (which are 
constrained only by $t\bar{t}t\bar{t}$ and $t\bar{t}b\bar{b}$ data)
exhibit three flat directions~\cite{Ethier:2021bye}. In order to tackle this, we remove the five four-heavy coefficients from the linear fit\footnote{In principle one could instead rotate to the principal component analysis (PCA) basis
  and constrain the two non-flat directions in the four-heavy subspace,
  but even so, the obtained constraints
  remain much looser in comparison with those obtained in the quadratic EFT fit~\cite{Ethier:2021bye}. 
  
  In our fits, we also keep the $t\bar{t} t\bar{t}$ and $t\bar{t} b \bar{b}$ datasets after removing the five four-heavy coefficients. We 
  have verified that including or excluding theses sets has no
  significant impact whatsoever on the remaining coefficients.
}
Hence, in our linear fit we have $N=20$ independent coefficients constrained
from the data, whereas in the quadratic fit, we fit all $N=25$ independent coefficients.

The linear EFT fits presented in this section are performed with the \simunet{}  methodology
in the fixed-PDF option, as described in Sect.~\ref{sec:methodology}; 
we explicitly verified that the \simunet{} methodology reproduces the posterior distributions 
provided by \smefit (using
either the NS (Nested Sampling) or MCfit options)
for a common choice of inputs, as explicitly demonstrated in App.~\ref{app:benchmark}.
However, in the case of the quadratic EFT fits we are unable to use the \simunet{} 
methodology due to a failure of the Monte-Carlo sampling method utilised in the \simunet{} and \smefit codes;
this is discussed in App.~\ref{app:quad} and a dedicated study of the problem will be the subject of future work.
For this reason, quadratic EFT fits in this section are carried out
with the public {\sc\small SMEFiT} code using the NS
mode~\cite{Giani:2023gfq}. 
To carry out these fits, the full dataset listed in
Tables~\ref{tab:input_datasets_toppair}--\ref{tab:input_datasets_4}, together
with the corresponding SM and EFT theory calculations described in Sect.~\ref{sec:theory},
have been converted to the {\sc\small SMEFiT} data format (this conversion
was also already used for the benchmarking of App.~\ref{app:benchmark}).

\paragraph{Fisher information.}
The sensitivity to the EFT operators of the various processes entering
the fit can be evaluated by means of the Fisher information, $F_{ij}$, which
quantifies the information carried by a given dataset on the EFT coefficients $c_i$~\cite{Brehmer:2016nyr}. 
In a linear EFT setting,
the Fisher information is given by:
\begin{equation}
\label{eq:fisherdef}
F_{ij} = L^{(i)T} ({\rm cov}_{\rm exp})^{-1} L^{(j)}
\end{equation}
where the $k$-th entry, $L^{(i)}_k$, of the vector $L^{(i)}$ is the
linear contribution multiplying $c_i$ in the SMEFT theory prediction
for the $k$-th data point, and ${\rm cov}_{\rm exp}$ is the
experimental covariance matrix. In particular, the Fisher information
is an $N \times N$ matrix, where $N$ is the number of EFT
coefficients, and it depends on the dataset. 
An important property of the Fisher information is that it is related the covariance matrix $C_{ij}$ of the maximum likelihood estimators by the \textit{Cramer-Rao bound}:
\begin{equation}
\label{eq:cramerrao}
C_{ij} \geq (F^{-1})_{ij},
\end{equation}
indicating that larger values of $F_{ij}$ will translate to tighter bounds on the
EFT coefficients.

\begin{figure}[t]
        \centering
        \includegraphics[scale=0.48]{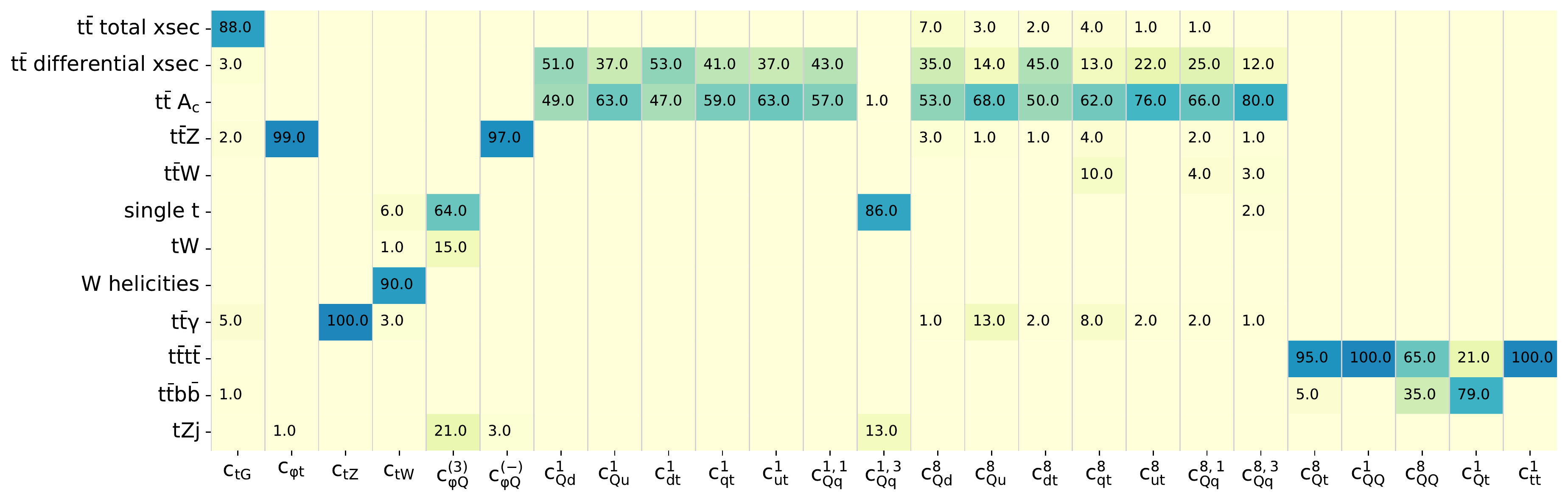}
	\caption{Relative constraining power (in \%) on each of the operators for each of the processes entering the fit, 
	as defined in Eq.~\eqref{eq:relativeconstrainingpower}.}
	\label{fig:fisher_diag}
\end{figure}

Before displaying the results of the fixed-PDF SMEFT analysis in
Sect.~\ref{subsec:smeftresults}, we use the Fisher information 
to assess the relative impact of each sector of top quark data on the
EFT parameter space; this is done in the linear analysis, including
${\cal O}(1/\Lambda^2)$ SMEFT corrections.  
In the quadratic case, once ${\cal O}(1/\Lambda^4)$ SMEFT corrections
are included, the dependence of $F_{ij}$ on the Wilson coefficients makes interpretation more difficult.
Writing $F_{ij}(D)$ for the
Fisher information matrix evaluated on the dataset $D$, we define the relative constraining power of the dataset $D$
via:
\begin{equation}
\label{eq:relativeconstrainingpower}
	\textrm{relative constraining power of }D\textrm{ on operator }c_i = F_{ii}(D) \bigg/ \displaystyle \sum_{\text{sectors }D'} F_{ii}(D').
\end{equation}
Since $F_{ii}(D)$ corresponds to the constraining power of the dataset $D$ in a one-parameter fit of the Wilson coefficient $c_i$,
this definition only quantifies how much a dataset impacts one-parameter fits of single Wilson coefficients in turn; however, this will give 
a general qualitative picture of some of the expected behaviour in the global fit too. 
We display the results of evaluating the relative constraining power of each top quark data sector on each of the
parameters in Fig.~\ref{fig:fisher_diag}, quoting the results in percent (\%).

As expected, $t \bar{t}$ total cross sections constitute the dominant source of constraints
on the coefficient $c_{tG}$.  Each of the four-fermion operators receive important constraints
from differential $t \bar{t}$ distributions and charge asymmetry measurements.
Note that this impact is magnified when we go beyond individual fits, in which
case measurements of charge asymmetries are helpful in breaking flat directions 
amongst the four-fermion operators~\cite{Brivio:2019ius,Ellis:2020unq}.
The coefficient $c_{Qq}^{1,3}$ is the exception as it 
is instead expected to be well-constrained by single top production.  
We note that the measurements of $W$ helicities are helpful in constraining the coefficient $c_{tW}$,
while $t \bar{t} Z$ measurements provide the dominant source of constraints on $c_{\varphi Q}^{(-)}$.
We observe that the neutral top coupling $c_{tZ}$ is entirely 
constrained by $t \bar{t} \gamma$, 
and that the effects of $t \bar{t} t \bar{t}$ and $t \bar{t} b \bar{b}$ are mostly restricted to the 4-heavy operators
$c_{Qt}^{8}$, $c_{QQ}^{1}$, $c_{QQ}^{8}$, $c_{Qt}^{1}$ and $c_{tt}^{1}$. 

\subsection{Fixed-PDF EFT fit results}
\label{subsec:smeftresults}

In this section, we present the results of the linear and quadratic fixed-PDF fits
with the settings described in Sect~\ref{subsec:smeftfitsettings}. 

\paragraph{Fit quality.}
We begin by discussing the fit quality of the global SMEFT determination, quantifying 
the change in the data-theory agreement relative to the SM in both the linear and quadratic SMEFT
scenarios.  Table~\ref{tab:smeft_fit_quality} provides the values of the $\chi^2$ per data point in the SM and in the case of 
the SMEFT at both linear and quadratic order in the EFT expansion for
each of the processes entering the fit.
Here, in order to ease the comparison of our results to those of
\smefit and \fitm, we quote the $\chi^2$ per data point computed
by using the covariance matrix defined in Eq.~\eqref{eq:expthcov},
which includes both the experimental uncertainty and the PDF
uncertainty. The corresponding values obtained by using the
experimental $\chi^2$ definition of Eq.~\eqref{eq:chi2exp}, along with
a fine-grained fit quality description are given in App.~\ref{app:fit_quality}.

We observe that in many sectors, the linear EFT fit improves the fit
quality compared to the SM; notably, the $\chi^2_{\rm exp+th}$ per data point for inclusive $t\bar{t}$ is vastly improved from 1.71 to 1.11. When
quadratic corrections are also considered, the fit quality is usually poorer compared to the linear fit. For example,
in inclusive $t\bar{t}$ the $\chi^2_{\rm exp+th}$ per data point deteriorates from 1.11 to 1.69. This is not unexpected, however,
since the flexibility of the quadratic fit is limited by the fact that 
for sufficiently large values of Wilson coefficients the EFT can only
make positive corrections.\footnote{This also has methodological implications.
Large quadratic corrections can negatively impact the Monte-Carlo sampling method
used by \simunet, as discussed in App.~\ref{app:quad}.}

\begin{table}[t!]
\begin{center}
\centering
\renewcommand{\arraystretch}{1.7}
\begin{tabularx}{\textwidth}{lcccc}
\toprule
Process  $\qquad $ & $n_{\rm dat}$ & $\qquad $$\chi^2_{\rm exp+th}$ [SM] & $\quad $ $\chi^2_{\rm exp+th}$ [SMEFT $\mathcal{O}(\Lambda^{-2})]$ 
& $\chi^2_{\rm exp+th}$ [SMEFT $\mathcal{O}(\Lambda^{-4})]$
\\ \midrule
	$t \bar{t}$ & 86 & 1.71 & 1.11 & 1.69 \\
	$t \bar{t}$ AC & 18 & 0.58 & 0.50 & 0.60 \\
	$W$ helicities & 4 & 0.71 & 0.45 & 0.47  \\
        \midrule
        $t\bar{t}Z$ & 12 & 1.19 & 1.17 & 0.94  \\ 
        $t\bar{t}W$ & 4 & 1.71 & 0.46 & 1.66 \\
        $t \bar{t} \gamma$ & 2 & 0.47 & 0.03 & 0.59 \\ 
        \midrule
        $t \bar{t} t \bar{t}$ \& $t \bar{t} b \bar{b}$ & 8 & 1.32 & 1.06 & 0.49  \\
        \midrule
        single top & 30 & 0.504& 0.33 & 0.37  \\ 
        $tW$ & 6 & 1.00 & 0.82 & 0.82 \\ 
        $tZ$ & 5 & 0.45 & 0.30 & 0.31 \\
        \midrule
        {\bf Total} & {\bf 175} & {\bf 1.24} & {\bf 0.84} & {\bf 1.14}  \\ 
\bottomrule
\end{tabularx}
\end{center}
\caption{The values of the $\chi^2$ per data point for the fixed-PDF EFT fits presented in this section,
  both for individual groups of processes, and for the total
  dataset. Here the $\chi^2$ is actually the $\chi^2_{\rm exp+th}$
  defined by using the theory covariance matrix defined in
  Eq.~\eqref{eq:expthcov}. 
  In each case we indicate the number of data points, the $\chi^2_{\rm exp+th}$ obtained using the baseline
  SM calculations, and the results of both the linear and quadratic EFT fits.
}
\label{tab:smeft_fit_quality}
\end{table}

It is also useful to calculate the goodness of fit, quantified by the
the $\chi^2$ per degree of freedom, $\chi^2/n_{\rm dof} = \chi^2/(n_{\rm dat} - n_{\rm param})$, 
which additionally accounts for the complexity of the models we are
using in each fit. In our case, we find $\chi^2_{\rm exp+th}/n_{\rm dof} = 1.25$ in the SM and $\chi^2_{\rm exp+th}/n_{\rm dof} = 0.95$ 
and $1.33$ in the linear
and quadratic EFT scenarios respectively.  
We see that while the EFT at quadratic order does not provide a better fit than the SM, neglecting quadratic EFT corrections leads to
a significant improvement in the overall fit quality.

\begin{figure}[tb!]
	\centering
	\includegraphics[scale=0.7]{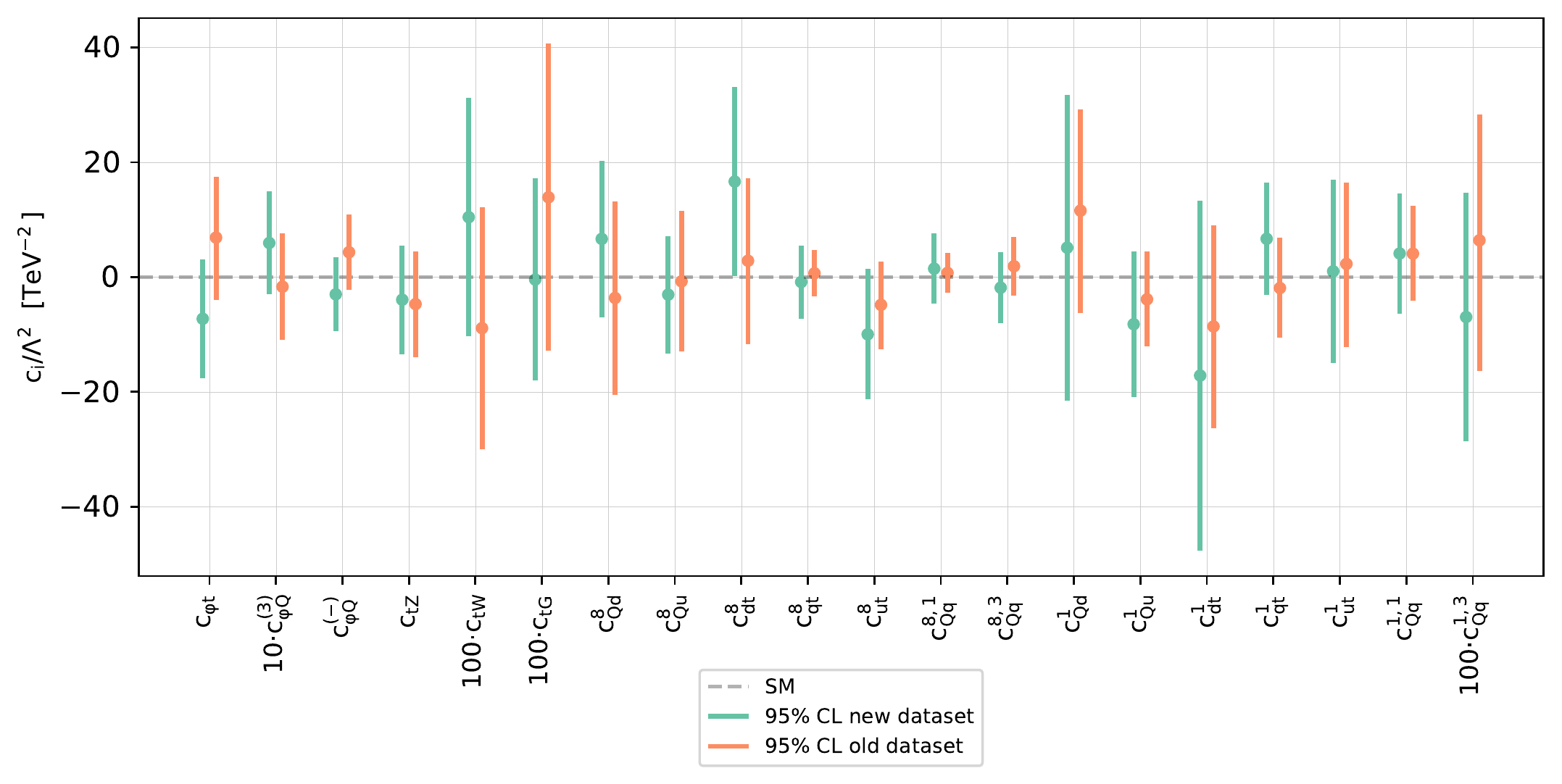}
	\caption{The 95\% CL intervals on the EFT coefficients entering the linear fit,
          evaluated with \simunet{} on the dataset considered in this work, and evaluated with \smefit on the
          top quark dataset entering the analysis of~\cite{Ethier:2021bye} 
          (note that at the linear EFT level, the results obtained with \simunet{}
          coincide with those provided by \smefit for the same dataset, as demonstrated in
          App.~\ref{app:benchmark}).
         Note also that the constraints on selected coefficients 
          are rescaled by the factors shown, for display purposes.
	  }
	\label{fig:linear_eft_constraints}
\end{figure}

\begin{figure}[tb!]
        \centering
        \includegraphics[scale=0.7]{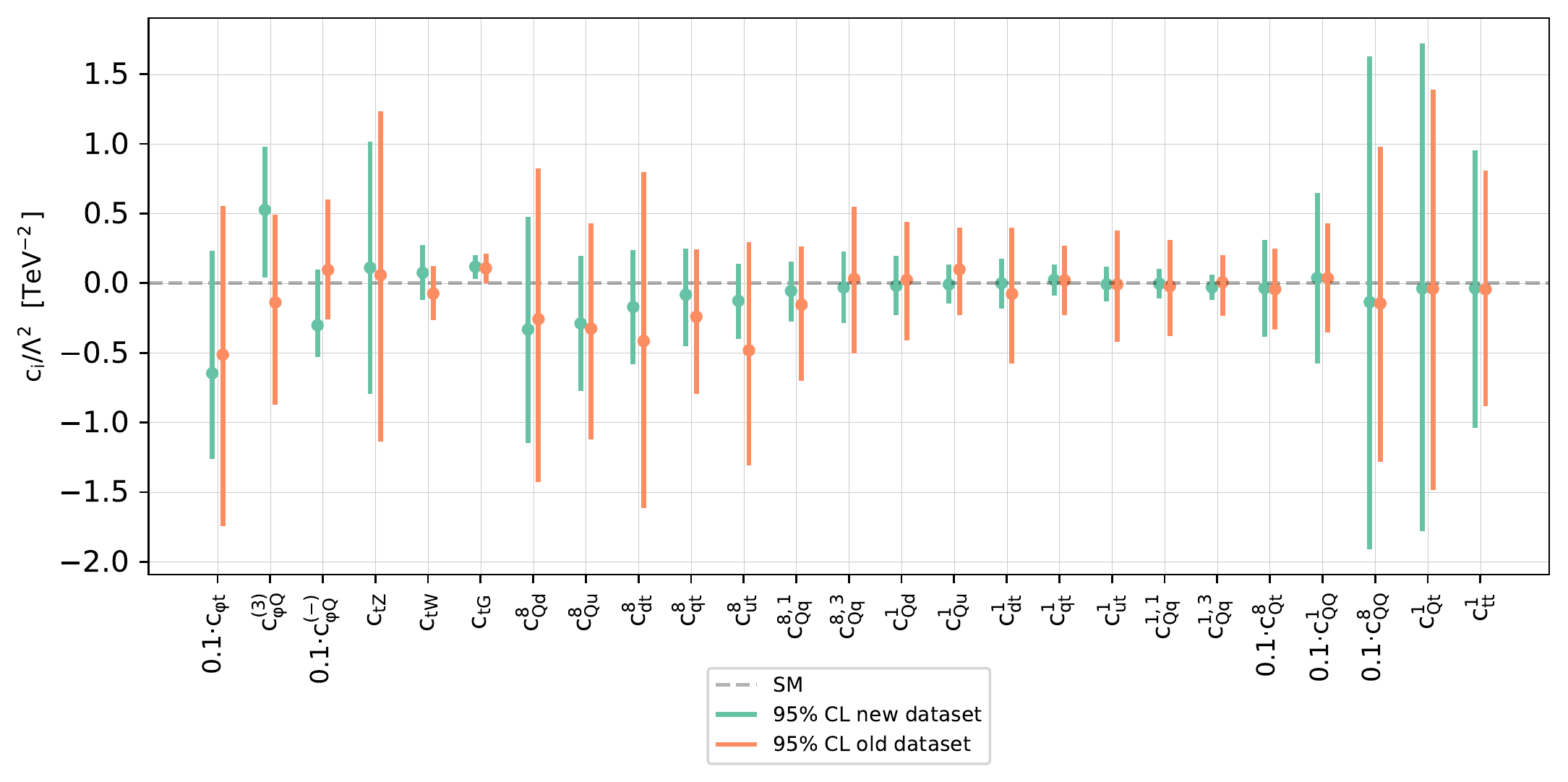}
	\caption{The 95\% CL intervals on the EFT coefficients entering the quadratic fit,
          evaluated with {\sc\small SMEFiT}.
          We compare the results based on the full top quark dataset with
          the corresponding results obtained from the subset of top quark
          measurements entering the analysis of~\cite{Ethier:2021bye}.  
         As in Fig.~\ref{fig:linear_eft_constraints}, the constraints on selected coefficients 
          are rescaled by the factors shown, for display purposes.        
        }
	\label{fig:quadratic_eft_constraints}
\end{figure}

\paragraph{Constraints on the EFT parameter space.}
Next, we present the constraints on the EFT parameter space. In
Fig.~\ref{fig:linear_eft_constraints}, we display the 95\% CL
constraints on the 20 Wilson coefficients entering the linear fit. Two
sets of constraints are shown; in green, we give the intervals
obtained from a fit to the 175 data points introduced in
Sect.~\ref{sec:exp}, whilst in orange, we give in the intervals
obtained from a fit to the older top quark dataset used in the global
analysis of Ref.~\cite{Ethier:2021bye}, obtained from a fit of 150
data points. This comparison allows us to quantify the information
gained from the latest Run II datasets, relative to those available to
previous analyses. The same comparison, this time at quadratic order in the EFT expansion,  
is shown in Fig.~\ref{fig:quadratic_eft_constraints} (note that in this plot we display constraints on 
all 25 coefficients, including the 4-heavy coefficients $c_{Qt}^{8}$, $c_{QQ}^{1}$, $c_{QQ}^{8}$, $c_{Qt}^{1}$ and $c_{tt}^{1}$).

We first note that Figures~\ref{fig:linear_eft_constraints} and~\ref{fig:quadratic_eft_constraints} both
demonstrate good agreement between the fits using old and new datasets,
and consistency between the new fit SMEFT bounds and the SM.
At the linear level, the most noticeable improvement concerns $c_{tG}$; its
 95\% C.L. bounds decrease from $[-0.13, 0.41]$ to $[-0.18, 0.17]$,
 thanks to the increased amount of information in the input dataset,
 coming in particular from $t \bar{t}$ data. This results in both a 
 tightening of the constraints by about 35\% and a shift in the best-fit point. 
For many of the other coefficients, the bounds are either stable
(e.g. $c_{tZ}$), or exhibit a shift in central value but no significant tightening (e.g. $c_{\varphi t}$, undergoes
a shift of -14.3, but a decrease in the size of the constraints 95\%
C.L. by only 1\%, and $c_{\varphi Q}^{(-)}$ undergoes a shift of
$-7.33$, but its bounds only tighten by 2\%). Finally, we note that some coefficients instead exhibit a broadening of constraints with the new dataset relative to the old
dataset (for example, some of the four-fermion operators). The increase in the size of the constraints could point to some
inconsistency within the new inclusive $t\bar{t}$ dataset; however, given that the bounds are very large anyway, at the edges of the 
intervals we are likely to approach a region where the EFT is no
longer valid in both cases, hence no definite conclusions may be
drawn. 

At the quadratic order in the EFT expansion, however, the impact of the latest Run II dataset is clear;
we see a marked improvement in many of the SMEFT constraints.
As shown in Fig.~\ref{fig:quadratic_eft_constraints}, the bounds on all 14 of the four-fermion operators become noticeably smaller as a result of 
the increase in precision in the $t \bar{t}$ sector.
The constraint on $c_{tZ}$ is improved by the inclusion of
measurements of the $t \bar{t} \gamma$ total cross sections, resulting
in a tightening of 24\%.
The addition of the $p_{T}^{\gamma}$ spectrum~\cite{Aad:2020axn} would
yield an even stronger constraint, as seen in~\cite{Ellis:2020unq}. We will make use of this observable in future work when 
unfolded measurements are made available.
Contrary to the linear fit, where we singled out $c_{\varphi t}$ and $c_{\varphi Q}^{(-)}$ as examples of coefficients which shift, but whose bounds are not improved, the constraints on $c_{\varphi t}$, $c_{\varphi Q}^{(-)}$ markedly tighten in the quadratic fit in the presence of new data; in particular, the size of the bounds on $c_{\varphi t}$, $c_{\varphi Q}^{(-)}$ decrease by 35\% and 28\%, respectively.
On the other hand, despite the addition of new $t\bar{t}t\bar{t}$,
$t\bar{t} b \bar{b}$ datasets, we find limited sensitivity to the five
four-heavy coefficients $c_{Qt}^{8}$, $c_{QQ}^{1}$, $c_{QQ}^{8}$,
$c_{Qt}^{1}$ and $c_{tt}^{1}$. In fact, with the new data we see a
broadening of the bounds. As with the linear fit, this could point to
either an inconsistency in the $t\bar{t}t\bar{t}$, $t\bar{t}b\bar{b}$ data, or
simply to the ambiguity associated to the EFT validity in that
particular region of the parameter space. 

\begin{figure}[htb!]
\centering
        \begin{subfigure}[b]{0.49\textwidth}
        \includegraphics[scale=0.35]{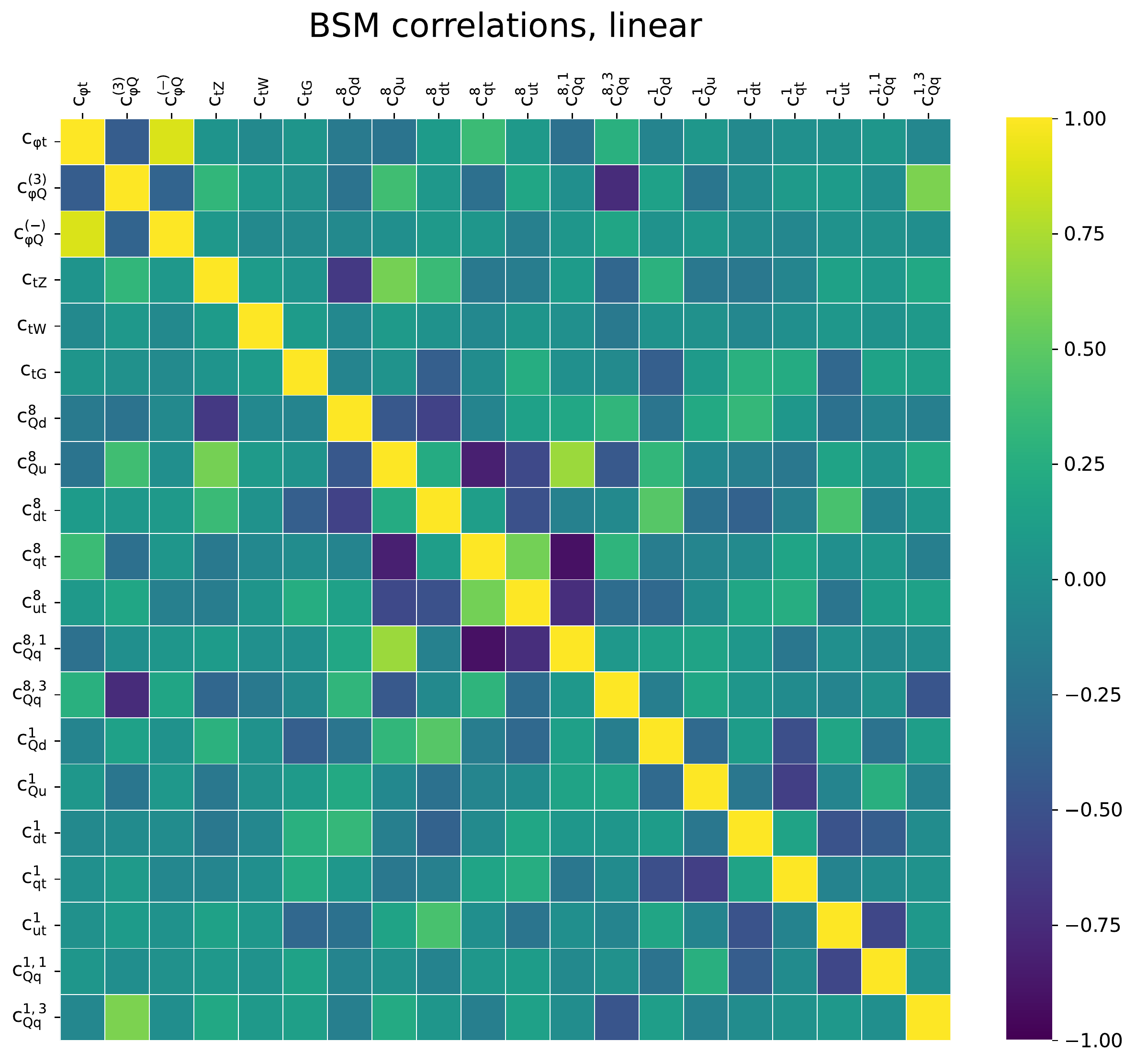}
        \end{subfigure}
        \begin{subfigure}[b]{0.49\textwidth}
        \includegraphics[scale=0.39]{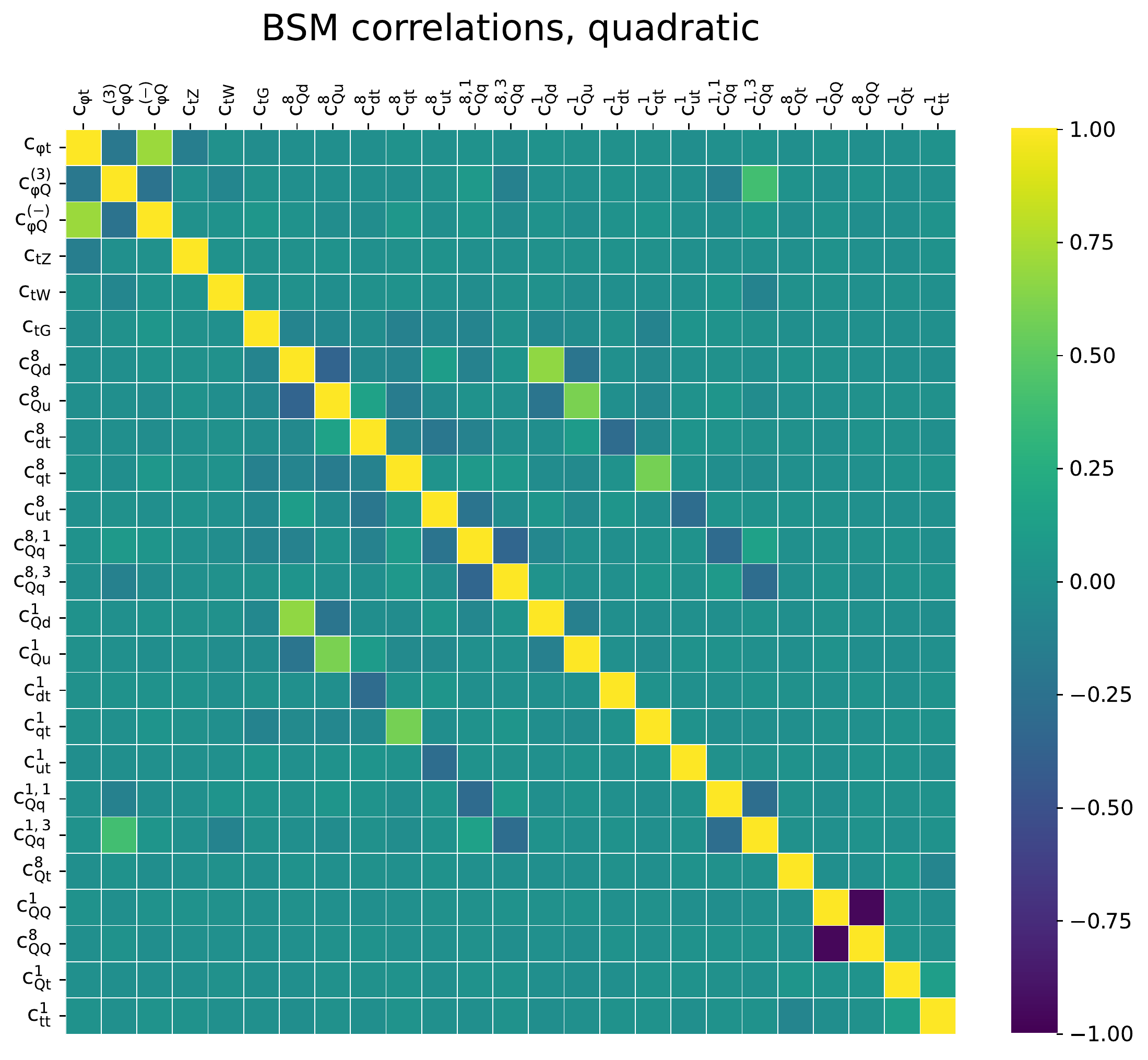}
        \end{subfigure}
	\caption{The values of the correlation coefficients evaluated
          between all pairs of Wilson coefficients entering the
          EFT fit at the  linear  (left) and quadratic  (right)  order.
          As explained in the next, the number of fitted DoFs is different
          in each case.
        }
\label{fig:bsmcorr}
\end{figure}

\paragraph{Correlations.}
Figure~\ref{fig:bsmcorr} shows the correlations the between Wilson
coefficients evaluated in this analysis both at the linear and the
quadratic order in the EFT expansion, shown on the left and right
panels respectively.
In the linear fit, we first note a number of large correlations amongst the octet four-fermion operators
which enter the $t \bar{t}$ production together.  
The singlet four-fermion operators are similarly correlated among themselves, although 
their correlations are comparatively suppressed.  
The coefficients $c_{\varphi t}$ and $c_{\varphi Q}^{(-)}$
exhibit a large positive correlation due to their
entering into $t \bar{t} Z$ production together,
while $c_{Qq}^{3,1}$ and $c_{\varphi Q}^{(3)}$ have positive
correlations through their contribution to single top production.
Further non-zero correlations are found, for example amongst the pairs $c_{Qq}^{8,3}$ \& $c_{\varphi Q}^{(3)}$,
$c_{Qd}^{8}$ \& $c_{tZ}$ and $c_{Qu}^{8}$ \& $c_{tZ}$.

At quadratic order, however, we observe that many of these correlations are suppressed,
as a result of the fact that the inclusion of quadratic corrections lifts many of the 
degeneracies in the fit.
We observe that the pairs $c_{\varphi t}$, $c_{\varphi Q}^{(-)}$
and $c_{Qq}^{1,3}$, $c_{\varphi Q}^{(3)}$ remain correlated though.
The 4-heavy operators are also included in this quadratic fit, and we find large anti-correlations between 
$c_{QQ}^{1}$ and $c_{QQ}^8$, indicating that they are poorly distinguished in the $t \bar{t} t \bar{t}$
and $t \bar{t} b \bar{b}$ processes.  Finally, note that we obtain subtle non-zero correlations
between the octet and singlet four-fermion operators constructed from the same fields, for example between $c_{Qd}^8$ and $c_{Qd}^1$.
This is a result of the fact that $t \bar{t}$ measurements provide the dominant source of constraints on these coefficients and are
very sensitive to
quadratic corrections, and at this order the contribution from these operators differs only by
a numerical factor.

\subsection{Study of the CMS 1D vs 2D distribution}
\label{subsec:cms1dvs2d}
In the dataset selection discussed in Sect.~\ref{sec:dataselection}, 
the double-differential distribution measurements performed by CMS at $\sqrt{s}=13$~TeV
in the $\ell+$jets channel~\cite{CMS:2021vhb} -- binned with respect to the top
quark pair invariant mass $m_{t\bar{t}}$ and the top quark pair rapidity $y_{t\bar{t}}$ -- is found to be poorly described by the
SM theory, with an experimental $\chi^2$ that is $22\sigma$ 
away from the median of the $\chi^2$ distribution of a
perfectly consistent dataset made of $n_{\rm dat}=34$.
Even by increasing the weight of this dataset in a weighted fit (see Sect.~\ref{sec:dataselection} for a more
detailed explanation), the $\chi^2_{\rm exp}$ per data point improves
only moderately to 4.56 and the  $\chi^2$-statistic of the other datasets deteriorates significantly, hinting
to both an internal incompatibility of the CMS 2D distribution and to an incompatibility with the rest of the data. 
For this reason, the dataset was excluded from our analysis, and replaced by the the single-differential
distribution in $m_{t\bar{t}}$ -- presented in the same publication~\cite{CMS:2021vhb}. The latter is well 
described by the SM theoretical predictions. 

 In this section we present a dedicated analysis to assess whether the SMEFT corrections can improve the
 theoretical description of this dataset. In particular, we compare a fixed-PDF fit including the 13 TeV
 CMS double-differential $(m_{t\bar{t}},y_{t\bar{t}})$  distribution (CMS 2D)
 to the default one including the 13 TeV single-differential $m_{t\bar{t}}$ distribution (CMS 1D).

 %
\begin{table}[thb]
\begin{center}
\centering
\renewcommand{\arraystretch}{1.7}
\begin{tabularx}{\textwidth}{l|c|cc|cc|cc}
\toprule
Process  & $\qquad n_{\rm dat}\qquad$ &
\multicolumn{2}{|c|}{$\chi^2_{\rm exp+th}$ [SM]} &
\multicolumn{2}{|c|}{$\chi^2_{\rm exp+th}$ [SMEFT $\mathcal{O}(\Lambda^{-2})]$}&
\multicolumn{2}{|c}{$\chi^2_{\rm exp+th}$ [SMEFT $\mathcal{O}(\Lambda^{-4})]$}\\
& & CMS 1D & CMS 2D & CMS 1D & CMS 2D & CMS 1D & CMS 2D\\
\midrule
	$t \bar{t}$ & 86 & 1.71 & 2.07 & 1.11 &  1.18 & 1.69& 1.87\\
	$t \bar{t}$ AC & 18 & 0.58 & 0.58 & 0.50 & 0.47 & 0.60& 0.60 \\
	$W$ helicities & 4 & 0.71 & 0.71& 0.45 & 0.45 & 0.48& 0.46  \\
        \midrule
	$t\bar{t}Z$ & 12 & 1.19 & 1.19 & 1.17 & 1.07  &0.94 & 0.95  \\ 
	$t\bar{t}W$ & 4 & 1.71 & 1.71& 0.46 & 0.46 &1.66 & 1.82\\
        $t \bar{t} \gamma$ & 2 & 0.47 & 0.47& 0.03 & 0.03& 0.58 & 0.18\\ 
        \midrule
        $t \bar{t} t \bar{t}$ \& $t \bar{t} b \bar{b}$ & 8 & 1.32 &
        1.32 & 1.06 & 1.28& 0.49 & 0.49  \\
        \midrule
        single top & 30 & 0.50 & 0.50 & 0.33 & 0.34 & 0.37 & 0.35\\ 
        $tW$ & 6 & 1.00 & 01.00& 0.82 & 0.86 & 0.82 & 0.84\\ 
        $tZ$ & 5 & 0.45 & 0.45& 0.30 & 0.30 & 0.31 & 0.30\\
        \midrule
        {\bf Total} & {\bf 175} & {\bf 1.24} & {\bf 1.48} & {\bf
          0.84} & {\bf 0.91} & {\bf 1.14}  & {\bf 1.29}\\ 
\bottomrule
\end{tabularx}
\end{center}
\caption{\label{tab:chi2-cms2d1d} Same as Table~\ref{tab:smeft_fit_quality}, now comparing the values of the $\chi^2_{\rm exp+th}$ per data point for
  the fixed-PDF EFT fits presented in this section, the default one including the 13 TeV single-differential $m_{t\bar{t}}$ distribution (CMS 1D) and the one including the 13 TeV  CMS double-differential $(m_{t\bar{t}},y_{t\bar{t}})$  distribution (CMS 2D), both for individual groups of processes, and for the total dataset.
}
\end{table}

 First of all, it is interesting to notice that the inclusion of
 quadratic SMEFT corrections in the fit does not improve much the
 quality of the fit of the CMS 2D distribution, with $\chi^2_{\rm exp+th}/n_{\rm
   dat}$ decreasing from 2.80 (in the SM) 
to $\chi^2_{\rm exp+th}/n_{\rm dat}=2.57$ including  SMEFT $\mathcal{O}(\Lambda^{-4})$ corrections.
 On the other hand, if SMEFT linear $\mathcal{O}(\Lambda^{-2})$ corrections are included, the fit quality of the CMS 2D distribution
 improves substantially with $\chi^2_{\rm exp+th}/n_{\rm dat}=1.22$. 

 In order to assess what is the effect of the inclusion of the CMS 2D distribution on the other top sector data,
 in Table~\ref{tab:chi2-cms2d1d} we compare the fit quality of the two (fixed-PDF) SMEFT fits, CMS 1D and CMS 2D,
 both for individual groups of processes, and for the total dataset.
 We observe that the fit quality deteriorates both in the SM and in
 the quadratic SMEFT fit once the CMS 2D distribution is fitted. The deterioration of the fit
 quality is mostly driven by a deterioration of the fit quality of the
 $t\bar{t}$ and $t\bar{t}W$ sectors. However at the level of linear
 SMEFT fit, the quality of the fit deteriorates only moderately and it
 is mostly driven by the slight deterioration in the inclusive
 $t\bar{t}$ sector and in the $t\bar{t}t\bar{t}\,\&\,
 t\bar{t}b\bar{b}$ one.

At the level of fit of the Wilson coefficients, the quadratic SMEFT
fits yields similar 95\% C.L. bounds on the EFT coefficients. This is
somewhat expected, given that the fit quality of the CMS 2D data does not improve once quadratic SMEFT
corrections are included, and the fit quality of the other datasets remains pretty much the same.
\begin{figure}[htb]
        \centering
        \includegraphics[scale=0.7]{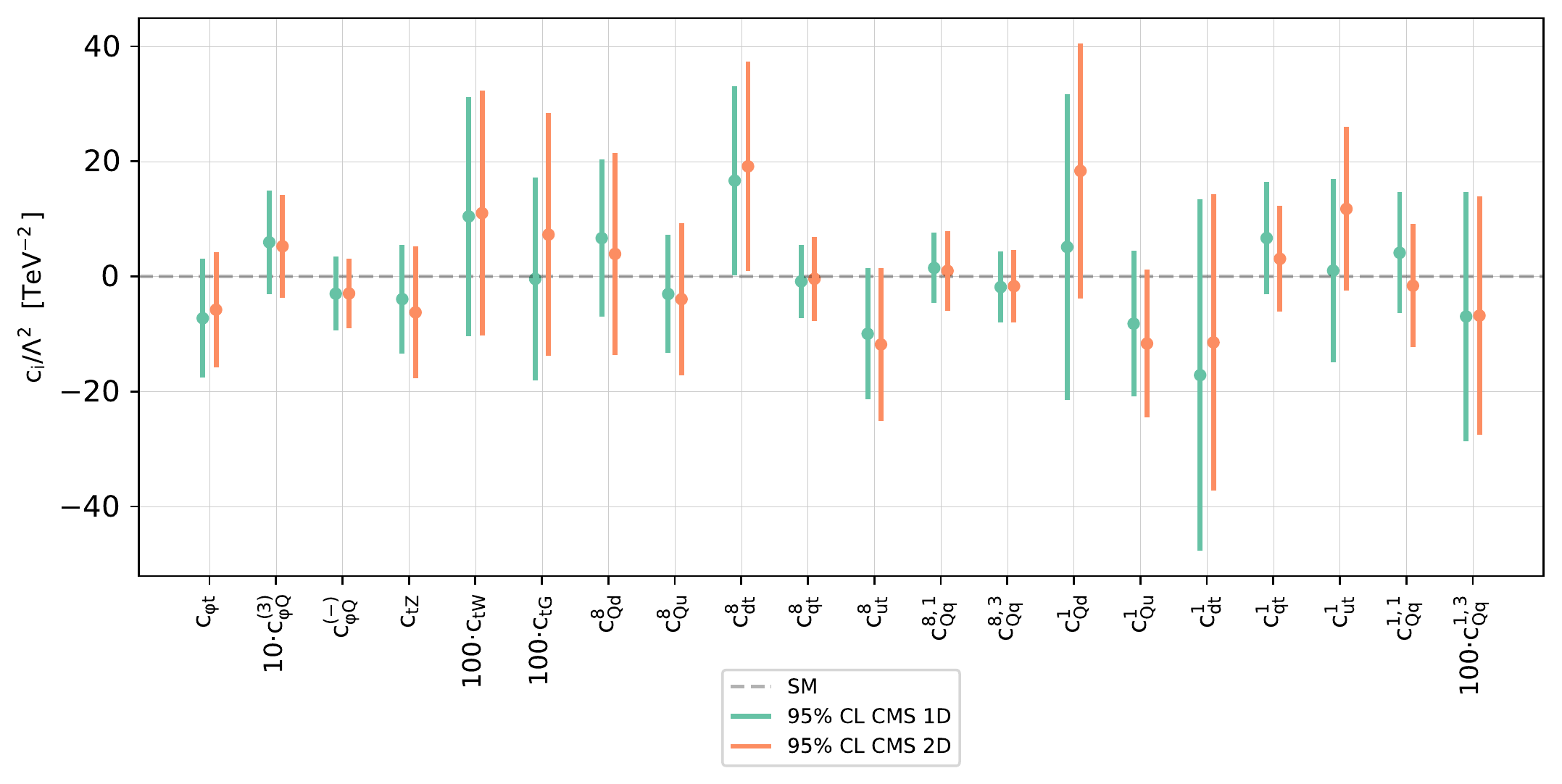}
	\caption{The 95\% CL intervals on the EFT coefficients entering the liner fits,
          evaluated with \simunet.
          We compare the results based on the default full top quark dataset (CMS 1D) with
          the corresponding results obtained from the same set of top quark
          measurements in which the single-differential $m_{t\bar{t}}$ 13 TeV CMS distribution is substituted by the
          double differential  $(m_{t\bar{t}},y_{t\bar{t}})$ 13 TeV CMS distribution (CMS 2D).        
        }
	\label{fig:linear_eft_cms2d_cms1d}
      \end{figure}
On the other hand the bounds obtained from a SMEFT linear fit change
more significantly depending on whether the CMS 1D or the CMS 2D
distribution is used in the fit.
In particular the central value of the $c_{tG}$ 95\% C.L. bounds is
shifted upwards by 1$\sigma$, as well as the bounds on $c^1_{Qd}$ and
$c^1_{ut}$, while the bounds on $c^1_{Qu}$ and
$c^1_{qt}$ shift downwards, as expected from the EFT coefficient
correlations displayed in Fig.~\ref{fig:bsmcorr}.

As a result of our analysis, we observe that the bounds on the EFT in
the linear case do depend on the input dataset quite significantly and
careful consideration has to be made to the overall dataset
compatibility and to the outcome of the fit once quadratic corrections
are included. The current analysis in particular shows that the
incompatibility of the 13 TeV CMS double differential distribution
cannot be entirely attributed to new physics effects parametrised by
the SMEFT expansion, rather there are internal experimental or
theoretical incompatibilities that affect the results.

\subsection{Correlations between PDFs and EFT coefficients}
\label{subsec:pdfeftcorr}

\begin{figure}[hbt!]
\centering
        \includegraphics[width=0.48\textwidth]{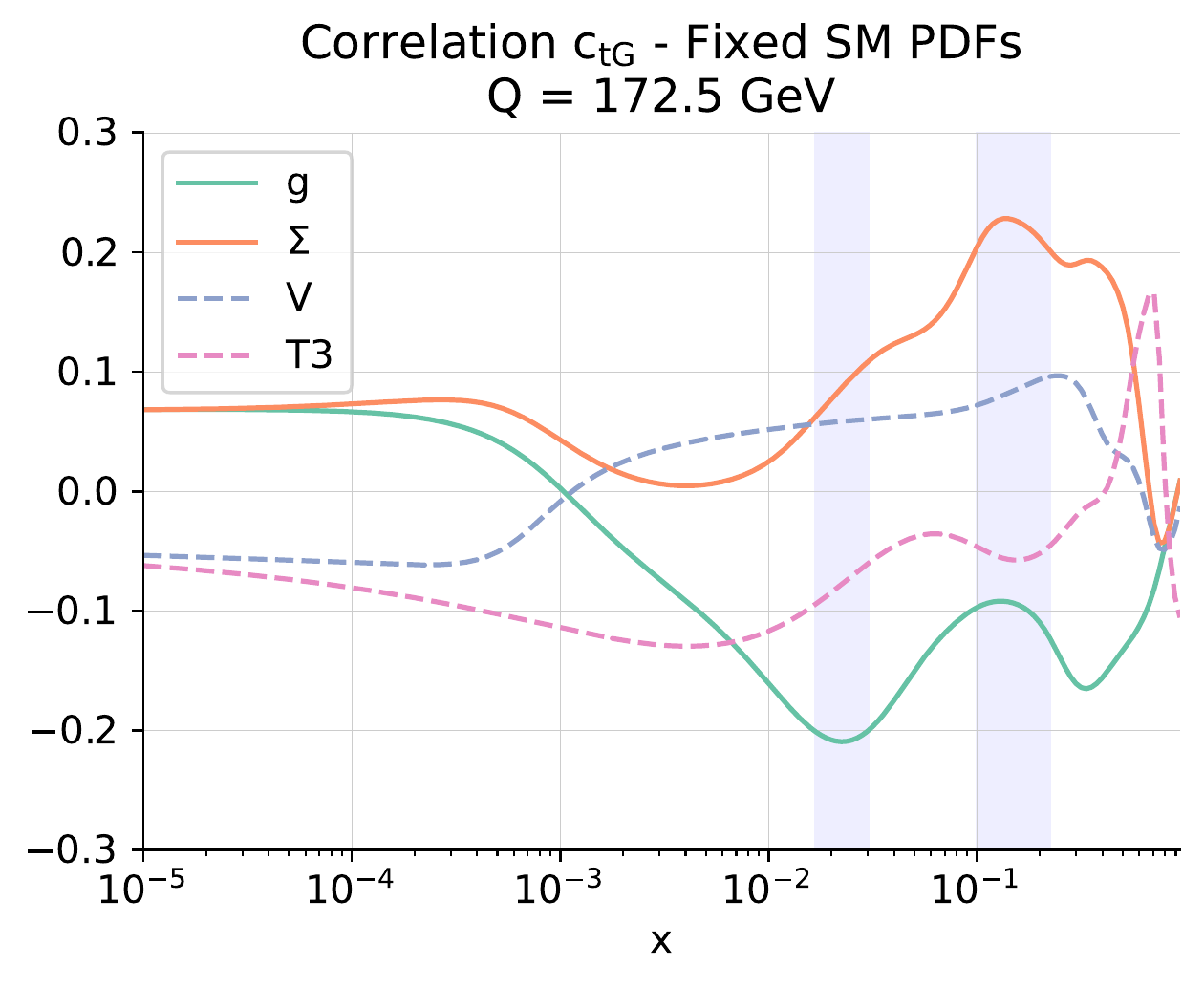}
        \includegraphics[width=0.48\textwidth]{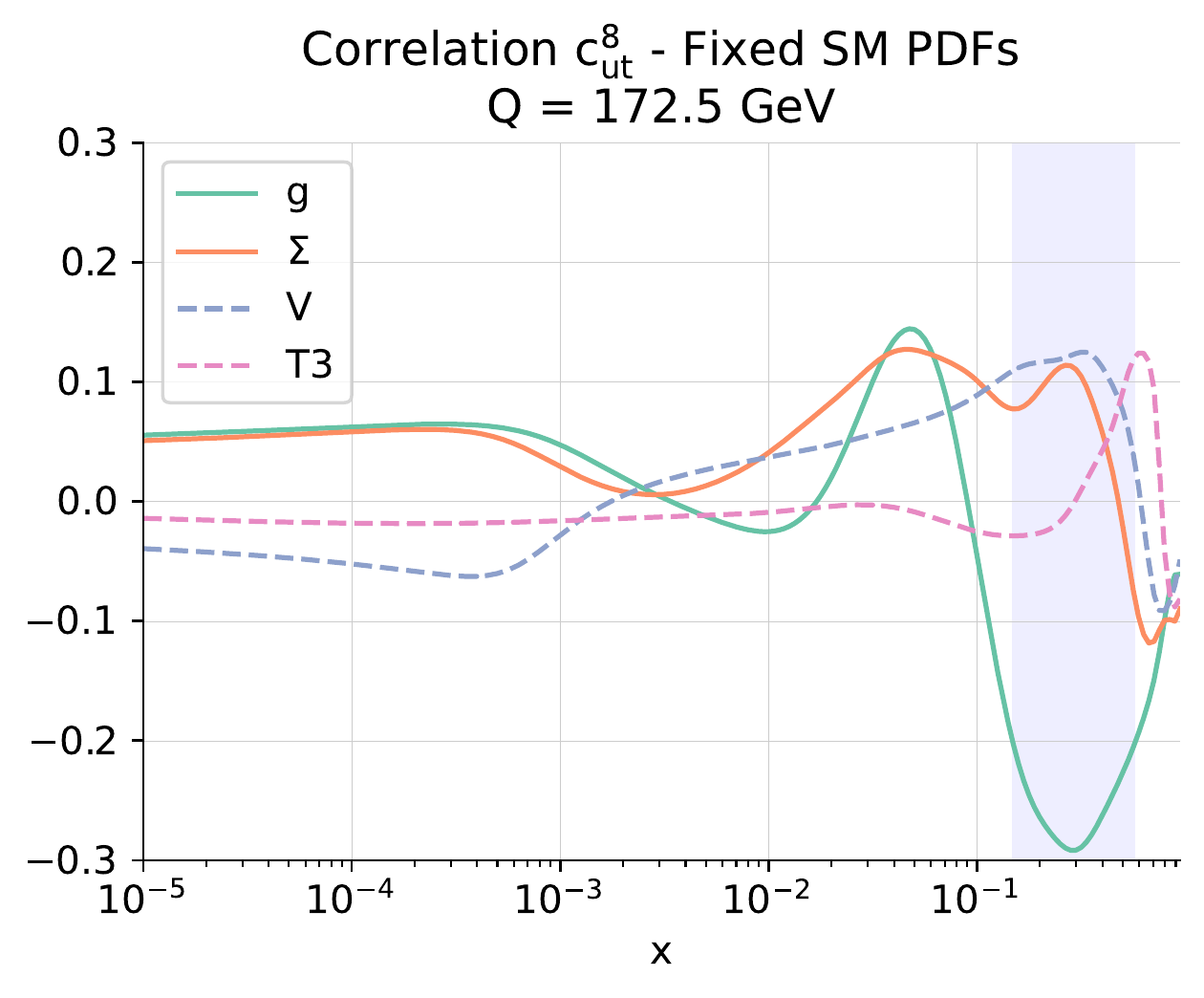}
	\caption{The value of the correlation coefficient $\rho$
          between the PDFs and selected EFT coefficients as a function of $x$
          and $Q=172.5$ GeV.
          We show the results for the gluon, the total singlet $\Sigma$, total valence $V$, and non-singlet
          triplet $T_3$ PDFs.
          We provide results for representative EFT coefficients, namely $c_{tG}$ and  $c_{ut}^{(8)}$.
        }
\label{fig:pdfbsmcorr}
\end{figure}

We conclude this section by discussing the correlations observed between the PDFs and Wilson coefficients.
The PDF-EFT correlation coefficient for a Wilson coefficient $c$ and a
PDF $f(x, Q)$ at a given $x$ and $Q^2$ is defined as
\begin{equation}
\label{eq:correlationL2CT}
\rho\lp c, f(x,Q^2)\rp=\frac{\la c^{(k)}f^{(k)}(x,Q^2)\ra_k - \la c^{(k)}\ra_k \la f^{(k)}(x,Q^2)\ra_k
}{\sqrt{\la (c^{(k)})^2\ra_k - \la c^{(k)}\ra_k^2}\sqrt{\la\left( f^{(k)}(x,Q^2)\right)^2\ra_k - \la f^{(k)}(x,Q^2)\ra_k^2}}  \, ,
\end{equation}
where $c^{(k))}$ is the best-fit value of the Wilson coefficient for
the $k$-th replica and $f^{(k)}$ is the $k$-th PDF replica computed at
a given $x$ and $Q$, and $\la \cdot \ra_k$ represents the average
over all replicas. We will compute the correlation between a SM PDF and the Wilson coefficients, both of which have been separately determined from 
the total dataset including all new top quark data.  
By doing so we hope to shed light on which Wilson coefficients,
and which PDF flavours and kinematical regions, are strongly impacted by the top quark data and therefore exhibit a potential for interplay
in a simultaneous EFT-PDF determination.
The EFT corrections will be
restricted to linear order in the EFT expansion.

Fig.~\ref{fig:pdfbsmcorr} displays a selection of the largest correlations.
We observe that the gluon PDF in the large-$x$ region is significantly correlated with the Wilson coefficients 
$c_{tG}$, $c_{ut}^{(8)}$.  On the other hand, relatively large correlations are observed between $c_{tG}$ and the
the total singlet $\Sigma$, while the total valence $V$, and non-singlet triplet $T_3$ PDFs show no relevant correlations
with the selected coefficients.
This is not surprising, given the impact of top quark pair production total cross sections and differential distributions
in constraining these PDFs and Wilson coefficients.  
Whilst these correlations are computed from a determination of the SMEFT in which the PDFs are fixed to SM PDFs, 
the emergence of non-zero correlations provides an indication of the potential for interplay between the PDFs and the SMEFT
coefficients; this interplay will be investigated in a simultaneous determination
in the following section.

\section{SMEFT-PDFs from top quark data}
\label{sec:joint_pdf_smeft}

In this section we present the main results of this work, namely
the simultaneous determination of the proton PDFs and the SMEFT Wilson coefficients
from the LHC Run II top quark data described in Sect.~\ref{sec:exp},
following the \simunet{} methodology summarised in Sect.~\ref{sec:methodology}.
This determination of the SMEFT-PDFs from top quark data is carried
out at the linear, $\mathcal{O}(1/\Lambda^2)$, level in the EFT expansion.
We do not perform simultaneous fits at the quadratic level due to
shortcomings of the Monte Carlo replica method, on which
\simunet{} is based; this is discussed in detail in App.~\ref{app:quad}, and
will also be the subject of future work.

\begin{figure}[t]
\centering
  \begin{subfigure}[b]{0.49\textwidth}
  \includegraphics[scale=0.5]{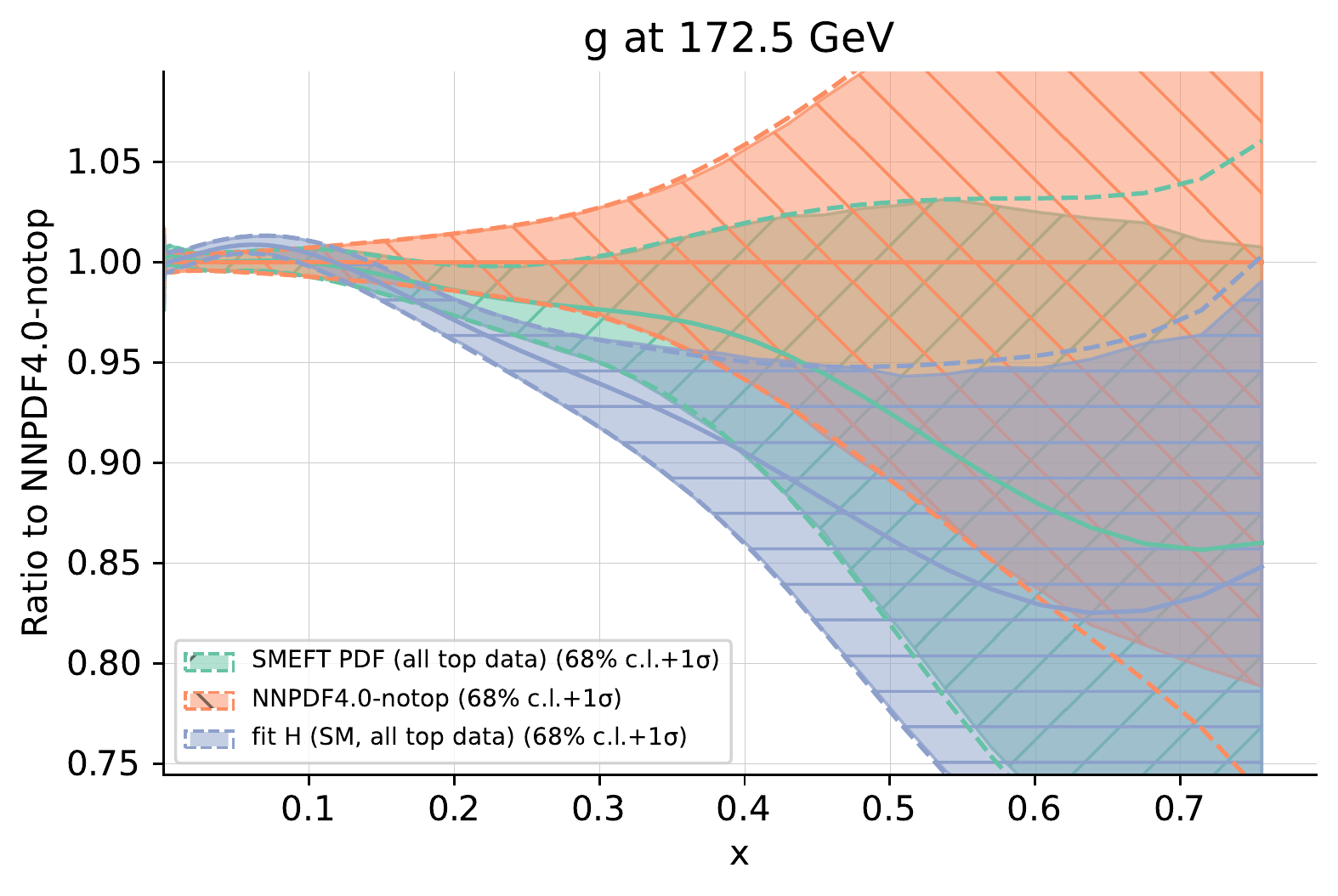}
  \end{subfigure}
  \begin{subfigure}[b]{0.49\textwidth}
    \includegraphics[scale=0.5]{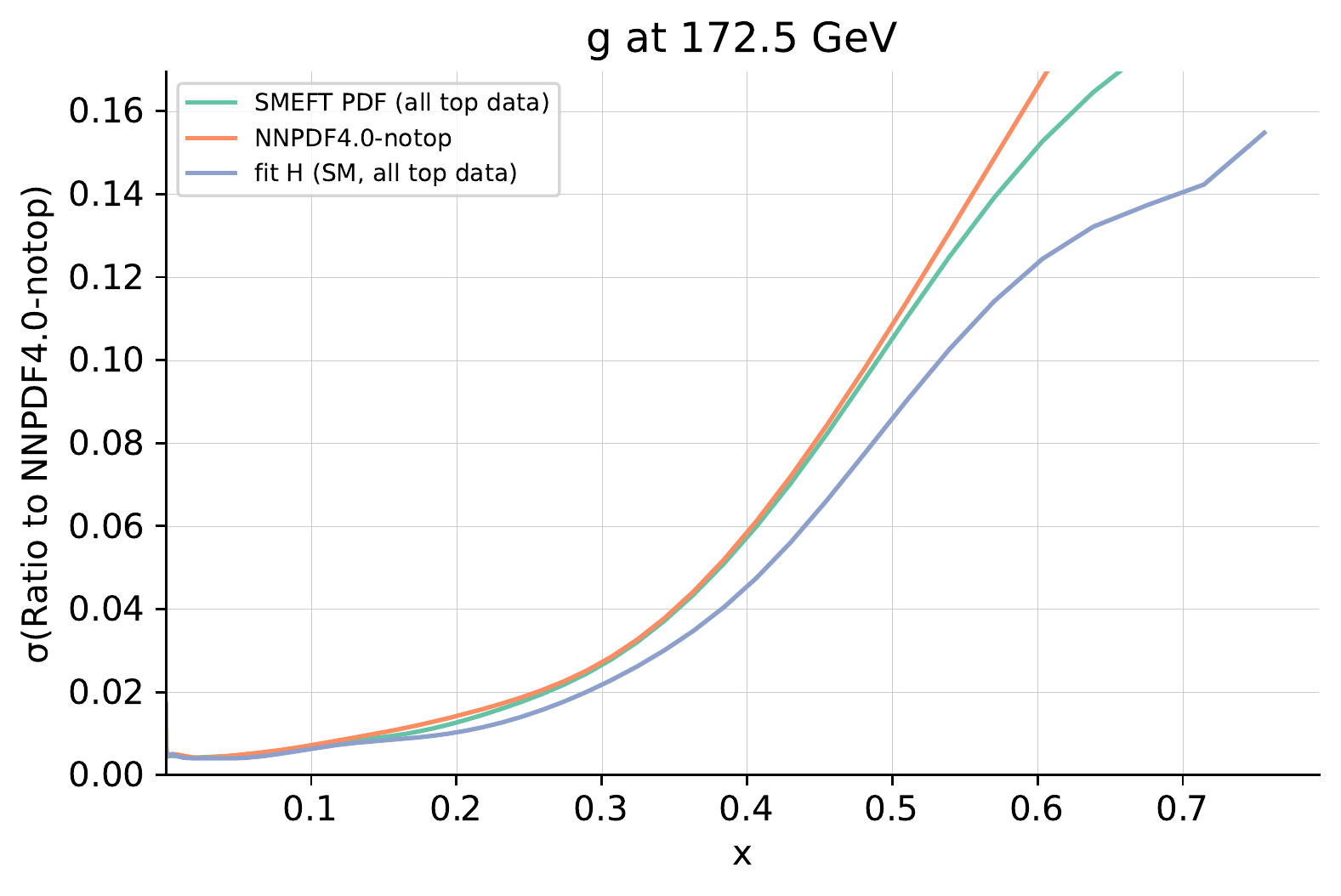}
  \end{subfigure}
   \begin{subfigure}[b]{0.49\textwidth}
  \includegraphics[scale=0.5]{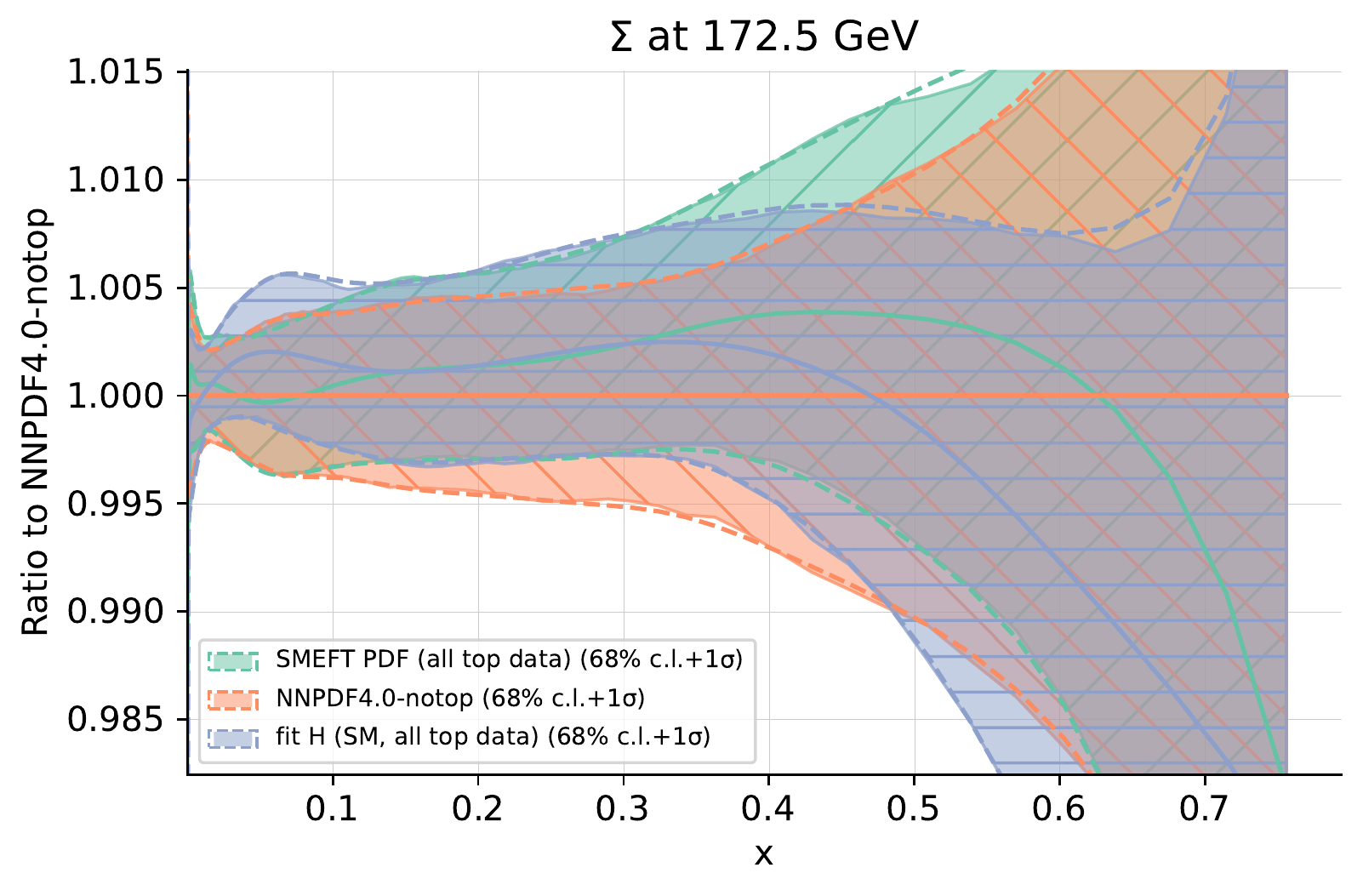}
  \end{subfigure}
  \begin{subfigure}[b]{0.49\textwidth}
  \includegraphics[scale=0.5]{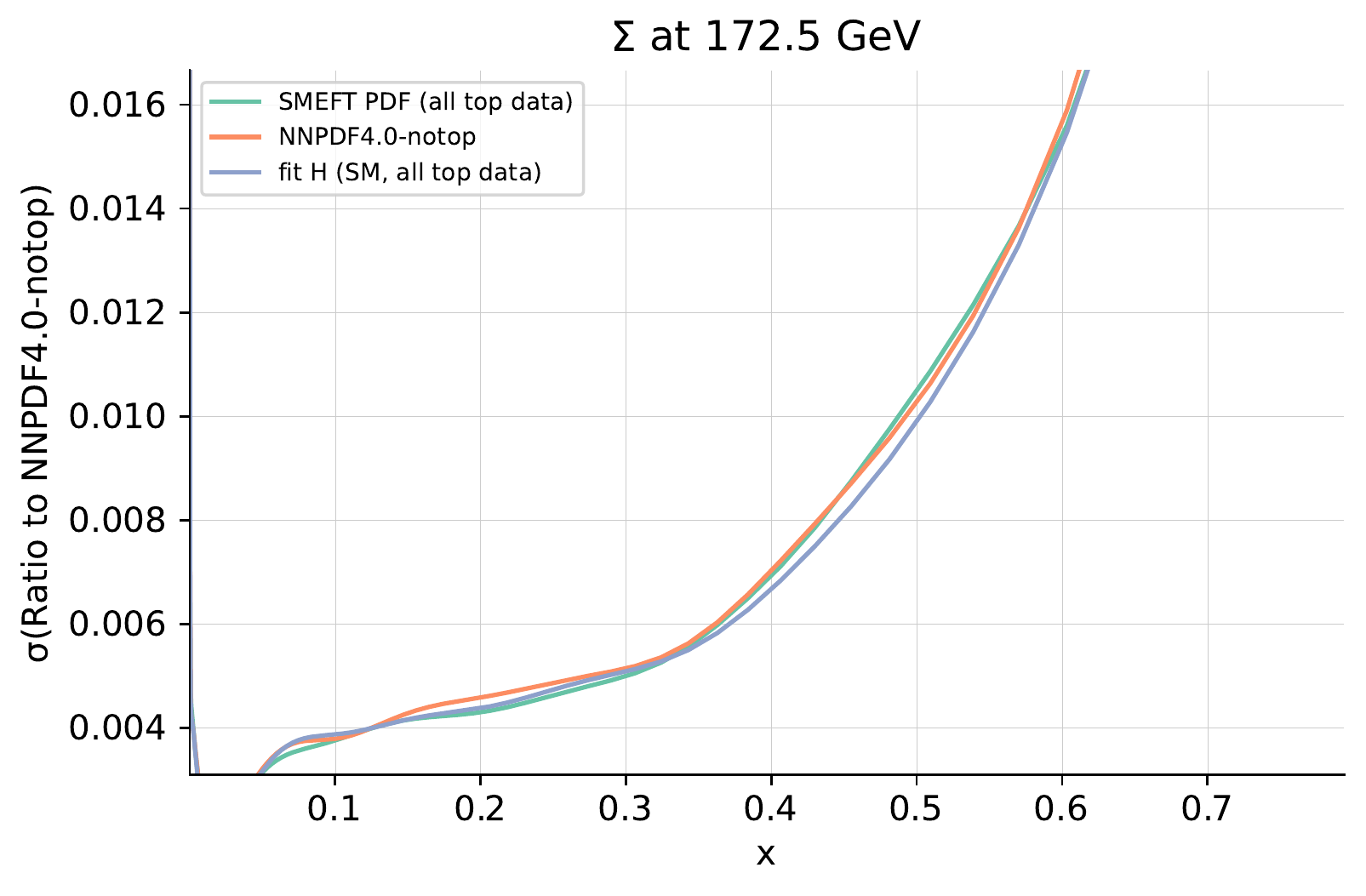}
  \end{subfigure}
        \caption{
    Left: a comparison between the gluon (upper panel)
    and quark singlet (lower panel) PDFs evaluated at $Q=m_t=172.5$ GeV
    in the large-$x$ region.
    We display the \nnpdfnotop baseline, the SM-PDFs of
    fit H in Table~\ref{tab:fit_list} which include
    the full top quark dataset, and their SMEFT-PDF counterparts.
    The results are normalised to the central value of the \nnpdfnotop fit.
    Right: the same comparison, but now for the relative $1\sigma$ PDF uncertainties.
}
\label{fig:SMEFTpdf_gluon}
\end{figure}

\paragraph{PDFs from a joint SMEFT-PDF fit.} We begin by discussing the PDFs obtained through a
joint fit of PDFs and Wilson coefficients
from the complete LHC top quark dataset considered in this work.
Simultaneously extracting the PDFs and the EFT coefficients
from top quark data has a marked impact on the former, as compared to a SM-PDF
baseline, but we shall see has much less impact on the latter, as compared to the results of the
corresponding fixed-PDF EFT analyses.
Fig.~\ref{fig:SMEFTpdf_gluon}
displays a comparison between the gluon and quark singlet PDFs, as
well as of their relative $1\sigma$ PDF uncertainties, for 
the \nnpdfnotop baseline, the SM-PDFs of
fit H in Table~\ref{tab:fit_list} which include
the full top quark dataset, and their SMEFT-PDF counterparts based
on the same dataset.
PDFs are compared in the large-$x$ region for   $Q=m_t=172.5$ GeV. In
the left panel they are normalised to the central value of the
\nnpdfnotop fit.

While differences are negligible for the quark singlet PDF, both in terms of
central values and uncertainties, they are more marked for the gluon
PDF. Two main effects are observed therein.
First, the central value of the SMEFT-PDF gluon moves upwards as compared
to the SM-PDF fit based on the same dataset, ending up halfway between the latter
and the \nnpdfnotop fit.
Second, uncertainties increase for the SMEFT-PDF determination as compared
to the SM-PDFs extracted from the same data, becoming close to the uncertainties
of the \nnpdfnotop fit except for $x\gsim 0.5$.
In both cases, differences are restricted to the large-$x$ region with $x\gsim 0.1$,
where the impact of the dominant top quark pair production measurements is
the largest.

The results of Fig.~\ref{fig:SMEFTpdf_gluon} for the gluon PDF therefore indicate that
within a simultaneous extraction of the PDFs and the EFT coefficients, the impact
of the top quark data on the PDFs is diluted, with the constraints it provides 
partially `reabsorbed' by the Wilson coefficients.
This said, there remains a pull of the top quark data as compared to
the no-top baseline fit
which is qualitatively consistent with the pull
obtained in a SM-PDF determination based on the same dataset, albeit of reduced magnitude.
Interestingly, as we show below, while the SMEFT-PDF gluon is significantly
modified in the joint fit as compared to a SM-PDF reference, much smaller
differences are observed at the level of the bounds on the EFT parameters themselves.

\begin{figure}[t]
\centering
\includegraphics[width=0.49\textwidth]{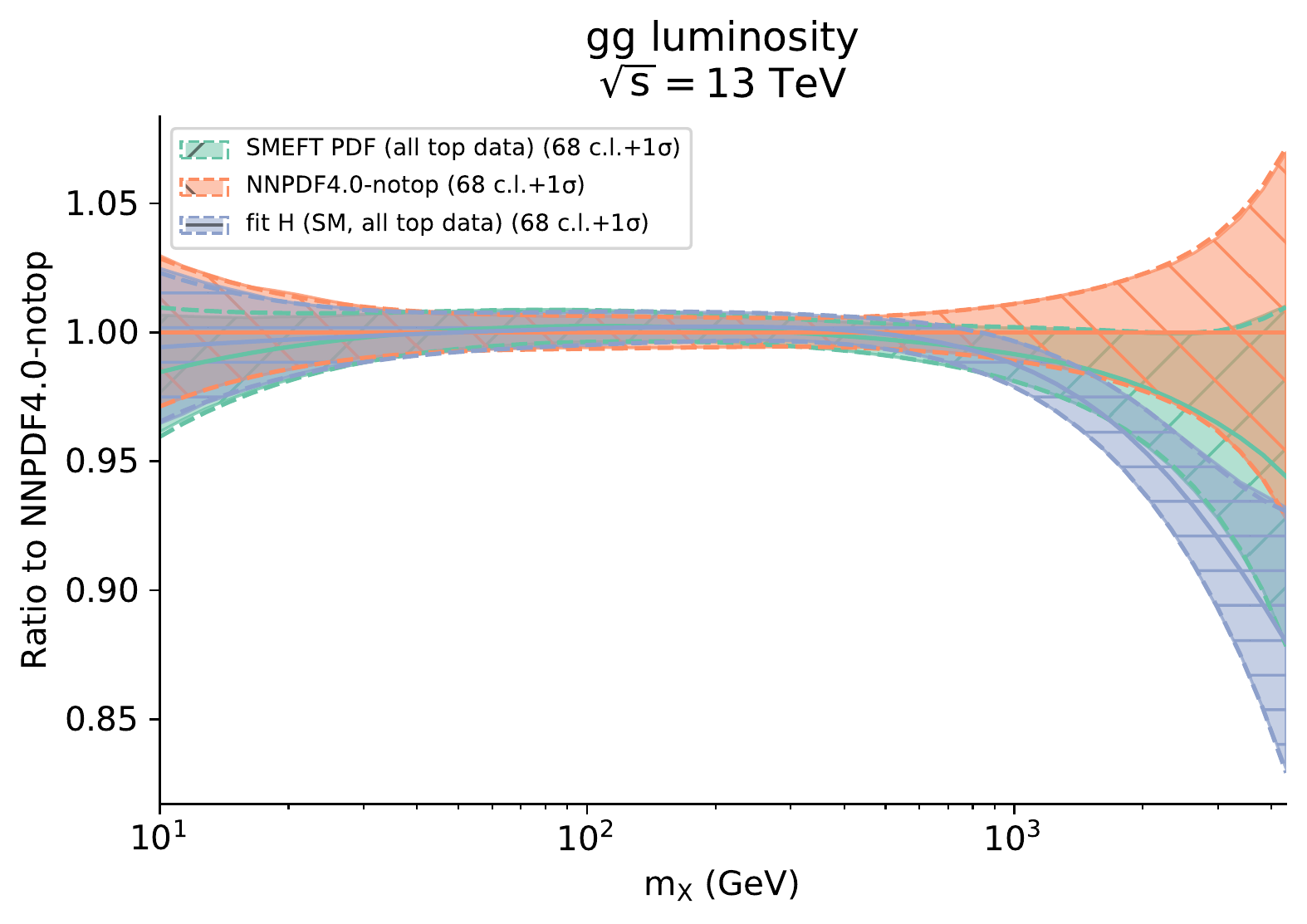}
\includegraphics[width=0.49\textwidth]{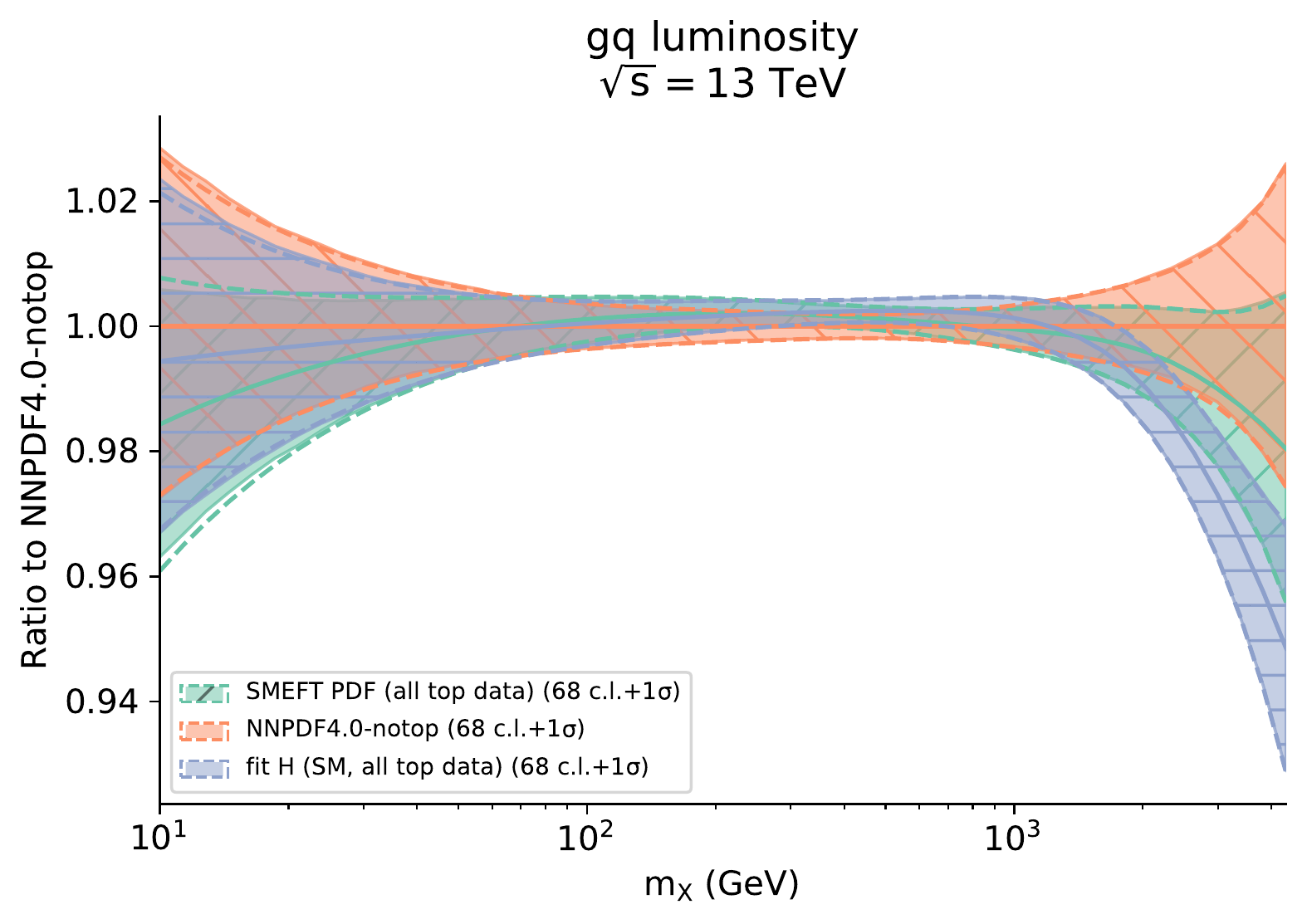}
\caption{The gluon-gluon (left panel) and quark-gluon (right panel)
  partonic luminosities at $\sqrt{s}=13$ TeV as a function of the final-state
  invariant mass $m_X$.
  We compare the\nnpdfnotop baseline fit with its SM-PDF counterpart
  including all top quark data considered (fit H in Table~\ref{tab:fit_list}) as well as with
  the SMEFT-PDF determination.
  Results are presented as the ratio to the central value of the
  \nnpdfnotop baseline.
}
\label{fig:simu_luminosities}
\end{figure}

Fig.~\ref{fig:simu_luminosities} displays the same comparison
as in Fig.~\ref{fig:SMEFTpdf_gluon} now for the case of the gluon-gluon and quark-gluon 
partonic luminosities at $\sqrt{s}=13$ TeV, as a function of the final-state
invariant mass $m_X$.
Consistently with the results obtained at the PDF level, one finds that
the three luminosities are almost identical for $m_X \le 1$ TeV, and
at higher invariant masses the SMEFT-PDF predictions are bracketed
between the \nnpdfnotop fit from above and the SM-PDF which includes
all top data (fit H in Table~\ref{tab:fit_list}) from below.
Hence, the net effect of simultaneously fitting the PDFs
and the EFT coefficients  is a dilution of constraints
provided by the top quark data on the former,
which translates into larger PDF uncertainties (which end up being rather similar to
those of \nnpdfnotop) and an increase
in the large-$m_X$ luminosity, e.g. of 5\% in the $gg$ case for $m_X\simeq 3$ TeV,
as compared to the SM-PDF luminosity.

This dilution arises because of an improved description of the top quark
data included in the fit, especially in the high $m_{t\bar{t}}$ bins.
In the SM-PDF case this can only be obtained by suppressing
the large-$x$ gluon, while in a SMEFT-PDF analysis this can also be achieved by shifting the EFT coefficients
from their SM values.
In other words, as compared to the \nnpdfnotop baseline, the gluon PDF experiences
a suppression at large-$x$ of up to 10\% when fitting the top quark data, and this
pull is reduced by approximately a factor two in the joint SMEFT-PDF determination due to the
coherent effect of the linear EFT cross-sections.

\begin{figure}[t]
\centering
\includegraphics[width=0.49\textwidth]{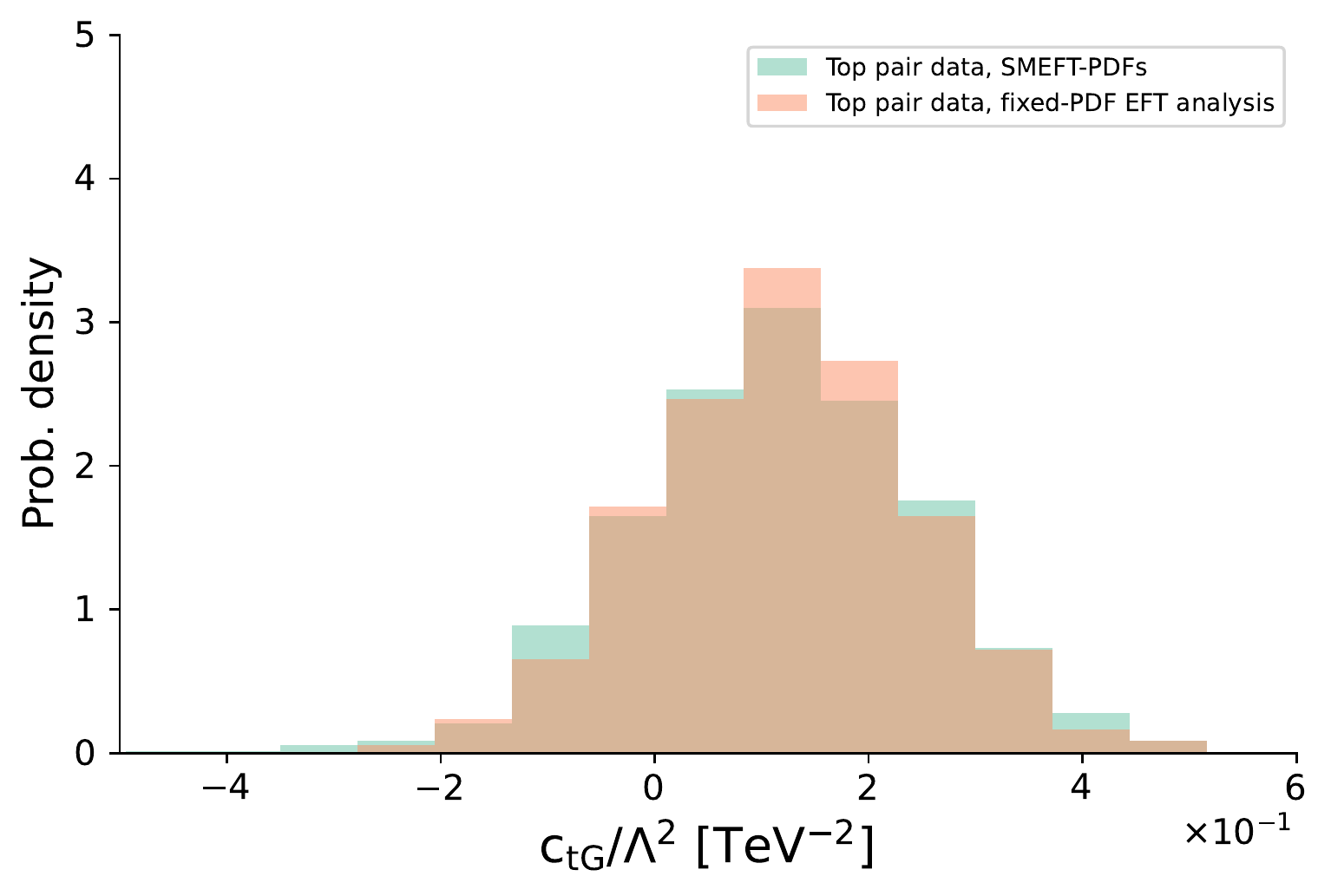}
\includegraphics[width=0.49\textwidth]{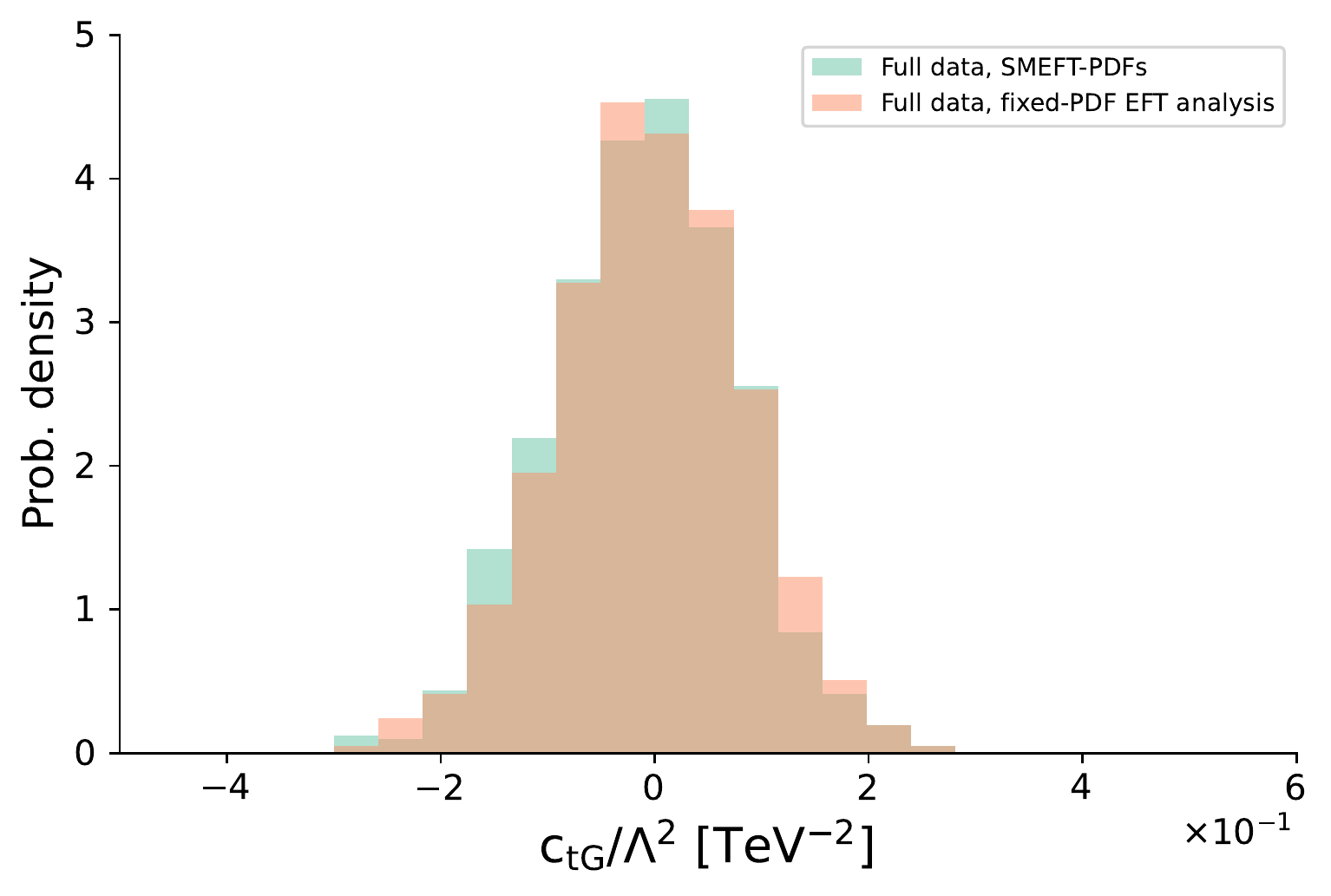}
\caption{Posterior distributions for the Wilson coefficient $c_{tG}$
  associated to the chromomagnetic operator in the joint SMEFT-PDF determination, compared
  with the corresponding results from the fixed-PDF EFT analysis whose settings are described in
  Sect.~\ref{sec:res_smeft}.
  We show results based on only top-quark pair production data (left) and
  in the whole top quark dataset considered in this work (right panel).
}
\label{fig:simultaneous_broadening}
\end{figure}

\paragraph{EFT coefficients from a joint SMEFT-PDF fit.} 
As opposed to the marked effect of the SMEFT-PDF interplay found for the large-$x$ gluon,
its impact is more moderate at the level of the bounds on the EFT coefficients,
and is restricted to mild shifts in the central values and a slight broadening
of the uncertainties.
This is illustrated by Fig.~\ref{fig:simultaneous_broadening},
showing the posterior distributions for the Wilson coefficient $c_{tG}$
  associated to the chromomagnetic operator in the joint SMEFT-PDF determination, compared
  with the corresponding results from the fixed-PDF EFT analysis whose settings are described in
  Sect.~\ref{sec:res_smeft}.
  The comparison is presented both for the fits which consider only top-quark pair production data
  and those based on the whole top quark dataset considered in this work.
  The leading effect of the chromomagnetic operator $\mathcal{O}_{tG}$ is to modify the
  total rates of 
  $t \bar{t}$ production without altering the shape of the differential distributions, and hence
  it plays an important role in a simultaneous SMEFT-PDF determination based on top quark data.

For fits based only on inclusive $t\bar{t}$ data, as shown in the left panel of
Fig.~\ref{fig:simultaneous_broadening}, the two posterior
distributions are similar; the distribution based on the SMEFT-PDF analysis
is slightly broader, approximately $10 \%$ so, as compared to the fixed-PDF EFT fit
to the same measurements.
This slight broadening is washed out in the fit to the full top quark dataset, as shown in the right panel of Fig.~\ref{fig:simultaneous_broadening}.
In both cases, the determination of $c_{tG}$ is consistent with the SM
at a 95\% CL, and the best-fit values of the coefficient
are the same in the SMEFT-PDF and fixed-PDF EFT analyses.
Hence, in the specific case of the chromomagnetic operator, the interplay between PDFs and EFT fits
is rather moderate and restricted to a broadening of at most $10\%$ in the 95\% CL bounds.
Similar comparisons have been carried out for other EFT coefficients as well as in the context
of fits to a subset of the data and/or to a subset of the coefficients.
We find that in general the impact of the SMEFT-PDF interplay
translates to a broadening of the uncertainties
in the EFT coefficients, which at most reaches $30\%$, and
alongside which the best-fit values remain stable\footnote{All results
  obtained with various subsets of the data and of the
  coefficients can be found at \url{https://www.pbsp.org.uk/research/topproject}}.

All in all, within  the global fit based on the best available theory predictions,
results for the EFT coefficients turn out to be very similar
in the fixed-PDF EFT and SMEFT-PDF fits.
This indicates that, provided a broad enough dataset and the most
up-to-date theory calculations are used, the PDF dependence on the cross-sections
entering an EFT interpretation of the LHC data is currently
subdominant and can be neglected (this is not the case for the
PDFs, see Fig.~\ref{fig:simu_luminosities}).
Nevertheless, this statement applies only to the dataset currently available,
and it is likely that the SMEFT-PDF interplay will become more significant in the future
once HL-LHC measurements become available~\cite{AbdulKhalek:2018rok,Azzi:2019yne},
as demonstrated in the case of high-mass Drell-Yan~\cite{Greljo:2021kvv}.

\begin{figure}[t!]
\centering
\includegraphics[width=0.99\textwidth]{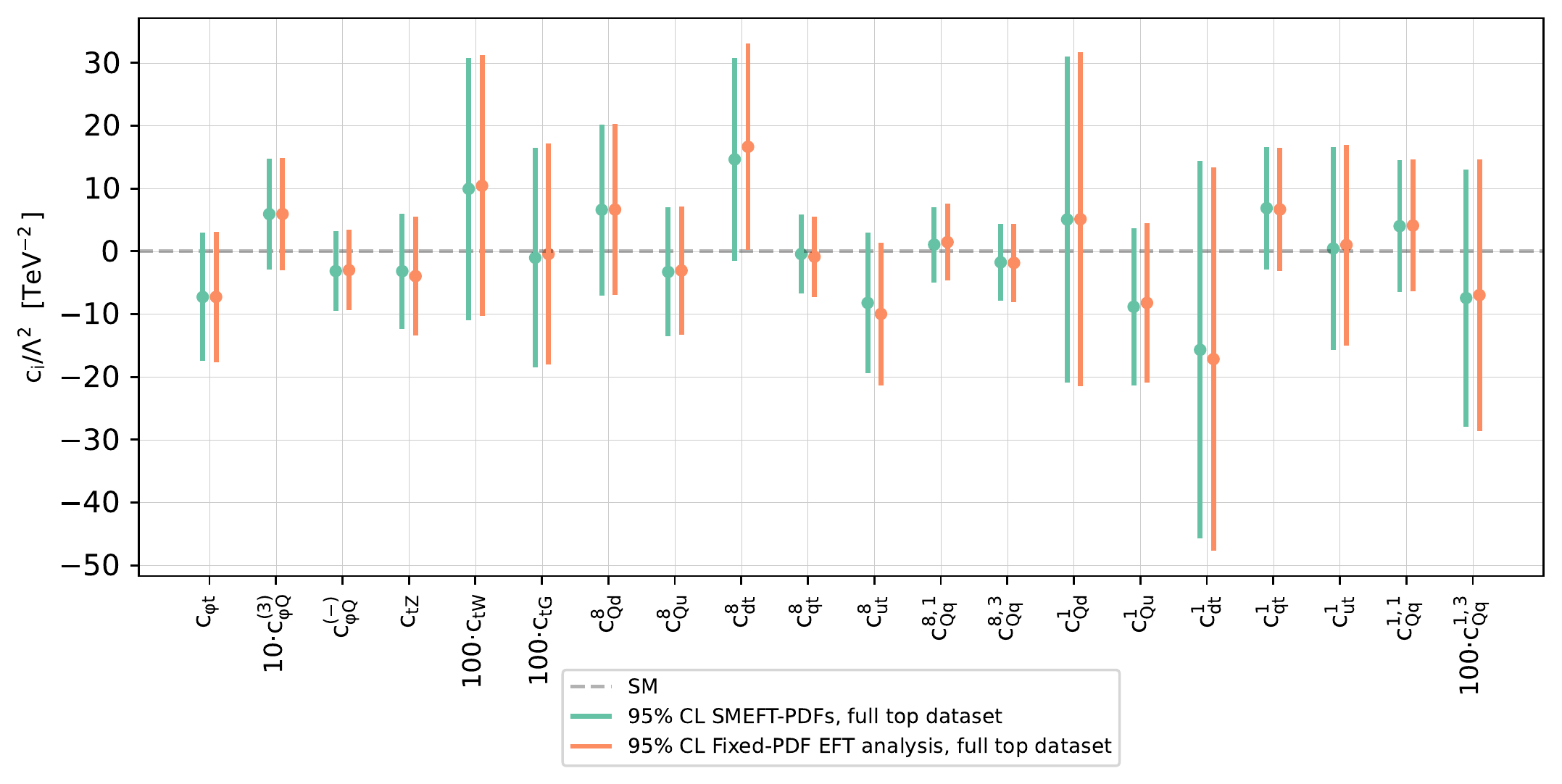}
\caption{
  Comparison of the $95 \%$ CL intervals on the 20 Wilson coefficients
  considered in this work (in the linear EFT case) between the outcome of the joint SMEFT-PDF determination
  and that of the fixed-PDF EFT analysis.
  The latter is based on SM and EFT calculations performed with
 \nnpdfnotop as input, see also  Sect.~\ref{sec:res_smeft}.
  In both cases, results are based on the full 
  top quark dataset being considered and EFT cross-sections
	are evaluated up  to linear, $\mathcal{O}\lp \Lambda^{-2}\rp$, corrections.
  The dashed horizontal line indicates the SM prediction, $c_k=0$.
  Note that some coefficients are multiplied by the indicated prefactor
  to facilitate the visualisation of the results. 
}
\label{fig:smeft_simu_bounds}
\end{figure}

The moderate impact of the SMEFT-PDF interplay on the Wilson coefficients 
for the full top quark dataset considered in this work is summarised in Fig.~\ref{fig:smeft_simu_bounds},
which 
compares the $95 \%$ CL intervals of the 20 fitted Wilson coefficients
relevant for the linear EFT fit obtained from the outcome of the joint SMEFT-PDF determination
and the fixed-PDF EFT analysis.
The latter is based on SM and EFT calculations performed with
\nnpdfnotop as input; see also the description of the settings in
Sect.~\ref{sec:res_smeft}.
The dashed horizontal line indicates the SM prediction,
and  some coefficients are multiplied by the indicated prefactor
to facilitate the visualisation of the results. 
Fig.~\ref{fig:smeft_simu_bounds} demonstrates that, other than slight
broadenings and shifts in the central values,
the results of the two analyses coincide.

\begin{figure}[t!]
\centering
\includegraphics[width=0.49\textwidth]{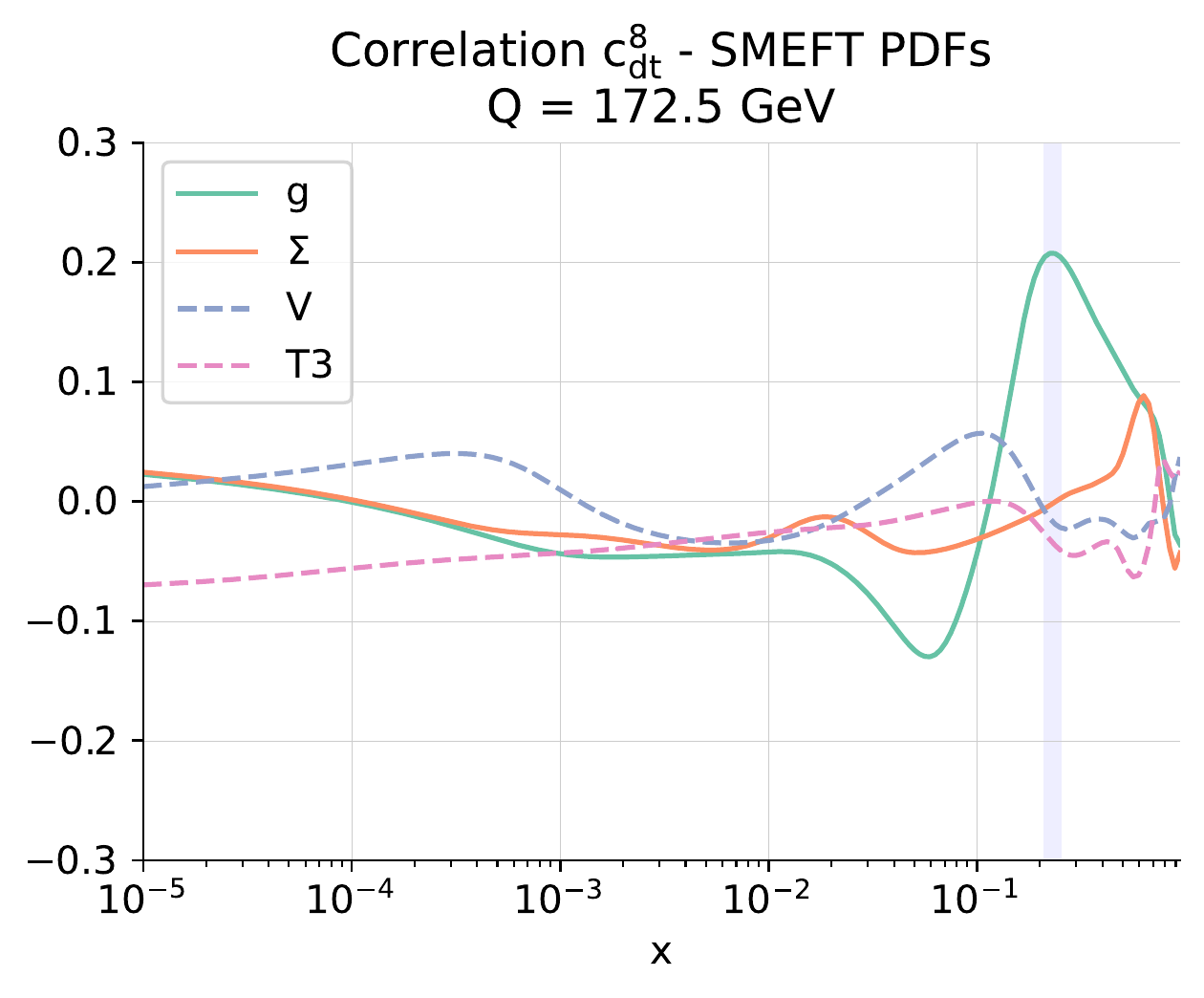}
\includegraphics[width=0.49\textwidth]{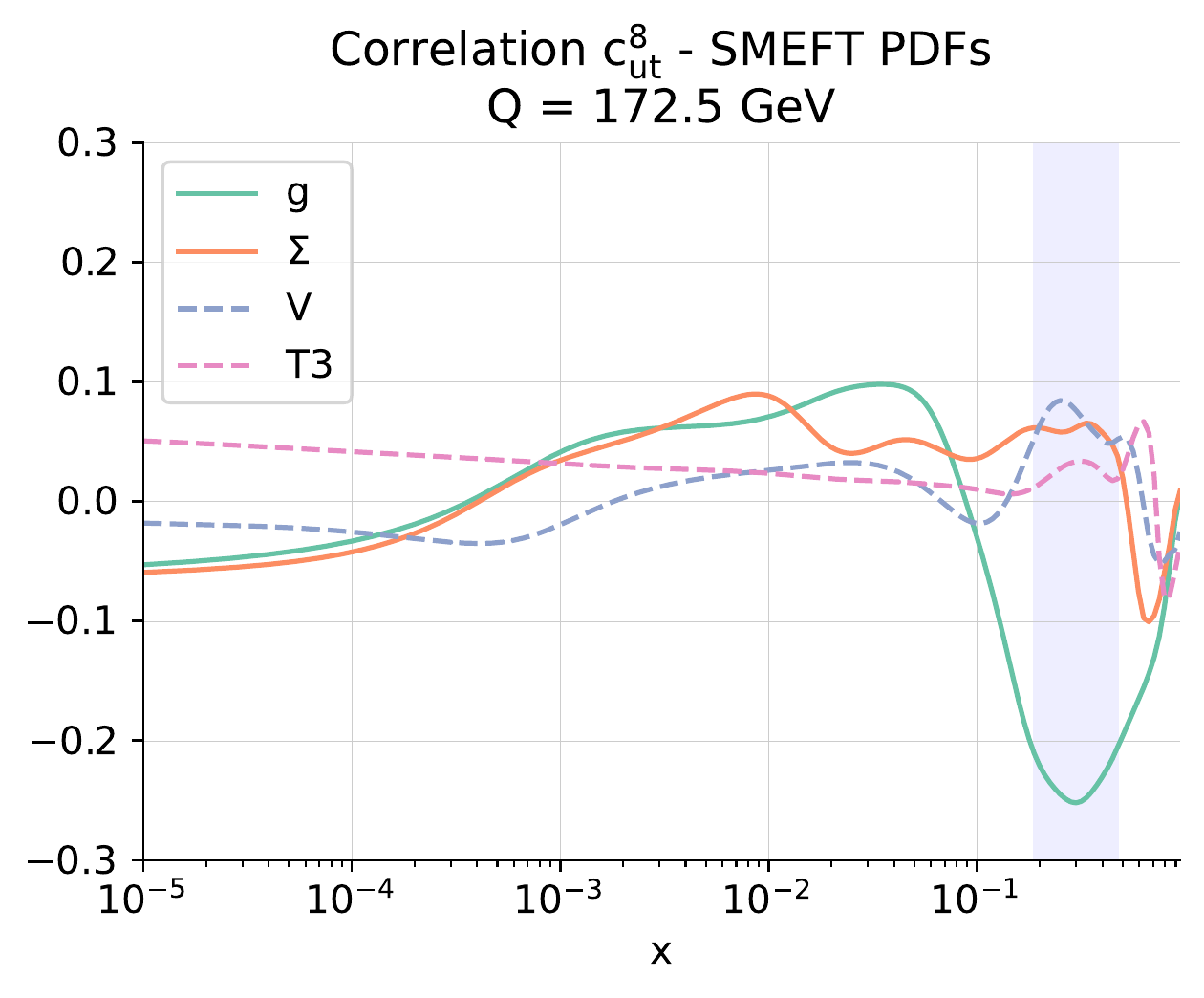}
\includegraphics[width=0.49\textwidth]{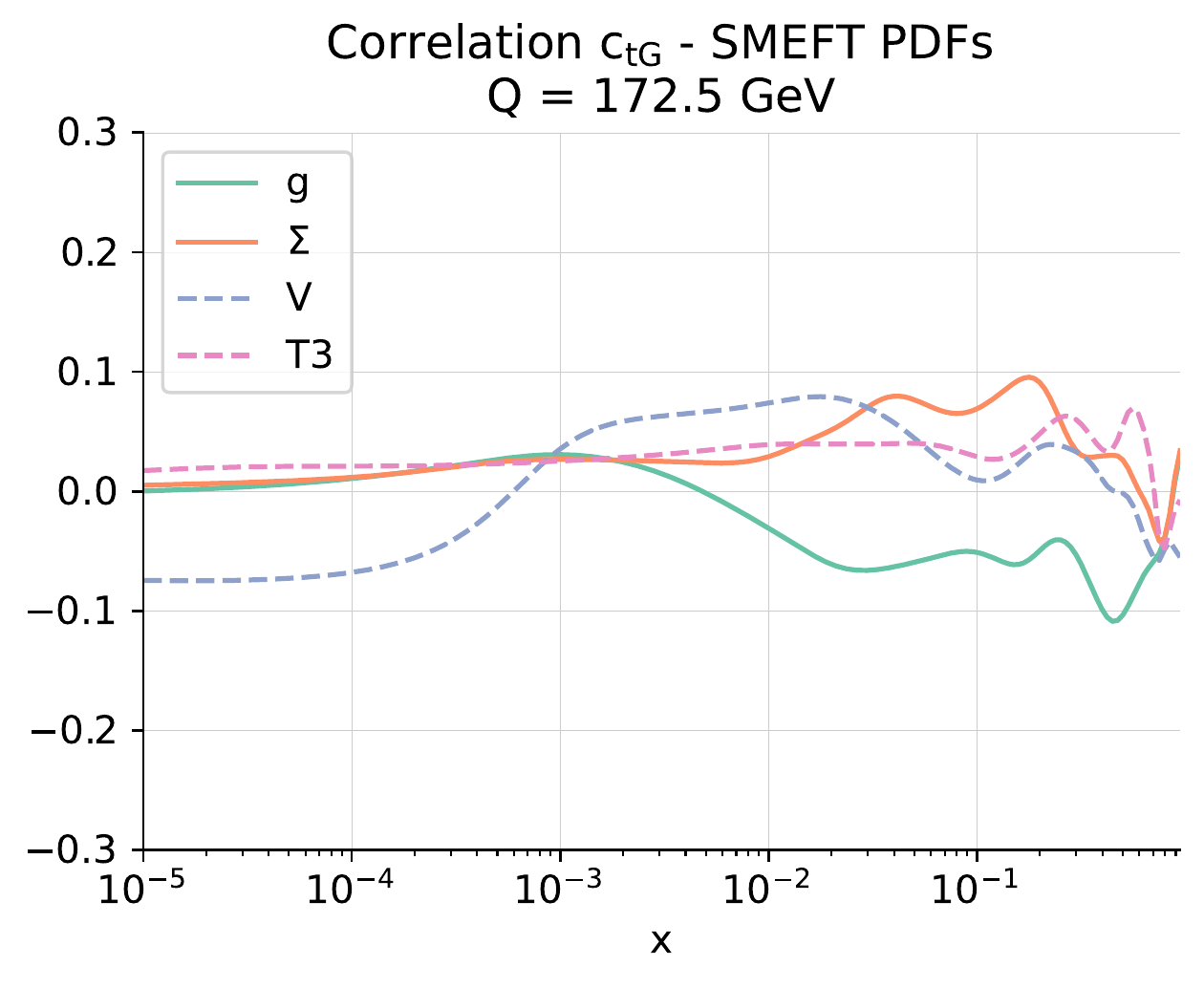}
\includegraphics[width=0.49\textwidth]{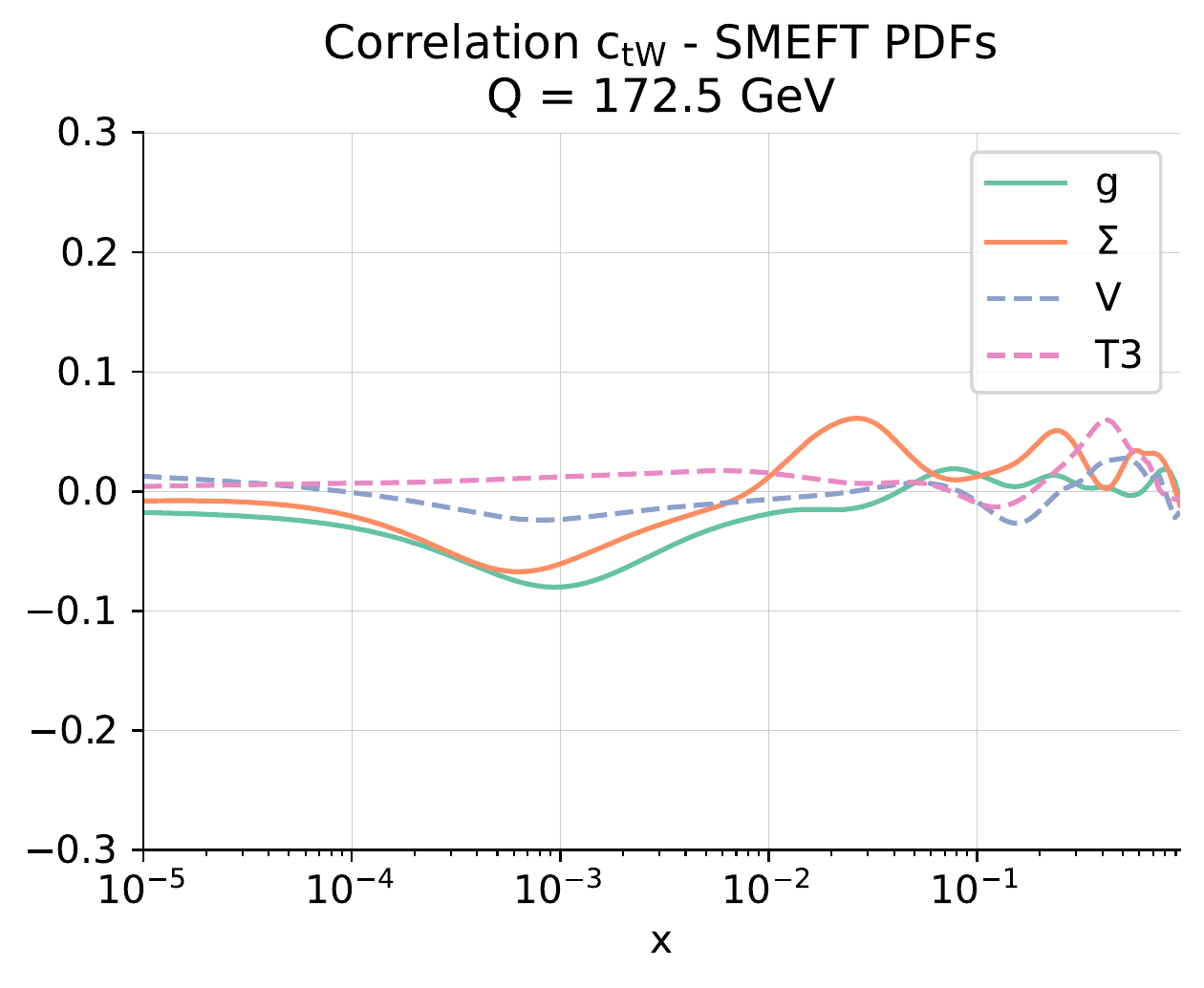}
\caption{The correlation coefficient $\rho(f_i, c_k)$
  between the SMEFT-PDFs $f_i$ and the Wilson coefficients $c_k$ evaluated at $Q=172.5$ GeV
  as a function of $x$.
  Each panel displays the correlations of the coefficient $c_k$
  with the gluon and the total singlet $\Sigma$, total valence $V$, and non-singlet
  triplet $T_3$ PDFs.
  We provide results for representative EFT coefficients, namely $c_{td}^{8}$,
  $c_{tu}^{8}$, $c_{tG}$, and $c_{tW}$.
  The largest correlations within the EFT coefficients  considered in this
  work are associated to four-fermion operators such as  $c_{td}^{8}$ and $c_{tu}^{8}$.
}
\label{fig:smeft_pdf_corr}
\end{figure}

\paragraph{Correlations.} Fig.~\ref{fig:smeft_pdf_corr} displays the
correlation coefficients~\cite{Ball:2010de}
between the SMEFT-PDFs and the Wilson coefficients evaluated at $Q=172.5$ GeV
as a function of $x$.
Each panel displays the correlations of the coefficient $c_k$
with the gluon and the total singlet $\Sigma$, total valence $V$, and non-singlet
triplet $T_3$ PDFs, and we consider four representative EFT coefficients, namely $c_{td}^{8}$,
$c_{tu}^{8}$, $c_{tG}$, and $c_{tW}$.
The largest correlations within the EFT coefficients  considered in this
work are associated to the gluon PDF and four-fermion operators such as  $c_{td}^{8}$ and $c_{tu}^{8}$
in the large-$x$ region, peaking at $x\simeq 0.3$.
Correlations for other values of $x$ and for the quark PDFs are negligible
for all operators entering the analysis.
We note that future data with an enhanced coverage of the high-$m_{t\bar{t}}$ region in top quark
pair-production
might alter this picture, given that for $m_{t\bar{t}}\gsim 3$ TeV
the  $q\bar{q}$ luminosity starts to become more relevant and eventually dominates over the $gg$
contribution.

\begin{figure}[t]
\centering
\includegraphics[width=0.98\textwidth]{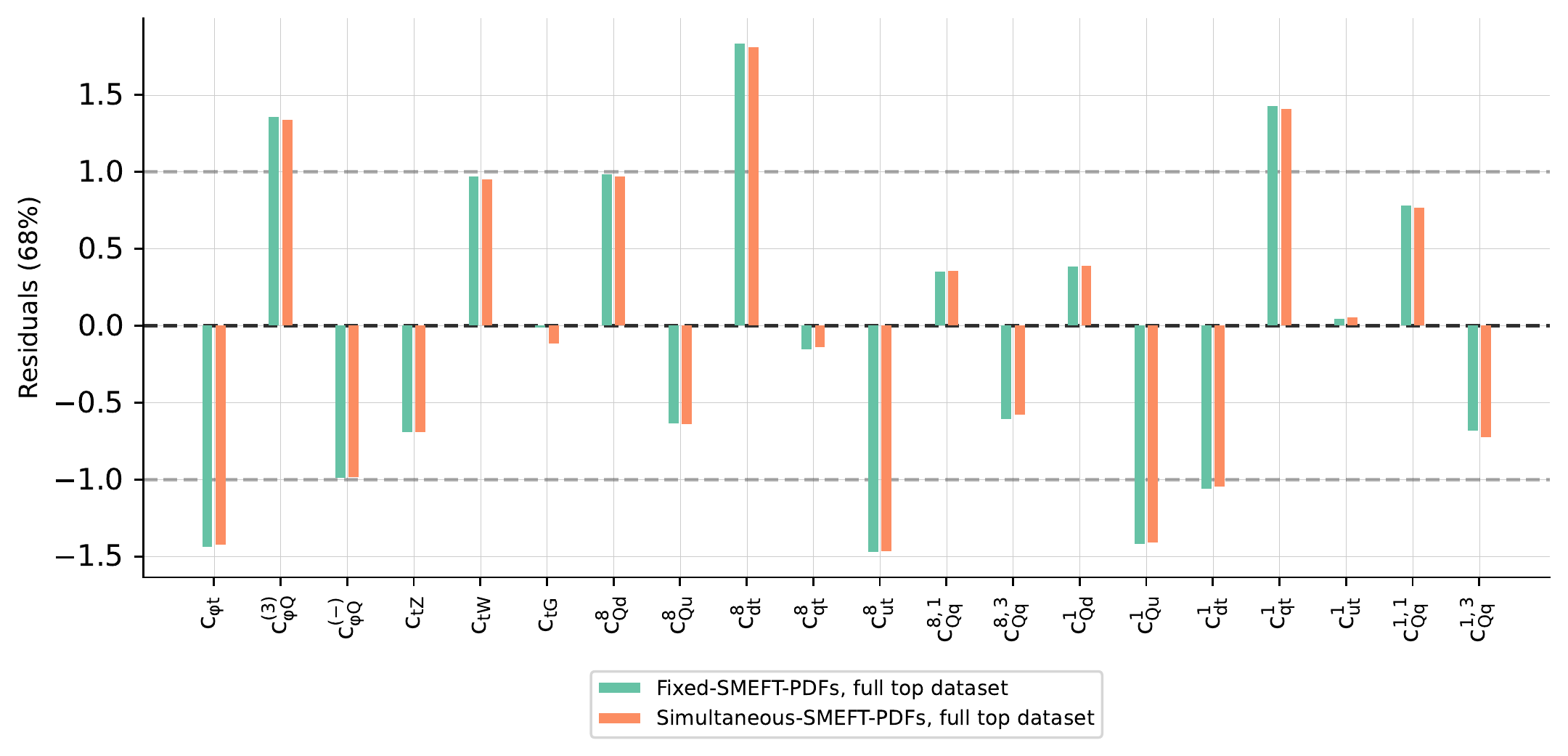}
\caption{The 68\% CL residuals, Eq.~(\ref{eq:residuals}), for the
  same Wilson coefficients displayed in Fig.~\ref{fig:smeft_simu_bounds},
  comparing the outcome of the joint SMEFT-PDF determination
  with that of a fixed-PDF EFT analysis.
  In the latter, we use as input for the theory calculations
  the  SMEFT-PDFs obtained in the joint fit rather than the
  \nnpdfnotop set
  used in Sect.~\ref{sec:res_smeft}.
  The horizontal dashed lines indicate the $\pm 1\sigma$ regions.
}
\label{fig:fixedsmeftpdf_res}
\end{figure}

\paragraph{Residuals.} Finally, Fig.~\ref{fig:fixedsmeftpdf_res} displays a similar comparison as
in Fig.~\ref{fig:smeft_simu_bounds} now at the level of the 68\% CL fit residuals defined as
\begin{equation}
  \label{eq:residuals}
R_n = \frac{c_n^*}{\sigma_n} \, ,
\end{equation}
where $c_n^*$ and $\sigma_n$ are the median value and the standard deviation of the Wilson coefficient
$c_n$ respectively, with $n=1,\ldots,N$, where $N$ is the number of operators.
The outcome of the joint SMEFT-PDF determination
is compared with that of a fixed-PDF EFT analysis
where we use as input for the theory calculations
the  SMEFT-PDFs obtained in the joint fit, rather than the \nnpdfnotop set.
That is, in both cases the information provided by the top quark data on the PDFs
and Wilson coefficients is accounted for, but in one case the cross-correlations
are neglected whereas they are accounted for in the other.
The residuals are similar in the two cases; they are slightly bigger (in absolute value)
in the fixed-PDF case in which the correlations between the SMEFT-PDFs and the EFT
coefficients are ignored.
This analysis further emphasises that, for the currently available top quark data, the cross-talk
between PDFs and EFT degrees of freedom does not significantly modify the posterior
distributions in the space spanned by the Wilson coefficients.\\

\noindent In summary, on the one hand we find that from the point of view of a PDF determination,
SM-PDFs and SMEFT-PDFs extracted from top quark data differ by an amount comparable
to their respective uncertainties in the case of the large-$x$ gluon.
On the other hand, at the level of Wilson coefficients the results are unchanged irrespective
of the PDF set used as input for the theory calculations; that is, bounds based
on \nnpdfnotop or the SMEFT-PDFs are almost the same.
Hence, while EFT interpretations of top quark data can safely ignore the PDF dependence,
at least for the settings adopted in this work, a global PDF fit could be significantly
distorted if BSM physics were to be present in the large-energy top quark distributions.
We use these findings in App.~\ref{app:recommendations} to provide recommendations
for interpretations of LHC top quark data in the context of either PDF or EFT analyses.

\section{Conclusions and outlook}
\label{sec:summary}

In this work we have carried out a comprehensive theoretical interpretation
of top quark data from the LHC, including all available measurements
based on the full integrated luminosity of Run II,
in terms of the proton PDFs and of Wilson coefficients in the SMEFT.
We have integrated a global determination of PDFs with
the fit of dimension-six Wilson coefficients modifying top quark interactions
into a simultaneous determination of the SMEFT-PDFs and the EFT coefficients.
The main outcome of our analysis is the assessment of the conditions
under which
the usual assumptions of SM-PDFs and fixed-PDF EFT analyses are valid,
and establishing when the SMEFT-PDF interplay cannot be neglected without
introducing a sizeable bias to the results.
Furthermore, we have provided strategies to disentangle eventual BSM signals from QCD effects
in the interpretation of these top quark measurements.

As a by-product of this determination of the SMEFT-PDFs, we have also
presented state-of-the-art SM-PDF and fixed-PDF
EFT interpretations of top quark data based on the broadest LHC experimental
dataset to date, therefore fully exploiting the information contained in the legacy Run II
top quark measurements.
From the SM-PDF study, we have quantified the impact of the recent Run II top quark production
data on the large-$x$ gluon, and assessed their compatibility with the
information provided by the datasets already included in previous global fits
such as \nnpdf.
From the fixed-PDF SMEFT analysis, we have benchmarked the \simunet{} performance
reproducing the \smefit results, determined the improved constraints
provided by the full-luminosity Run II data, and compared our findings with the
results presented from other groups.
All in all, we demonstrate the unparalleled sensitivity of Run II top quark
data both to probe the proton structure and to search for signatures of
new physics arising as anomalous top quark properties and interactions.

The results presented in this work could be extended in a number of directions.
One of the most pressing matters is the provision of a simultaneous determination
of PDFs with EFT coefficients entering at \textit{quadratic order}; the difficulties associated
with using quadratics with our current \simunet{} methodology are described in App.~\ref{app:quad}.
This will involve significant modification to the Monte-Carlo replica method used by \simunet{},
and will be the subject of future work.

On a different note, top quark measurements at the LHC have also been used
to determine the strong coupling $\alpha_s(m_Z)$ as well
as the top quark mass $m_t$, both independently and simultaneously
with the PDFs~\cite{Cooper-Sarkar:2020twv}.
It could hence be interesting to extend the \simunet{} framework
to account for the determination of other SM parameters from top quark
data in addition to the PDFs,
as it is sketched in the original \simunet publication~\cite{Iranipour:2022iak}, and to study their interplay with eventual
new physics effects.
Additionally, it would be interesting to extend future versions of the \simunet{} methodology to
account for potential renormalisation group effects on the
Wilson coefficients~\cite{Jenkins:2013zja,Jenkins:2013wua,Alonso:2013hga,Aoude:2022aro}.
One may also want to assess the SMEFT-PDF interplay for other types
of processes beyond those considered so far, and in particular
study inclusive jet
and di-jet production, which are instrumental for PDF fits in the same
region constrained by the top data~\cite{AbdulKhalek:2020jut} and also  provide
information on a large number of poorly-known SMEFT operators.
In the longer term, even processes that are never used for PDF fits, such as Higgs
or gauge boson pair
production, may reach a precision that makes them competitive, and in this case
the only option to account for this information is by means of the simultaneous
determination of PDFs and EFT coefficients.

Another possible development of the  \simunet{} methodology
would be to establish the interplay between PDFs and model parameters such as masses
and couplings
in specific UV-complete BSM scenarios  by using the SMEFT as a matching bridge~\cite{Dawson:2022ewj}. 
It would also be interesting to 
 extend the joint SMEFT-PDF determination as implemented
in \simunet{} to novel types of observables with enhanced or even
optimal sensitivity to either the PDFs of the EFT coefficients, such as
the ML-assisted unbinned multivariate observables introduced in Ref.~\cite{GomezAmbrosio:2022mpm}.
Finally, following along the lines of the HL-LHC projections
for high-mass $W$ and $Z$ production presented in~\cite{Greljo:2021kvv},
one can study the projected SMEFT-PDF synergies at future colliders,
certainly at the HL-LHC but also at the Electron Ion Collider~\cite{Boughezal:2022pmb}, sensitive
to directions in the SMEFT parameter space not covered by LHC data, and at the
Forward Physics Facility~\cite{Feng:2022inv,Anchordoqui:2021ghd},
where proton and nuclear structure can be probed 
while also obtaining information on anomalous neutrino interactions parametrised by extensions
of the SMEFT.

All results presented in this paper, as well as a broader selection of
results, including fits performed with various theory settings, on
subsets of datasets and/or operators are available at the webpage:
\begin{center}
\url{https://www.pbsp.org.uk/research/topproject}.
\end{center}
Readers are encouraged to contact the authors, should they need any
specific SM-PDF or SMEFT-PDF fits that can be obtained with our
methodology.

In order to facilitate that studies similar to those presented in this paper
are carried out by the LHC experimental collaborations with their own data, we plan
to release the open source \simunet{} framework together
with documentation and user-friendly examples in a future
publication, in a way that can be seamlessly integrated
with the NNPDF fitting code~\cite{NNPDF:2021uiq}. 
The availability of  \simunet{} as open source code   would allow
the LHC collaborations to study the SMEFT-PDF interplay in their own internal analyses, and
in particular to modify the input datasets and EFT operator basis to match their own settings.
In the meantime, we provide in App.~\ref{app:recommendations}
usage recommendations concerning
the joint interpretation of LHC measurements in terms of
both PDFs and EFT coefficients based on the findings of this work.

As the LHC approaches its high-luminosity era, it becomes more and more important
to develop novel analysis frameworks that make possible the full exploitation of
the information contained in the legacy LHC measurements.
The results presented in this work contribute to this program by demonstrating
how to achieve a consistent simultaneous determination
of the PDFs and EFT parameters from the same top quark dataset,
bypassing the need for the assumptions required for the SM-PDF and fixed-PDF EFT
interpretations.


\section*{Acknowledgments}
We thank Emanuele Roberto Nocera for his insight into the generation of the SM
predictions via APPLgrids.
We are grateful to Rene Poncelet for his support with the production
of NNLO QCD $C$-factors, and for assistance
in understanding the data-theory agreement for the CMS $13$ TeV double-differential 
$t\bar{t}$ distribution in the $\ell+$jets channel.
We thank the phenomenology working group at the University of Cambridge for useful
discussion during the project.
We are grateful to Jun Gao for supplying the QCD $C$-factors for the CMS single-top differential distribution at
$13$ TeV. 

M.~U., L.~M., M.~M., M~.M~.A. and Z.~K. are supported by the European Research Council under the
European Union's Horizon 2020 research and innovation Programme (grant agreement n.950246).
M.~U. is  partially supported by the STFC grant ST/T000694/1 and by the
Royal Society grant RGF/EA/180148.
The work of M.~U. is also funded by the Royal Society grant DH150088.
The work of J.~M. is supported by the Sims Fund Studentship.
The work of J.~R. is partially supported by NWO (Dutch Research Council)
and by an ASDI (Accelerating Scientific Discoveries) grant
from the Netherlands eScience Center.

\appendix
\section{Recommendations to quantify SMEFT-PDF interplay}
\label{app:recommendations}

The strategy presented here, based on 
carrying out a simultaneous determination of the SMEFT-PDFs and the Wilson coefficients, makes possible a
quantitative assessment of the interplay between the PDF and SMEFT
sensitivity when interpreting a given set of LHC measurements. 
As discussed in Sect.~\ref{sec:summary}, the \simunet{} framework
will be made public only at a later stage.
In the meantime, researchers interested in quantifying this interplay in their
own analysis can use available open-source platforms like {\sc\small xFitter},
as done by the CMS collaboration
in their QCD and EFT analyses of double-differential inclusive jet cross sections
at 13 TeV~\cite{CMS:2021yzl}.
In several cases, however, a joint  analysis may not be necessary,
and one can estimate the possible role of the SMEFT-PDF interplay by means
of a simplified approach.
The discussion is presented here for the case of top quark measurements,
but it can be straightforwardly extended to any other class of measurements.

\begin{itemize}
\item The most conservative strategy is to introduce a hard boundary between those processes
  used for PDF fits and those entering SMEFT interpretations.
  For instance, for a SMEFT analysis of LHC top quark measurements one can use
  PDF fits without top data, such as the \nnpdfnotop variant.
  We recall that with the NNPDF open-source code~\cite{NNPDF:2021uiq} one can produce fit variants
  with arbitrary input datasets.

\item A somewhat less  conservative assumption would be to introduce a hard boundary,
  not at the process level, but rather at the kinematic level.
  In the case of inclusive top quark pair production data, one can introduce a threshold on the
  top quark pair invariant mass $m_{t\bar{t}}$ such that data points below the threshold
  are used for the PDF fit and above it for the EFT interpretation, thus
  benefiting from the increased sensitivity to BSM effects in the high-energy tails
  of the LHC distributions.
  Again, using public PDF fitting codes one can produce fit variants
  with tailored kinematical cuts to separate the ``PDF'' region from the ``EFT'' region.
  One drawback of this approach is that there is never a clear-cut separation
  between these regions; depending on the EFT operators considered,
  the high-energy region may not be the dominant one.

\item As demonstrated in this work, many measurements of relevance for EFT
  interpretations have limited sensitivity to PDFs and vice-versa.
  In such cases, the SMEFT-PDF interplay can be safely neglected.
  One can determine under which settings this condition is satisfied by
  adding the dataset under consideration either to a global PDF fit
  (using for example the NNPDF fitting code~\cite{NNPDF:2021uiq}) or to a global SMEFT
  fit (using the \smefit or \fitm frameworks).
  If the results of either the PDF or EFT fits are unchanged, one can safely neglect
  the SMEFT-PDF interplay in this case.

\item If instead the above analysis shows that a given dataset provides non-trivial information 
  on both the PDFs and on the EFT Wilson coefficients, the only options for a consistent
  theoretical interpretation are either introducing a hard boundary in the analysis (either
  at the process level or at the kinematic level) or to carry out
  the joint SMEFT-PDF interpretation of the full dataset using 
  \simunet{} or other available tools.

\end{itemize}

Furthermore, it should be emphasised that, as opposed to the PDF case,
the sensitivity to EFT coefficients of
a given measurement depends in general on the choice of EFT operator considered.
Hence, the above considerations assume a specific choice of operators and could be
different if this choice is varied.

\section{EFT operator basis}
\label{sec:operators}

Table~\ref{tab:ops} lists the dimension-six
SMEFT operators relevant for this analysis, together with the corresponding
degrees of freedom (DoF) entering the fit.
In terms of the flavour assumptions, we follow the LHC top quark working group
recommendations~\cite{Aguilar-Saavedra:2018ksv},
which were also adopted in~\cite{Ethier:2021bye,Ellis:2020unq}.
The flavour symmetry group is
given by U(3)$_l \times$ U(3)$_e \times$ U(3)$_d \times$ U(2)$_u \times$ U(2)$_q$, i.e. one singles out operators that contain
top quarks (right-handed $t$ and SU(2) doublet $Q$).
This means that we work in a 5-flavour scheme, where the only massive fermion in the theory is the top quark.
The upper part in Table~\ref{tab:ops} defines the relevant two-fermion 
operators modifying the interactions of the third-generation
quarks.
We also indicate the notation used for the associated Wilson coefficients; those
in brackets are not degrees of freedom entering the fit,
and instead the two additional DoF defined in the middle table are used.
The bottom table defines the four-fermion DoF entering the fit,
expressed in terms of the corresponding four-fermion Wilson coefficients
associated to dimension-six SMEFT operators in the Warsaw basis.

EFT cross-sections at the linear and quadratic order in the EFT expansion
are computed for the datasets considered in this analysis
and for the DoF defined in Table~\ref{tab:ops}, both
at LO and NLO in perturbative QCD.
Furthermore we use the $m_W$-scheme,  meaning that the four EW inputs are
$\{m_W, G_F, m_h, m_Z\}$.
In particular, the electric charge $e$ becomes a dependent parameter and is shifted by 
the effects of higher-dimensional operators.

\begin{table}[htbp]
  \begin{center}
    \renewcommand{\arraystretch}{1.62}
    {\small
    \begin{tabular}{lll}
      \toprule
      Operator $\qquad$ & Coefficient & Definition \\
                \midrule
    $\Op{\varphi Q}^{(1)}$ & --~~($c_{\varphi Q}^{(1)}$) & $i\big(\varphi^\dagger\lra{D}_\mu\,\varphi\big)
 \big(\bar{Q}\,\gamma^\mu\,Q\big)$ \\\hline
    $\Op{\varphi Q}^{(3)}$ & $c_{\varphi Q}^{(3)}$  & $i\big(\varphi^\dagger\lra{D}_\mu\,\tau_{\sss I}\varphi\big)
 \big(\bar{Q}\,\gamma^\mu\,\tau^{\sss I}Q\big)$ \\ \hline
    $\Op{\varphi t}$ & $c_{\varphi t}$& $i\big(\varphi^\dagger\,\lra{D}_\mu\,\,\varphi\big)
 \big(\bar{t}\,\gamma^\mu\,t\big)$ \\ \hline
      $\Op{tW}$ & $c_{tW}$ & $i\big(\bar{Q}\tau^{\mu\nu}\,\tau_{\sss I}\,t\big)\,
 \tilde{\varphi}\,W^I_{\mu\nu}
 + \text{h.c.}$ \\  \hline
 $\Op{tB}$ & --~~($c_{tB}$) &
 $i\big(\bar{Q}\tau^{\mu\nu}\,t\big)
 \,\tilde{\varphi}\,B_{\mu\nu}
 + \text{h.c.}$ \\\hline
    $\Op{t G}$ & $c_{tG}$ & $i\,\big(\bar{Q}\tau^{\mu\nu}\,T_{\sss A}\,t\big)\,
 \tilde{\varphi}\,G^A_{\mu\nu}
 + \text{h.c.}$ \\
 \bottomrule\\
 \toprule
      DoF $\qquad$  & Definition \\
                \midrule
$c_{\varphi Q}^{(-)}$ &  $c_{\varphi Q}^{(1)} - c_{\varphi Q}^{(3)}$ \\
\midrule
$c_{tZ}$ &   $-\sin\theta_W c_{tB} + \cos\theta_W c_{tW} $\\
  \bottomrule\\
 \toprule
          DoF $\qquad$ &  Definition (Warsaw basis) \\
          \midrule
      $c_{QQ}^1$    &   $2\ccc{1}{qq}{3333}-\frac{2}{3}\ccc{3}{qq}{3333}$ \\ \hline
    $c_{QQ}^8$       &         $8\ccc{3}{qq}{3333}$\\  \hline
 $c_{Qt}^1$         &         $\ccc{1}{qu}{3333}$\\   \hline
 $c_{Qt}^8$         &         $\ccc{8}{qu}{3333}$\\   \hline
  $c_{tt}^1$         &     $\ccc{}{uu}{3333}$  \\    \hline
 $c_{Qq}^{1,8}$       &      $\ccc{1}{qq}{i33i}+3\ccc{3}{qq}{i33i}$     \\   \hline
  $c_{Qq}^{1,1}$         &   $\ccc{1}{qq}{ii33}+\frac{1}{6}\ccc{1}{qq}{i33i}+\frac{1}{2}\ccc{3}{qq}{i33i} $   \\    \hline
   $c_{Qq}^{3,8}$         &   $\ccc{1}{qq}{i33i}-\ccc{3}{qq}{i33i} $   \\   \hline
  $c_{Qq}^{3,1}$          &     $\ccc{3}{qq}{ii33}+\frac{1}{6}(\ccc{1}{qq}{i33i}-\ccc{3}{qq}{i33i}) $   \\     \hline
   $c_{tq}^{8}$         &  $ \ccc{8}{qu}{ii33}   $ \\    \hline
   $c_{tq}^{1}$       &   $  \ccc{1}{qu}{ii33} $\\    \hline
   $c_{tu}^{8}$      &   $2\ccc{}{uu}{i33i}$  \\     \hline
    $c_{tu}^{1}$        &   $ \ccc{}{uu}{ii33} +\frac{1}{3} \ccc{}{uu}{i33i} $ \\   \hline
    $c_{Qu}^{8}$         &  $  \ccc{8}{qu}{33ii}$\\     \hline
    $c_{Qu}^{1}$     &  $  \ccc{1}{qu}{33ii}$  \\     \hline
    $c_{td}^{8}$        &   $\ccc{8}{ud}{33jj}$ \\    \hline
    $c_{td}^{1}$          &  $ \ccc{1}{ud}{33jj}$ \\     \hline
    $c_{Qd}^{8}$        &   $ \ccc{8}{qd}{33jj}$ \\     \hline
    $c_{Qd}^{1}$         &   $ \ccc{1}{qd}{33jj}$\\
         \bottomrule
\end{tabular}
}
\end{center}
  \caption{Upper table:  definition of the two-fermion dimension-six
    SMEFT operators relevant for this analysis.
    These operators modify the interactions of the third-generation
    quarks.
    We also indicate the notation for the associated Wilson coefficients; those
    in brackets are not degrees of freedom entering the fit.
    Middle table: the two additional degrees of freedom
    used in the fit involving two-fermion operators, defined in terms
    of the coefficients of the upper table.
    Bottom table:
    the four-fermion degrees of freedom considered here,
    expressed in terms of the corresponding four-fermion Wilson coefficients
    of dimension-six SMEFT operators in the Warsaw basis.
    \label{tab:ops}}
\end{table}

\section{Benchmarking with  SMEFiT}
\label{app:benchmark}

Here we  benchmark the performance of \simunet{} when operating
as a fixed-PDF EFT fitter, by means of a tuned comparison with
the {\sc\small SMEFiT} framework when identical theory and experimental inputs
are used in both cases.
\smefit currently provides two strategies to determine posterior
distributions in the EFT parameter space. The first, known as MCfit, is based on the Monte Carlo
replica method followed by parameter optimisation (as similarly used in \simunet{}).
The second is based on Nested Sampling (NS), a purely Bayesian approach for parameter inference.
At the linear EFT level, \smefit results based on MCfit and NS are identical~\cite{Ethier:2021bye}.
At the quadratic level~\cite{Giani:2023gfq},
however, the MCfit approach is affected by the potential pitfalls described in App.~\ref{app:quad},
which also prevent the use of \simunet{} for joint SMEFT-PDF fits in the presence
of quadratic EFT corrections.
We demonstrate now that, for a  linear EFT determination and a common choice of inputs, 
the results obtained using the two frameworks display excellent agreement.

\begin{figure}[t]
        \centering
        \includegraphics[width=\linewidth]{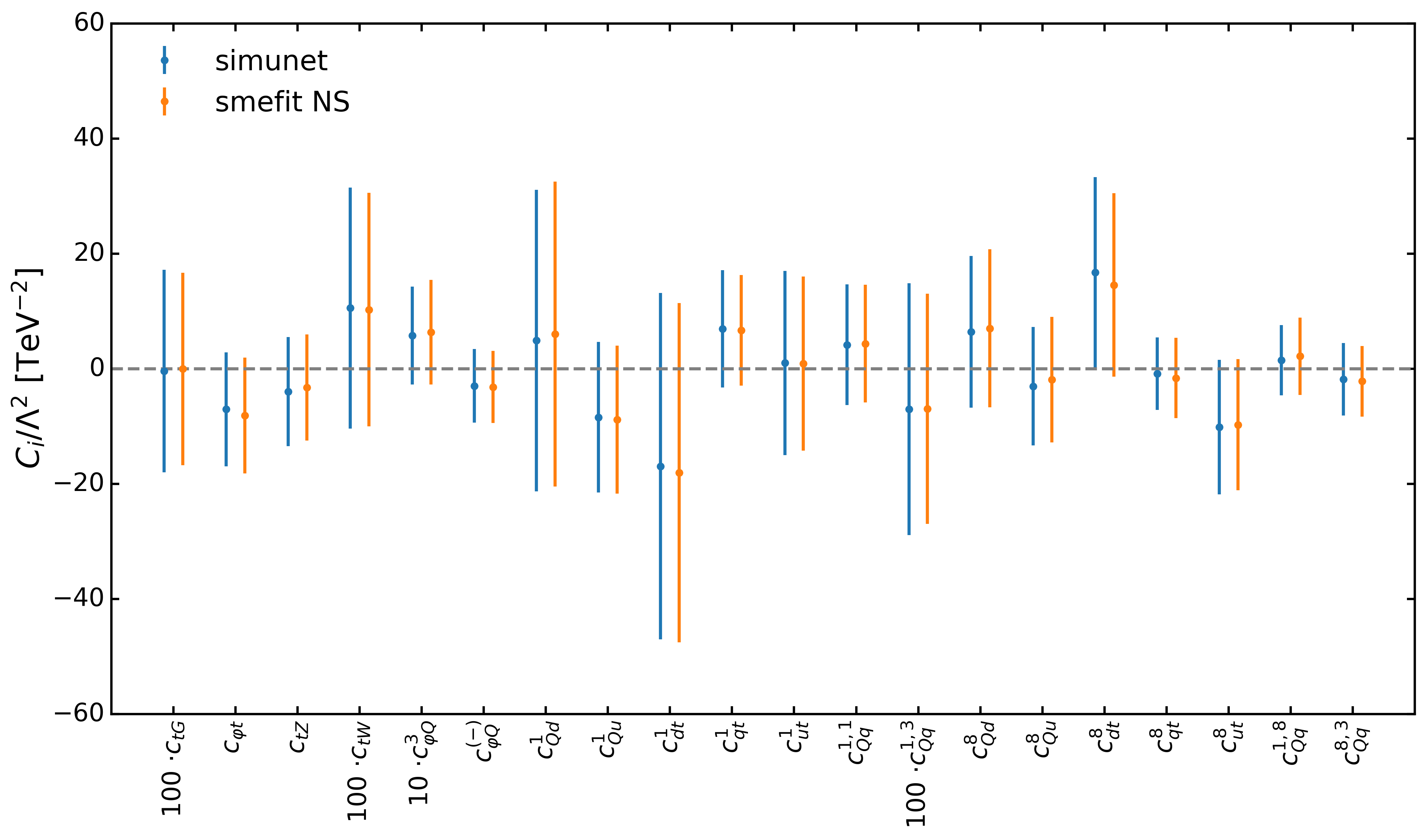}
	\caption{Comparison between the 95\% CL intervals
          obtained with \simunet{} in the fixed-PDF linear EFT case (full top quark dataset)
          with the corresponding results based on the {\sc\small SMEFiT}
          framework (with the Nested Sampling option).
          In both cases 
          the same experimental and theoretical inputs are adopted, see text for more details.
          These \simunet{} results coincide with those
          displayed in Fig.~\ref{fig:smeft_simu_bounds} in the fixed-PDF EFT analysis.
        }
	\label{fig:simunet-vs-smefit-linear}
\end{figure}

Fig.~\ref{fig:simunet-vs-smefit-linear} displays this comparison between the
marginalised 95\% CL intervals of \simunet{}
and {\sc\small SMEFiT} (operating in the Nested Sampling mode) for a linear EFT fit that takes as input
in both cases the full dataset discussed in this work as well as the same SM and EFT
theory calculations (and hence also the same operator basis).
To ensure identical inputs, a parser has been written that automatically converts data and theory files
from the \simunet{} to the  {\sc\small SMEFiT} standard formats.
 The \simunet{} results displayed here coincide with those
 displayed in Fig.~\ref{fig:smeft_simu_bounds} in the fixed-PDF EFT analysis.
 Satisfactory agreement between the two fitting frameworks is obtained,
 and similar benchmarks have been successfully performed for other variants
 of the fixed-PDF EFT fits presented in this work.

 This benchmark study ensures that the optimisation
 algorithm adopted by \simunet{} is suitable both for PDF determinations (since it is based
 on the NNPDF4.0 settings) as well as for the determination of the EFT coefficients.
 The optimisation settings that provide the best performance
 of the SM-PDF and fixed-PDF EFT analyses are then combined
 for the  simultaneous SMEFT-PDF extraction, where all weights of the \simunet{} network
 are allowed to be constrained by the data.

\FloatBarrier

\section{Fit quality}
\label{app:fit_quality}

Here we summarise the $\chi^2$ (computed using the experimental
definition, Eq.~\ref{eq:chi2exp}) for the key PDF and EFT analyses presented in this work.
Tables~\ref{tab:chi2-baseline},~\ref{tab:chi2-top} and~\ref{tab:chi2-single-top} display
the values of the $\chi^2$ per data point for  representative PDF and EFT fits,
where for each dataset we also indicate the number of data points $n_{\rm dat}$.
Specifically, we consider: (i) three SM-PDF fits: \nnpdfnotop, \nnpdf, 
and Fit H in Table~\ref{tab:fit_list} (using our full top quark dataset);
(ii) two fixed-PDF EFT fits, one based on \nnpdfnotop as input,
and the other using Fit H as input; (iii) the outcome
of the simultaneous SMEFT-PDF determination.
We show sequentially the non-top datasets,
the $t\bar{t}$ inclusive and associated production datasets,
and the single-top inclusive and associated production datasets.
We also provide the total $\chi^2$ for separate groups of processes,
including the  full non-top and top datasets, and well as for their total sum.
    
In Tables~\ref{tab:chi2-top} and~\ref{tab:chi2-single-top},
entries in italics indicate datasets 
that do not enter the corresponding SM-PDF fit; for these datasets,
we evaluate the associated
$\chi^2$ values \textit{a posteriori} using the resulting PDFs.
For instance, all top data is removed from the \nnpdfnotop fit,
and for some top quark observables such as four-heavy-quark production,
PDF dependence is neglected.
Furthermore, we note that the $\chi^2$ values for the \nnpdfnotop fit
and for the fixed-PDF EFT fit based on \nnpdfnotop are identical
in Table~\ref{tab:chi2-baseline},
since the latter uses only top data as input.
For the same reason, the entries in the columns for the SM-PDF Fit H
and the fixed-PDF EFT fits based on Fit H are the same.
   
\begin{table}[htbp]
  \begin{center}
  \renewcommand{\arraystretch}{1.50}
\tiny
\begin{tabular}{ l | c| C{1.4cm} | C{1.4cm} | C{1.3cm} | C{1.4cm} | C{1.4cm} | C{1.9cm} }
 \toprule
 \multirow{3}{*}{Dataset}    & \multirow{3}{*}{$n_{\rm dat}$}   &  \multicolumn{6}{c}{$\chi^2/n_{\rm dat}$}  \\[1.5ex]
 &   & \multicolumn{3}{c|}{\bf SM-PDF fits} & \multicolumn{2}{c|}{\bf Fixed-PDF EFT fits}   & {\bf SMEFT-PDFs}    \\
  &   & NNPDF4.0 (no top)  & NNPDF4.0 & Fit H & NNPDF4.0 (no top) & Fit H & Joint fit   \\
 \midrule
 \midrule
 SLAC      & 67   & 0.768   & 0.758 & 0.753 & 0.768 & 0.753  & 0.761  \\
 BCDMS   & 581 &  	1.265  & 1.247 & 1.251 & 1.265  & 1.251  & 1.245   \\
 NMC       & 325 & 1.298   & 1.306 & 1.304  & 1.298  & 1.304  & 1.292    \\
 CHORUS  & 832 & 0.899    & 0.900 & 0.901  & 0.899  & 0.901  & 0.898     \\
  NuTeV     & 76  & 0.398 & 0.425  & 0.423 & 0.398  & 0.423  & 0.393    \\
 HERA  & 1208 & 1.204  & 1.203 & 1.204  & 1.204  & 1.204  & 1.200    \\
 \midrule
     {\bf Total DIS}  & {\bf 3089} &  {\bf 1.122 }  & {\bf 1.119 } & {\bf 1.119 } & {\bf 1.122 } & {\bf 1.119 } & {\bf 1.114 } \\
     \midrule
     \midrule
 E886    $\sigma^d_{\rm DY}/\sigma^p_{\rm DY}$ & 15  & 0.527    & 0.524  & 0.536  & 0.527 & 0.536   & 0.544   \\
 E886    $\sigma^p_{\rm DY}$                               & 89  & 1.566  & 1.559  & 1.615  & 1.566 & 1.615  & 1.602   \\
 E605   $\sigma^p_{\rm DY}$                                & 85  & 0.456   & 0.456  & 0.460 & 0.456  & 0.460  & 0.464    \\
 E906   $\sigma^d_{\rm DY}/\sigma^p_{\rm DY}$ &  6  &  0.827   &    0.885  & 0.886   & 0.827 & 0.886  &  0.955    \\
\midrule
{\bf Total fixed-target DY} & {\bf 195} & {\bf 0.981 } & {\bf 0.979 }   & {\bf 1.008 } & {\bf 0.981} & {\bf 1.008 } & {\bf 1.006 }  \\
 \midrule
 \midrule
 CDF    $d\sigma_Z/dy_Z$                                    & 28 &  1.234  & 1.280  &  1.275 & 1.234  & 1.275  & 1.188    \\
 D0     $d\sigma_Z/dy_Z$                                     & 28 &   0.637  & 0.644 &  0.640 & 0.637  & 0.640 & 0.632   \\
 D0    	$W\to \mu \nu$ asy.                       & 9  &   1.958 & 1.929 & 1.905  & 1.958  & 1.905 & 1.641     \\
 \midrule
 ATLAS low-mass DY 7 TeV               & 6 &  0.875  & 0.879 & 0.883  & 0.875  & 0.883 & 0.888    \\
 ATLAS high-mass DY 7 TeV               & 5 & 1.694 & 1.691 & 1.661 & 1.694  & 1.661 & 1.629  \\
 ATLAS $W, Z$ 7 TeV ($\mathcal{L} = 35$ pb${}^{-1}$) & 30   &  1.000 & 0.988 & 1.000 & 1.000 & 1.000  & 0.991    \\
 ATLAS  $W,Z$ 7 TeV ($\mathcal{L} = 4.6$ fb${}^{-1}$) & 61  & 1.689 & 1.686  & 1.685 & 1.685  & 1.685 & 1.599    \\
ATLAS low-mass DY 2D 8 TeV            & 60 & 1.203 &1.216 & 1.225 & 1.203  & 1.225 & 1.207   \\
ATLAS high-mass DY 2D 8 TeV           & 48 & 1.123  & 1.110 & 1.118 & 1.123  & 1.118 & 1.147   \\
ATLAS $\sigma_{W,Z}^{\rm tot}$ 13 TeV    & 3 & 0.727 & 0.778  & 0.706 & 0.727 & 0.706 & 0.471 \\
 \midrule
 CMS $W$ electron asymmetry 7 TeV             & 11 & 0.836 & 0.833 & 0.836 & 0.836 & 0.836 & 0.892   \\
CMS $W$ muon asymmetry 7 TeV              & 11 & 1.728 & 1.714 & 1.733 & 1.728  & 1.733 &  1.699  \\
CMS DY 2D 7 TeV                                        & 110 & 1.367 & 1.361 & 1.353 & 1.367 & 1.353 &  1.339 \\
CMS $W$ rapidity 8 TeV                              & 22 & 1.375 & 1.361 & 1.365 & 1.375  & 1.365 & 1.162  \\
 \midrule
 LHCb $Z \rightarrow ee$ 7 TeV                       & 9  &  1.630  & 1.648 & 1.636 & 1.630 & 1.636 & 1.669 \\
 LHCb $Z \rightarrow ee$ 8 TeV                        & 17  & 1.272   & 1.325 & 1.334 & 1.272 &  1.334 & 1.266 \\
 LHCb  $W,Z \rightarrow \mu$  7 TeV                 & 29  &   2.032 &  1.948 & 1.902 & 2.032 & 1.902 & 1.896   \\
 LHCb  $W,Z \rightarrow \mu$   8 TeV                & 30  & 1.494   & 1.426 & 1.417 & 1.494  & 1.417 & 1.386  \\
  LHCb  $Z \rightarrow ee$   13 TeV                & 15  & 1.701 & 1.726  & 1.717  & 1.701  & 1.717 & 1.640   \\
  LHCb  $Z \rightarrow \mu\mu$   13 TeV                & 16  & 0.970 & 0.993 & 0.978  & 0.970  & 0.978 & 0.928   \\
\midrule
{\bf Total inclusive gauge boson production} & {\bf 548} &  {\bf 1.344}  & {\bf 1.339}  & {\bf 1.335}  & {\bf 1.344}  & {\bf 1.335} & {\bf 1.297}    \\
 \midrule
 \midrule
ATLAS $W^{\pm}+$jet 8 TeV                      & 30 & 0.954 & 0.959 & 0.957  & 0.954  & 0.957 & 0.973 \\
ATLAS $Z$ $p_T$ 8 TeV ($p_T, m_{\ell\ell}$)      & 44 & 0.911 & 0.905 & 0.904 & 0.911 & 0.904 & 0.910 \\
ATLAS $Z$ $p_T$ 8 TeV ($p_T, y_Z$)                 & 48 & 0.909 & 0.898 & 0.896 & 0.909 & 0.896 & 0.924\\
ATLAS incl. jets 8 TeV, $R=0.6$                           & 171 & 0.682 & 0.687 & 0.687 & 0.682  & 0.687 & 0.667  \\
ATLAS dijets 7 TeV, $R=0.6$                               & 90 & 2.184  & 2.149  & 2.125  & 2.184 & 2.125 &. 2.177  \\
ATLAS isolated $\gamma$ prod. 13 TeV             & 53 & 0.811  & 0.763 & 0.753 & 0.811  & 0.753 & 0.813  \\
 \midrule
CMS $Z$ $p_T$ 8 TeV                                 & 28 & 1.447 & 1.401 & 1.407  & 1.447 & 1.407 & 1.473  \\
CMS incl. jets 8 TeV                                     & 185 & 1.140 & 1.183 & 1.200 & 1.140  & 1.200 & 1.118   \\
CMS dijets 7 TeV                                          & 54 & 1.788 & 1.810  & 1.809 & 1.788 & 1.809  & 1.800  \\
 \midrule
     {\bf Total jets, $Z$ $p_T$ and isolated photon}                      & {\bf 703} &  {\bf 1.154}  & {\bf 1.157} & {\bf 1.158}  & {\bf 1.154}  & {\bf 1.158}  & {\bf 1.148}  \\
      \midrule
 \midrule
     {\bf Total non-top data}                      & {\bf 4535} & {\bf 1.148}  &  {\bf 1.145} & {\bf 1.146} & {\bf 1.148} & {\bf 1.146} & {\bf 1.137} \\
 \bottomrule
\end{tabular}
\end{center}
  \caption{\small \label{tab:chi2-baseline} The values of the $\chi^2$ per data point (using the experimental
    definition of Eq.~\eqref{eq:chi2exp}) for key PDF and EFT analyses presented in this work.
    For each dataset, we indicate the number of data points $n_{\rm dat}$
    and the the $\chi^2$ values for {\it i)} three SM-PDF fits:
    \nnpdfnotop, \nnpdf,
    and Fit H in Table~\ref{tab:fit_list} (full top quark dataset);
    {\it ii)} two fixed-PDF EFT fits: one based on \nnpdfnotop as input,
    and the other using Fit H as input; and finally  {\it iii)} the outcome
    of the simultaneous SMEFT-PDF determination.
    Here we restrict ourselves to the non-top datasets considered in this
    analysis, the corresponding results for the top quark observables
    are provided in Tables~\ref{tab:chi2-top} and~\ref{tab:chi2-single-top}.
    We also provide the total $\chi^2$ for separate groups of processes as well
    as for the full non-top dataset.
    The $\chi^2$ values for the \nnpdfnotop fit
    and for the fixed-PDF EFT fit based on \nnpdfnotop are identical,
    since the latter uses only top data as input.
    For the same reason, the entries in the columns for the SM-PDF Fit H
    and the fixed-PDF EFT fits based on Fit H are the same.
}
\end{table}

\begin{table}[htbp]
  \begin{center}
  \renewcommand{\arraystretch}{1.60}
  \tiny
\begin{tabular}{ l | c| C{1.4cm} | C{1.4cm} | C{1.3cm} | C{1.4cm} | C{1.4cm} | C{1.9cm} }
 \toprule
 \multirow{3}{*}{Dataset}    & \multirow{3}{*}{$n_{\rm dat}$}   &  \multicolumn{6}{c}{$\chi^2/n_{\rm dat}$}  \\[1.5ex]
 &   & \multicolumn{3}{c|}{\bf SM-PDF fits} & \multicolumn{2}{c|}{\bf Fixed-PDF EFT fits}   & {\bf SMEFT-PDFs}    \\
  &   & NNPDF4.0 (no top)  & NNPDF4.0 & Fit H & NNPDF4.0 (no top) & Fit H & Joint fit   \\
 \midrule
 \midrule
    ATLAS $\sigma(t\bar{t})$, dilepton, 7 TeV    & 1    &   \textit{4.100}  & 4.540 & 4.599 & 1.733 & 1.824 & 1.974     \\
    ATLAS $\sigma(t\bar{t})$, dilepton, 8 TeV & 1 & \textit{0.005} & 0.021 & 0.023 & 0.405 & 0.382 & 0.307 \\
    ATLAS $1/\sigma d\sigma/dm_{t\bar{t}}$, dilepton, 8 TeV & 5 & \textit{0.262} & \textit{0.284} & 0.291 & 0.305 & 0.309 & 0.310  \\
    ATLAS $\sigma(t\bar{t})$, $\ell+$jets, 8 TeV & 1 & \textit{0.218} & \textit{0.270} & 0.277 & 0.002 & 0.001  & 0.000  \\
    ATLAS $1/\sigma d\sigma/d|y_t|$, $\ell+$jets, 8 TeV & 4 & \textit{6.266} & 3.219 & 2.827 & 1.178 & 1.021 & 1.144 \\
    ATLAS $1/\sigma d\sigma/d|y_{t\bar{t}}|$, $\ell+$jets, 8 TeV & 4 & \textit{7.978} & 3.725 & 3.332 & 2.883 & 2.105 & 2.646  \\
    ATLAS $\sigma(t\bar{t})$, dilepton, 13 TeV           & 1    &  \textit{1.388}  & \textit{1.397}  & 1.374 & 0.004 & 0.004 & 0.002    \\
    ATLAS $\sigma(t\bar{t})$, hadronic, 13 TeV         & 1    & \textit{0.231}   &  \textit{0.231} & 0.230 & 0.082 & 0.082  & 0.092    \\
    ATLAS $1/\sigma d^2\sigma/d|y_{t\bar{t}}|dm_{t\bar{t}}$, hadronic, 13 TeV & 10 & \textit{2.276}  & \textit{2.007}  & 1.928 & 1.915 & 1.824 & 1.857  \\
    ATLAS $\sigma(t\bar{t})$, $\ell+$jets, 13 TeV & 1 & \textit{0.489} & 0.492 & 0.485 & 0.004 & 0.004 & 0.013 \\
    ATLAS $1/\sigma d\sigma/dm_{t\bar{t}}$, $\ell+$jets, 13 TeV & 8 & \textit{1.607} & \textit{1.788}  & 1.835 & 2.087 & 2.072 & 2.090  \\
    \midrule
    CMS $\sigma(t\bar{t})$, combined, 5 TeV & 1 & \textit{0.442}  & 0.542 & 0.555 & 0.211 & 0.236 & 0.239  \\
    CMS $\sigma(t\bar{t})$, combined, 7 TeV & 1 & \textit{0.795} & 1.032 & 1.065 & 0.004 & 0.010 & 0.027  \\
    CMS $\sigma(t\bar{t})$, combined, 8 TeV & 1 & \textit{0.170} & 0.247 & 0.258 & 0.174 & 0.156 & 0.103  \\
    CMS $1/\sigma d^2\sigma/d|y_{t\bar{t}}|dm_{t\bar{t}}$, dilepton, 8 TeV & 16 & \textit{1.094}  & \textit{0.994} & 0.963  & 0.519 & 0.580 & 0.514  \\
    CMS $1/\sigma d\sigma/d|y_{t\bar{t}}|$, $\ell+$jets, 8 TeV & 9 & \textit{2.127} &1.245 & 1.131 & 0.928 & 0.995 & 0.935  \\
    CMS $\sigma(t\bar{t})$, dilepton, 13 TeV & 1 & \textit{0.064}  & 0.063 & 0.066 & 0.680 & 0.680 & 0.600  \\
    CMS $1/\sigma d\sigma/dm_{t\bar{t}}$, dilepton, 13 TeV & 5 & \textit{2.760}  & 2.550 & 2.485 & 2.246 & 2.232 & 2.194  \\
    CMS $\sigma(t\bar{t})$, $\ell+$jets, 13 TeV & 1 & \textit{0.230} & \textit{0.227} & 0.234 & 2.153 & 2.155  & 1.907  \\
    CMS $1/\sigma d\sigma/dm_{t\bar{t}}$, $\ell+$jets, 13 TeV & 14 & \textit{1.829}  & \textit{1.481} & 1.393 & 0.962 & 0.911 & 0.912  \\
    \midrule
    ATLAS charge asymmetry, 8 TeV & 1 & \textit{0.679}  & \textit{0.678} & 0.675 & 0.567 & 0.587 & 0.571 \\
    ATLAS charge asymmetry, 13 TeV & 5 & \textit{1.012} & \textit{0.997} & 0.989 & 0.872 & 0.786 & 0.836 \\
    CMS charge asymmetry, 8 TeV & 3 & \textit{0.052} & \textit{0.052} & 0.053 & 0.072 & 0.069  & 0.071\\
    CMS charge asymmetry, 13 TeV & 3 & \textit{0.277} & \textit{0.283} & 0.286 & 0.443 & 0.510 & 0.451  \\
    ATLAS \& CMS combined charge asy., 8 TeV & 6 & \textit{0.603}  &\textit{0.602} & 0.602 & 0.646 & 0.659 & 0.651  \\
    \midrule
    ATLAS $W$-hel., 13 TeV & 2 & \textit{0.370} & \textit{0.370} & \textit{0.370} & 0.037 & 0.037 & 0.038   \\
    ATLAS \& CMS combined $W$-hel., 8 TeV & 2 & \textit{1.046} & \textit{1.046} & \textit{1.046} & 0.853 & 0.853 & 0.854  \\
       \midrule
   {\bf Total inclusive $t\bar{t}$}  & {\bf 108}  &  {\bf 1.700 } &  {\bf 1.328 } & {\bf 1.271 } & {\bf 1.032 } & {\bf 0.994} & {\bf 1.002 }  \\
     \midrule
     \midrule
    ATLAS $\sigma(t\bar{t}Z)$, 8 TeV & 1 & \textit{0.264}  &   \textit{0.232}  & 0.235 & 1.331 & 1.161 & 1.258 \\
    ATLAS $\sigma(t\bar{t}W)$, 8 TeV & 1 &  \textit{2.461} &  \textit{2.482} & 2.430 & 0.751 & 0.708 & 0.722   \\
    ATLAS $\sigma(t\bar{t}Z)$, 13 TeV & 1 &  \textit{0.702} &  \textit{0.747} & 0.742 & 0.001 & 0.000 & 0.000  \\
    ATLAS $1/\sigma d\sigma(t\bar{t}Z)/dp_T^Z$, 13 TeV & 5 &  \textit{1.961}  &  \textit{1.940} & 1.933 & 1.870 & 1.840 & 1.860  \\
    ATLAS $\sigma(t\bar{t}W)$, 13 TeV & 1 &  \textit{1.436} &  \textit{1.456} & 1.417 & 0.000 & 0.002 & 0.000 \\
    \midrule
    ATLAS $\sigma(t\bar{t}\gamma)$, 8 TeV & 1 & \textit{0.426} & \textit{0.426} & \textit{0.426} & 0.037 & 0.009 & 0.034  \\
    \midrule
    ATLAS $\sigma(t\bar{t}t\bar{t})$, multi-lepton, 13 TeV & 1 & \textit{3.655} & \textit{3.655} & \textit{3.655} & 4.289 & 3.784 & 4.145  \\
    ATLAS $\sigma(t\bar{t}t\bar{t})$, single lepton, 13 TeV & 1 & \textit{0.872} & \textit{0.872} & \textit{0.872} & 0.997 & 0.898 & 0.969  \\
    ATLAS $\sigma(t\bar{t}b\bar{b})$, $\ell+$jets, 13 TeV & 1 & \textit{2.062} & \textit{2.062} & \textit{2.062} & 1.093 & 1.316 & 1.152  \\
       \midrule
    CMS $\sigma(t\bar{t}Z)$, 8 TeV & 1 &  \textit{0.432}  &  \textit{0.470} & 0.466 & 0.005 & 0.020 & 0.010 \\
    CMS $\sigma(t\bar{t}W)$, 8 TeV & 1 &  \textit{2.281}  &  \textit{2.298} & 2.255 & 0.808 & 0.770 & 0.783  \\
    CMS $\sigma(t\bar{t}Z)$, 13 TeV & 1 &  \textit{1.185}  &  \textit{1.238} & 1.233 & 0.085 & 0.113 & 0.110 \\
    CMS $1/\sigma d\sigma(t\bar{t}Z)/dp_T^Z$, 13 TeV & 3 &  \textit{0.628}  &  \textit{0.598} & 0.590 & 0.936 & 0.756 & 0.898 \\
    CMS $\sigma(t\bar{t}W)$, 13 TeV & 1 &  \textit{0.667} &  \textit{0.682} & 0.652 & 0.297 & 0.329 & 0.307  \\
    \midrule
    CMS $\sigma(t\bar{t}\gamma)$, 8 TeV & 1 & \textit{0.508} & \textit{0.508} & \textit{0.508} & 0.022 & 0.045 & 0.024 \\
    \midrule
    CMS $\sigma(t\bar{t}t\bar{t})$, multi-lepton, 13 TeV & 1 & \textit{0.027} & \textit{0.027} & \textit{0.027} & 0.124 & 0.041 & 0.096  \\
    CMS $\sigma(t\bar{t}t\bar{t})$, single lepton, 13 TeV & 1 & \textit{0.231} & \textit{0.231} & \textit{0.231} & 0.262 & 0.238 & 0.255 \\
    CMS $\sigma(t\bar{t}b\bar{b})$, all-jet, 13 TeV & 1 & \textit{1.878} & \textit{1.878} & \textit{1.878} & 1.334 & 1.466 & 1.369  \\
    CMS $\sigma(t\bar{t}b\bar{b})$, dilepton, 13 TeV & 1 & \textit{0.962} & \textit{0.962} & \textit{0.962} & 0.312 & 0.447 & 0.347  \\
    CMS $\sigma(t\bar{t}b\bar{b})$, $\ell+$jets, 13 TeV & 1 & \textit{0.900} & \textit{0.900} & \textit{0.900} & 0.135 & 0.269 & 0.167  \\
    \midrule
   {\bf Total associated $t\bar{t}$}  & {\bf 26}  &  {\bf 1.255 } &  {\bf 1.255 } & {\bf 1.246 } & {\bf 0.925 } & {\bf 0.888 } & {\bf 0.913 } \\
 \bottomrule
\end{tabular}
\end{center}
  \caption{\small \label{tab:chi2-top}
    Same as Table~\ref{tab:chi2-baseline} now
    for the  inclusive and associated $t\bar{t}$ production datasets.
     Values in italics indicate datasets 
    that do not enter the corresponding SM-PDF fit: for those processes,
    we evaluate a posteriori the physical observables and the associated
    $\chi^2$ values using the resulting PDFs from the fit.
    For instance, top data is removed from the \nnpdfnotop,
    and for some top quark observables like $t\bar{t}t\bar{t}$ their
    PDF dependence is neglected.
      The last row indicates the total values adding together the non-top
    and top data contributions to the $\chi^2$.
}
\end{table}

\begin{table}[htbp]
  \begin{center}
  \renewcommand{\arraystretch}{1.70}
    \tiny
\begin{tabular}{ l | c| C{1.4cm} | C{1.4cm} | C{1.3cm} | C{1.4cm} | C{1.4cm} | C{1.9cm} }
 \toprule
 \multirow{3}{*}{Dataset}    & \multirow{3}{*}{$n_{\rm dat}$}   &  \multicolumn{6}{c}{$\chi^2/n_{\rm dat}$}  \\[1.5ex]
 &   & \multicolumn{3}{c|}{\bf SM-PDF fits} & \multicolumn{2}{c|}{\bf Fixed-PDF EFT fits}   & {\bf SMEFT-PDFs}    \\
  &   & NNPDF4.0 (no top)  & NNPDF4.0 & Fit H & NNPDF4.0 (no top) & Fit H & Joint fit   \\
 \midrule
 \midrule
     ATLAS $t$-channel $\sigma(t)$, 7 TeV & 1 & \textit{0.785} & \textit{0.756} & 0.757 & 0.179 & 0.135 & 0.155  \\
    ATLAS $t$-channel $\sigma(\bar{t})$, 7 TeV & 1 & \textit{0.304}  & \textit{0.296} & 0.282 & 0.003 & 0.005 & 0.004  \\
    ATLAS $t$-channel $1/\sigma d\sigma(tq)/dy_t$, 7 TeV & 3 & \textit{0.909} & 0.959 & 0.971 & 0.885 & 0.946 & 0.900  \\
     ATLAS $t$-channel $1/\sigma d\sigma(\bar{t}q)/dy_{\bar{t}}$, 7 TeV & 3 & \textit{0.061}  & 0.062 & 0.062 & 0.061 & 0.063 & 0.062  \\
     ATLAS $t$-channel $\sigma(t)$, 8 TeV & 1 & \textit{0.795}  & \textit{0.739} & 0.739 & 0.000 & 0.017 & 0.007 \\
     ATLAS $t$-channel $\sigma(\bar{t})$, 8 TeV & 1 & \textit{2.314} & \textit{2.277}  & 2.221 & 0.379 & 0.337 & 0.356  \\
    ATLAS $t$-channel $1/\sigma d\sigma(tq)/dy_t$, 8 TeV & 3 & \textit{0.288} & 0.246 & 0.241 & 0.290 & 0.245 & 0.283  \\
     ATLAS $t$-channel $1/\sigma d\sigma(\bar{t}q)/dy_{\bar{t}}$, 8 TeV & 3 & \textit{0.196} & 0.190 & 0.187 & 0.198 & 0.189 & 0.192  \\
     ATLAS $s$-channel $\sigma(t + \bar{t})$, 8 TeV & 1 & \textit{0.211} & \textit{0.203} & 0.216 & 0.000 & 0.000 & 0.000  \\
    ATLAS $t$-channel $\sigma(t)$, 13 TeV & 1 & \textit{0.738} & \textit{0.722} & 0.720 & 0.261 & 0.224 & 0.235  \\
    ATLAS $t$-channel $\sigma(\bar{t})$, 13 TeV & 1 & \textit{0.400} & \textit{0.393} & 0.384 & 0.057 & 0.050 & 0.051  \\
     ATLAS $s$-channel $\sigma(t + \bar{t})$, 13 TeV & 1 & \textit{0.700} & \textit{0.688} & 0.703 & 0.127 & 0.138 & 0.118  \\
    \midrule
    CMS $t$-channel $\sigma(t) + \sigma(\bar{t})$, 7 TeV & 1 & \textit{0.765}  & 0.729  & 0.719 & 0.010 & 0.001 & 0.004  \\
    CMS $t$-channel $\sigma(t)$, 8 TeV & 1 & \textit{0.010} & \textit{0.006} & 0.005 & 0.487 & 0.628 & 0.566  \\
    CMS $t$-channel $\sigma(\bar{t})$, 8 TeV & 1 & \textit{0.079} & \textit{0.084} & 0.092 & 0.886 & 0.934  & 0.911 \\
    CMS $s$-channel $\sigma(t + \bar{t})$, 8 TeV & 1 & \textit{1.108}  & \textit{1.112} & 1.105 & 1.348 & 1.338 & 1.350 \\
    CMS $t$-channel $\sigma(t)$, 13 TeV & 1 & \textit{0.904} & \textit{0.883} & 0.880 & 0.281 & 0.235 & 0.248  \\
    CMS $t$-channel $\sigma(\bar{t})$, 13 TeV & 1 & \textit{0.136} & \textit{0.132} & 0.125 & 0.010 & 0.014 & 0.013 \\
   CMS $t$-channel $1/\sigma d\sigma/d|y^{(t)}|$, 13 TeV & 4 & \textit{0.412} & \textit{0.387} & 0.383 & 0.405 & 0.378 & 0.396  \\
       \midrule
   {\bf Total single top}  & {\bf 30}  & {\bf 0.509 }  &  {\bf 0.498 } & {\bf 0.495 } & {\bf 0.332 } & {\bf 0.330 } & {\bf 0.331 }  \\
     \midrule
     \midrule
    ATLAS $\sigma(tW)$, dilepton, 8 TeV & 1 & \textit{0.510} & \textit{0.529} & 0.538 & 0.144  & 0.168 & 0.167 \\
    ATLAS $\sigma(tW)$, single-lepton, 8 TeV & 1 & \textit{0.695} & \textit{0.708} & 0.713 & 0.428 & 0.449 & 0.448  \\
    ATLAS $\sigma(tW)$, dilepton, 13 TeV & 1 & \textit{1.150} & \textit{1.149} & 1.148 & 0.840 & 0.847 & 0.849  \\
    ATLAS $\sigma_{\text{fid}}(tZj)$, dilepton, 13 TeV & 1 & \textit{0.115} & \textit{0.115} & \textit{0.115} & 0.871 & 0.611 & 0.795  \\
    \midrule
    CMS $\sigma(tW)$, dilepton, 8 TeV & 1 & \textit{0.359} & \textit{0.371} & 0.376 & 0.134 & 0.150 & 0.149 \\
    CMS $\sigma(tW)$, dilepton, 13 TeV & 1 & \textit{0.434} & \textit{0.437} & 0.438 & 1.612 & 1.573 & 1.563 \\
    CMS $\sigma_{\text{fid}}(tZj)$, dilepton, 13 TeV & 1 & \textit{1.041} & \textit{1.041} & \textit{1.041} & 0.279 & 0.427 & 0.316  \\
    CMS $d\sigma_{\text{fid}}(tZj)/dp_T^{t}$, dilepton, 13 TeV & 3 & \textit{0.362} & \textit{0.362} & \textit{0.362} & 0.130 & 0.156 & 0.134  \\
   CMS $\sigma(tW)$, single-lepton, 13 TeV & 1 & \textit{2.837} & \textit{2.833} & 2.832 & 1.817 & 1.839 & 1.846  \\
    \midrule
   {\bf Total associated single top}  & {\bf 11}  & {\bf 0.748 }  & {\bf 0.752 } &  {\bf 0.753 }  & {\bf 0.592 } & {\bf 0.594 } & {\bf 0.594 } \\
     \midrule
     \midrule
   {\bf Total top}  & {\bf 175}  & {\bf 1.370}   & {\bf 1.139} & {\bf 1.102} & {\bf 0.868} & {\bf 0.839} & {\bf 0.848}    \\
     \midrule
     \midrule
     {\bf Total dataset}   &  {\bf 4710} & {\bf 1.156}  & {\bf 1.145} & {\bf 1.144} & {\bf 1.138} & {\bf 1.135} & {\bf 1.126}  \\
 \bottomrule
\end{tabular}
\end{center}
  \caption{\small \label{tab:chi2-single-top}
     Same as Table~\ref{tab:chi2-top} now
    for the  inclusive and associated single-top production datasets.
    Values in italics indicate datasets 
    that do not enter the corresponding fits: for those processes,
    we evaluate a posteriori the physical observables and the associated
    $\chi^2$ values using the resulting PDFs from the fit.
    The last row indicates the total values adding together the non-top
    and top data contributions to the $\chi^2$,
    and the next-to-last row the total top quark contribution.
}
\end{table}


Several observations can be drawn
from the results presented in Tables~\ref{tab:chi2-baseline},~\ref{tab:chi2-top} and~\ref{tab:chi2-single-top}.
First of all, we note that the description of the non-top data is improved in the simultaneous
SMEFT-PDF analysis, with $\chi^2=1.137$ to be compared with the analogous SM-PDF analysis (Fit H)
which leads to $\chi^2=1.146$.
Given that we have $n_{\rm dat}=4535$ no-top data points, this
improvement corresponds to 40 units in the absolute $\chi^2$.
This improvement cannot traced back to a specific measurements or group
of processes.
In turn, this improvement is also reflected for the full dataset,
whose $\chi^2$ is the lowest (1.126 for $n_{\rm dat}=4710$ points) in the SMEFT-PDF analysis.
This lower $\chi^2$ as compared to the SM-PDF baseline fits is not surprising,
taking into account the fact that (as discussed in Sect.~\ref{sec:res_smeft}) we now have more degrees
of freedom in the fit, and in particular the inclusion of linear EFT corrections
is instrumental in improving the $\chi^2$ to the top quark data as compared
to the SM-PDF fits.
We also find that the fit quality to the top datasets is similar in the joint SMEFT-PDF analysis
and in the fixed-PDF EFT fits with either \nnpdfnotop or Fit H as input
PDFs, with $\chi^2 = 0.848, 0.868, 0.839$ respectively in each case for
$n_{\rm dat}=175$.
The small differences between the three cases (less than five units
in absolute $\chi^2$) are consistent with the reported independence of the EFT interpretation
of top quark data with respect to the choice of input PDF in the calculation.

Finally, we note the only class of top quark production process for which the $\chi^2$ values
are markedly different in the SM-PDF fits is inclusive top quark pair production.
Indeed, the $\chi^2$ values for the complete top quark dataset are $1.370, 1.139$, and $1.102$ in the
SM-PDF fits without top data, with \nnpdf, and with the full top quark
dataset (Fit H) respectively; this improvement arises mostly from the $t\bar{t}$ group, whose $\chi^2$
values are $1.700, 1.328$, and $1.271$ for the same three fits respectively.
This observation is consistent with the findings of Sect.~\ref{sec:baseline_sm_fits}, that
the dominant PDF sensitivity in Fit H arises from inclusive $t\bar{t}$ production.
We also note that the $\chi^2$ for top quark data is similar between \nnpdf and Fit H, again
in agreement with the fact that the additional top quark measurements considered
here as compared to those in \nnpdf have a consistent pull on the PDFs, in particular
on the large-$x$ gluon.

\section{Pitfalls of the Monte-Carlo replica method for quadratic EFT fits}
\label{app:quad}

This analysis was originally prepared with the intention of performing a fully simultaneous SMEFT-PDF
fit using NLO QCD theory, including \textit{quadratic}, $\mathcal{O}\lp \Lambda^{-4}\rp$, contributions from the SMEFT.
To this end, the capabilities of the \simunet{} framework were extended, as discussed in Sect.~\ref{sec:new_simunet}, and
all necessary SMEFT $K$-factors needed for the quadratic predictions were produced.
However, whilst benchmarking our code, we noticed significant disagreement between quadratic SMEFT-only fits
produced using the \simunet{} methodology and the \smefit Nested Sampling option;\footnote{This disagreement is not present at the linear level; in this case, \simunet{} using the fixed-PDF option and \smefit using either the Nested Sampling or MCfit options perfectly coincide. See App.~\ref{app:benchmark} for more details.} on the other hand, we note perfect agreement
between our \simunet{} quadratic SMEFT-only fits and the \smefit MCfit option.

This disagreement can be traced back to 
a deficiency in the Monte-Carlo sampling method used to propagate experimental error to the SMEFT coefficients, which
currently prevents us from applying the
the \simunet{} framework to joint SMEFT-PDF
fits with quadratic EFT calculations. 
In this Appendix, we describe our current understanding of these limitations within the context of a toy model, 
and give a more realistic example from this work; further work on the topic is deferred to a future publication..

\subsection{A toy model for quadratic EFT fits}
In the following subsection, we consider a toy scenario involving a single data point $d$ and only one
Wilson coefficient $c$ (we ignore all PDF-dependence).
We suppose that our observed experimental data point $d$ is a random variable drawn from a normal distribution
centred on the underlying quadratic theory prediction, and with experimental variance $\sigma^2$:
\begin{equation}
\label{eq:data_gaussian}
d \sim N(t(c), \sigma^2).
\end{equation}
We assume that the theory prediction is quadratic in the SMEFT Wilson coefficient $c$, taking the form:
\begin{equation}
t(c) = t^{\text{SM}} + c t^{\text{lin}} + c^2 t^{\text{quad}} \, ,
\end{equation}
where we set $\Lambda=1$ TeV for convenience. Recall that $t^{\text{quad}} > 0$, since it corresponds
to a squared amplitude.
Given the observed data $d$, we would like to construct interval estimates for the parameter
$c$ (usually \textit{confidence intervals} in a frequentist setting or \textit{credible intervals} in a Bayesian setting).
Here, we shall describe the analytical construction of two interval estimates: first, using the Bayesian method, and second, using the
Monte-Carlo replica method.

\paragraph{Bayesian method.}
In the Bayesian approach, $c$ is treated as a random variable with its own distribution. By Bayes'
theorem, we can write the probability distribution of $c$, given the observed data $d$, up to a proportionality constant (given by $1/\mathbb{P}(d)$, where $\mathbb{P}(d)$ is called the \textit{Bayes' evidence}) as:
\begin{equation}
\mathbb{P}(c | d) \propto \mathbb{P}(d | c) \mathbb{P}(c),
\end{equation}
\noindent
where $\mathbb{P}(c|d)$ is called the \textit{posterior distribution} of $c$, given the observed data $d$,
and  $\mathbb{P}(c)$ is called the \textit{prior distribution} of $c$ - this is our initial `best guess' of the distribution of $c$ before the observation
of the data takes place.
This distribution 
is often taken to be uniform in SMEFT fits; we shall assume this here.

Given a value of $c$, the distribution $\mathbb{P}(d|c)$ of the data $d$ is assumed to be Gaussian, as specified in Eq.~(\ref{eq:data_gaussian}). In particular, we can deduce that the posterior distribution of $c$ obeys the following proportionality relation:\footnote{Technically, truncated according to the end-points of the uniform prior.}
\begin{equation}
\label{eq:bayesian_posterior}
\mathbb{P}(c|d) \propto \exp\left( -\frac{1}{2\sigma^2} \left( d - t(c) \right)^2 \right).
\end{equation}
\noindent This posterior distribution can be used to place interval estimates on the parameter $c$. One way of doing this is to construct \textit{highest density intervals}. These are computed as follows.
For a $100\alpha$\% credible interval, we determine the constant $p(\alpha)$ satisfying:
\begin{equation}
\int\limits_{\{c : ~\mathbb{P}(c|d) > p(\alpha)\}} \mathbb{P}(c|d)\ dc = \alpha.
\end{equation}
\noindent An interval estimate for $c$ is then given by $\{c : \mathbb{P}(c|d) > p(\alpha)\}$. In order to obtain such intervals then, we must construct the posterior $\mathbb{P}(c|d)$; efficient sampling from the posterior
is guaranteed by methods such as Nested Sampling~\cite{Feroz:2013hea,Feroz:2007kg}.

\paragraph{The Monte-Carlo replica method.}
This method takes a different approach in order to produce a posterior distribution for the parameter $c$.
Given the observed central data value $d$, one samples repeatedly from the normal distribution $N(d, \sigma^2)$ to generate a collection of \textit{pseudodata replicas}, which we shall denote as $d^{(1)}, ..., d^{(N_{\text{rep}})}$, where $N_{\text{rep}}$ is the total number of replicas.
Given a pseudodata replica $d^{(i)}$, one obtains a corresponding best-fit value of the Wilson coefficient parameter $c^{(i)}$ by minimising the $\chi^2$ of the theory to the pseudodata:
\begin{equation}
c^{(i)} = \argmin_{c} \chi^2(c,d^{(i)}) = \argmin_{c} \left( \frac{(d^{(i)} - t(c))^2}{\sigma^2} \right).
\end{equation}
In this toy scenario, we can determine an analytical formula for $c^{(i)}$:
\begin{equation}
  c^{(i)} = \begin{cases} \displaystyle -\frac{t^{\text{lin}}}{2t^{\text{quad}}}, & \text{if $d^{(i)} \leq \lp t^{\text{SM}} - (t^{\text{lin}})^2/4t^{\text{quad}}\rp $ ;} \label{eq:app1} \\[2ex]
    \displaystyle \frac{-t^{\text{lin}} \pm \sqrt{(t^{\text{lin}})^2 - 4t^{\text{quad}}(t^{\text{SM}} - d^{(i)})}}{2t^{\text{quad}}}, & \text{if $d^{(i)} \geq \lp t^{\text{SM}} - (t^{\text{lin}})^2/4t^{\text{quad}}\rp $  \, .} \end{cases}
\end{equation}
The first case arises when the $\chi^2$ to the pseudodata $d^{(i)}$ has a single minimum, whilst the second case arises when the $\chi^2$  has two minima.
The two cases are split based on the value of
\be
\label{eq:crosssec_minimum}
t_{\text{min}} = t^{\text{SM}} - (t^{\text{lin}})^2/4t^{\text{quad}} \, ,
\ee
which is the minimum value of the quadratic theory prediction $t(c) = t^{\text{SM}} + c t^{\text{lin}} + c^2 t^{\text{quad}}$. Note that for data replicas such that $d^{(i)}\le t_{\text{min}}$, the best-fit value $c^{(i)}$
becomes independent of $d^{(i)}$ and depends only on the ratio between linear and quadratic
EFT cross-sections.

Now, given that $d^{(i)}$ is a random variable drawn from the normal distribution $N(d,\sigma^2)$, one can infer the corresponding distribution of the random variable $c^{(i)}$, which is a function of the pseudodata $d^{(i)}$.
For a real random variable $X$ with associated probability density $P_X(x)$, a function $f : \mathbb{R} \rightarrow \mathbb{R}$ of the random variable has the distribution:
\begin{equation}
P_{f(X)}(y) = \int\limits_{-\infty}^{\infty} dx\ P_X(x) \delta(y - f(x)).
\end{equation}
In our case, $c^{(i)}$ is a multi-valued function of $d^{(i)}$ given the two square roots, but the formula is easily generalised to this case.
Recalling that the pseudodata replicas are generated according
to a Gaussian distribution around the central measurement $d$ with variance $\sigma^2$,
we find that the probability density function for the Wilson coefficient replica
$c^{(i)}$ is given (up to a proportionality constant) by:
\begin{align}
P_{c^{(i)}}(c) &\propto \int\limits_{-\infty}^{t_{\text{min}}} dx\ \delta\left( c + \frac{t^{\text{lin}}}{2t^{\text{quad}}} \right) \exp\left( -\frac{1}{2\sigma^2} (x - d)^2 \right) \notag \\[1.5ex]
&\qquad + \int\limits_{t_{\text{min}}}^{\infty} dx\  \delta\left( c - \left(\frac{-t^{\text{lin}} + \sqrt{(t^{\text{lin}})^2 - 4t^{\text{quad}}(t^{\text{SM}} - x)}}{2t^{\text{quad}}}\right)\right)\exp\left( -\frac{1}{2\sigma^2} (x - d)^2 \right) \notag\\[1.5ex]
& \qquad + \int\limits_{t_{\text{min}}}^{\infty} dx\ \delta\left( c - \left(\frac{-t^{\text{lin}} - \sqrt{(t^{\text{lin}})^2 - 4t^{\text{quad}}(t^{\text{SM}} - x)}}{2t^{\text{quad}}}\right)\right) \exp\left( -\frac{1}{2\sigma^2} (x - d)^2 \right).
\end{align}
Simplifying the delta functions in the second and third integrals, we find:
\begin{equation}
\label{eq:mc_posterior}
P_{c^{(i)}}(c) \propto \delta\left( c + \frac{t^{\text{lin}}}{2t^{\text{quad}}} \right) \int\limits_{-\infty}^{t_{\text{min}}} dx\ \exp\left( -\frac{1}{2\sigma^2} (x - d)^2 \right) + \frac{2}{|2ct^{\text{quad}}  + t^{\text{lin}}|} \exp\left( -\frac{1}{2\sigma^2} (d - t(c))^2 \right) \, .
\end{equation}
This result is different from the posterior distribution obtained by the Bayesian method in Eq.~(\ref{eq:bayesian_posterior}). Notable features of the posterior distribution $P_{c^{(i)}}(c)$ are: (i) the distribution has a Dirac-delta peak at $c = -t^{\text{lin}}/2t^{\text{quad}}$; (ii) elsewhere, the distribution is given by the Bayesian posterior distribution rescaled by a prefactor dependent on $c$.
Therefore, the Monte Carlo
replica optimisation method will not in general reproduce the Bayesian posterior.

However, one can note that in an appropriate limit, the Bayesian posterior \textit{is} indeed recovered.
In particular, suppose that the quadratic EFT cross-section is subdominant
compared to the linear term, $t^{\text{lin}} \gg t^{\text{quad}}$; in this case, we have that
$t_{\text{min}} \rightarrow -\infty$ so that the first term in Eq.~\eqref{eq:mc_posterior} vanishes
and the prefactor of the second term can be approximated with $2ct^{\text{quad}} + t^{\text{lin}} \approx t^{\text{lin}}$. 
Thus, the Bayesian posterior from Eq.~(\ref{eq:bayesian_posterior}) is indeed recovered, and we 
see that the two methods are formally identical for a linear EFT analysis.
Further, it is possible to show analytically that for multiple SMEFT parameters and multiple correlated data points, 
if only linear theory is used the two distributions agree exactly.


This calculation demonstrates that, for quadratic EFT fits, the Monte-Carlo replica method
will not in general reproduce the Bayesian posteriors that one would obtain from, say, a 
nested sampling approach; agreement will only occur provided quadratic EFT corrections 
are sufficiently subdominant in comparison with the linear ones.
For this reason, in this work we restrict the SMEFT-PDF fits based on \simunet{} (which rely
on the use of the Monte-Carlo replica method) to linear EFT calculations;
we defer the further investigation of the use of the Monte-Carlo replica method, and how
it might be modified for use in \simunet{}, to future works. 


\subsection{Application to one-parameter fits}
\label{subsec:mc_examples}

\begin{figure}[t]
        \centering
        \includegraphics[width=0.82\linewidth]{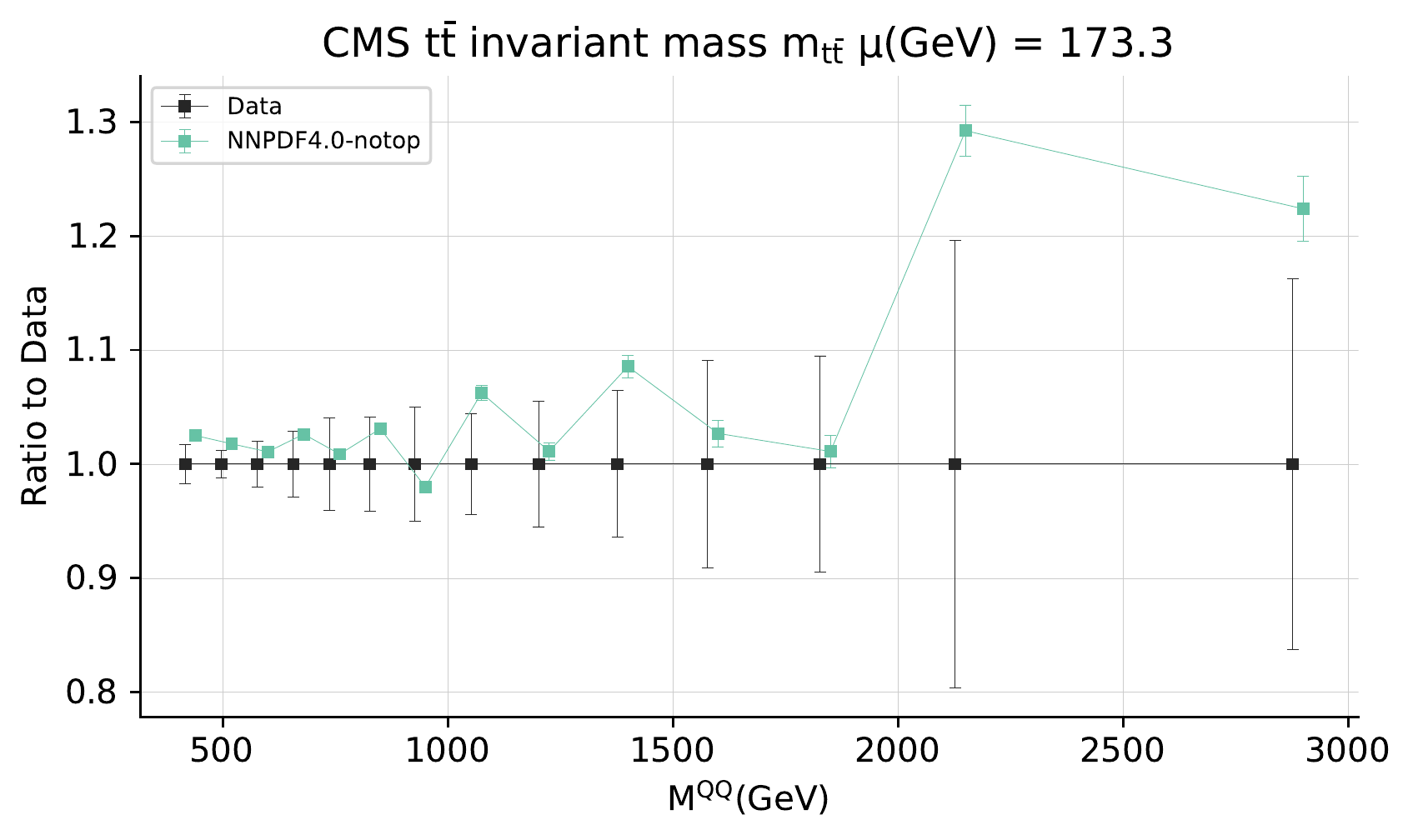}
        \caption{Comparison between the experimental data
          for the top-quark pair invariant mass $m_{t\bar{t}}$ distribution
          from the $\ell+$jets CMS measurement at 13 TeV~\cite{CMS:2021vhb}, with the corresponding
          SM theory calculations at NNLO using the NNPDF4.0 (no top) PDF
          set as input.
          For the latter, the error band indicates the PDF uncertainties, and
          for the former the diagonal entries of the experimental covariance matrix.
          Results are shown as ratios to the central value of the data.
          The SM theory predictions overshoot the data, especially in the high $m_{t\bar{t}}$
          regions, where energy-growing effects enhance the EFT corrections.
        }
    \label{fig:data-theory-CMS}
\end{figure}

As demonstrated above, the Monte-Carlo replica method will lead to
posterior distributions differing from their Bayesian counterparts whenever quadratic
EFT corrections dominate over linear ones.
Here, we show the numerical impact of these differences in a model case, namely
the one-parameter fit of the coefficient $c_{dt}^8$ to
the CMS 13 TeV $t\bar{t}$ invariant mass distribution measurement
based on the $\ell+$jets final-state~\cite{CMS:2021vhb}.
Fig.~\ref{fig:data-theory-CMS} compares the experimental data
from this measurement with the corresponding
SM theory calculations at NNLO using the NNPDF4.0 (no top) PDF
set as input.
We observe that the SM theory predictions overshoot the data, especially in the high $m_{t\bar{t}}$
regions, where energy-growing effects enhance the EFT corrections.
Given that the pseudodata replicas $d^{(i)}$ are fluctuated around the central value $d$,
the configuration where the SM overshoots the data
potentially enhances the contribution of the upper solution in Eq.~(\ref{eq:app1})
leading to the Dirac delta peak in the posterior Eq.~(\ref{eq:mc_posterior}).     

For the case of the $c_{dt}^8$ coefficient, the quadratic EFT corrections dominate over the linear ones
and hence the net effect of a non-zero coefficient is typically an upwards shift
of the theory prediction.
Indeed, we have verified that for this coefficient
the biggest negative correction one can obtain is of order $\sim 2\%$.
For the last $m_{t\bar{t}}$ bin, the minimum of the  theory cross
section $t_{\rm min}$ in Eq.~(\ref{eq:crosssec_minimum}) is obtained for a value
$c_{dt}^8 \approx -0.2$, while for the second to last bin instead $t_{\rm min}$
is minimised by $c_{dt}^8 \approx -0.3$.
The combination of these two features (a dominant quadratic EFT term, and a SM prediction overshooting
the data) suggests that the Monte-Carlo replica method's
posterior will be enhanced for $c_{dt}^8 \in (-0.3,-0.2)$ as compared
to the Bayesian posterior.

\begin{figure}[htb]
        \centering
        \includegraphics[width=0.49\linewidth]{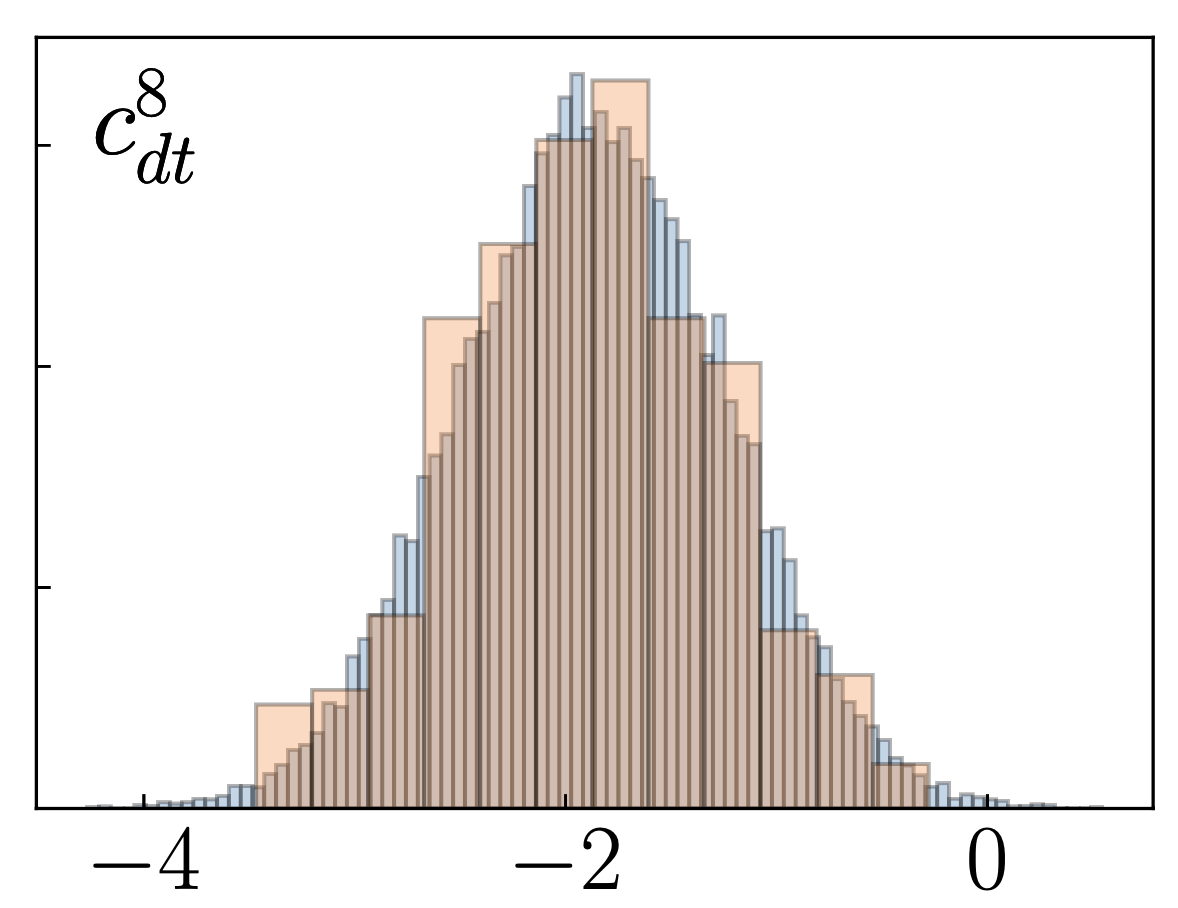}
        \includegraphics[width=0.49\linewidth]{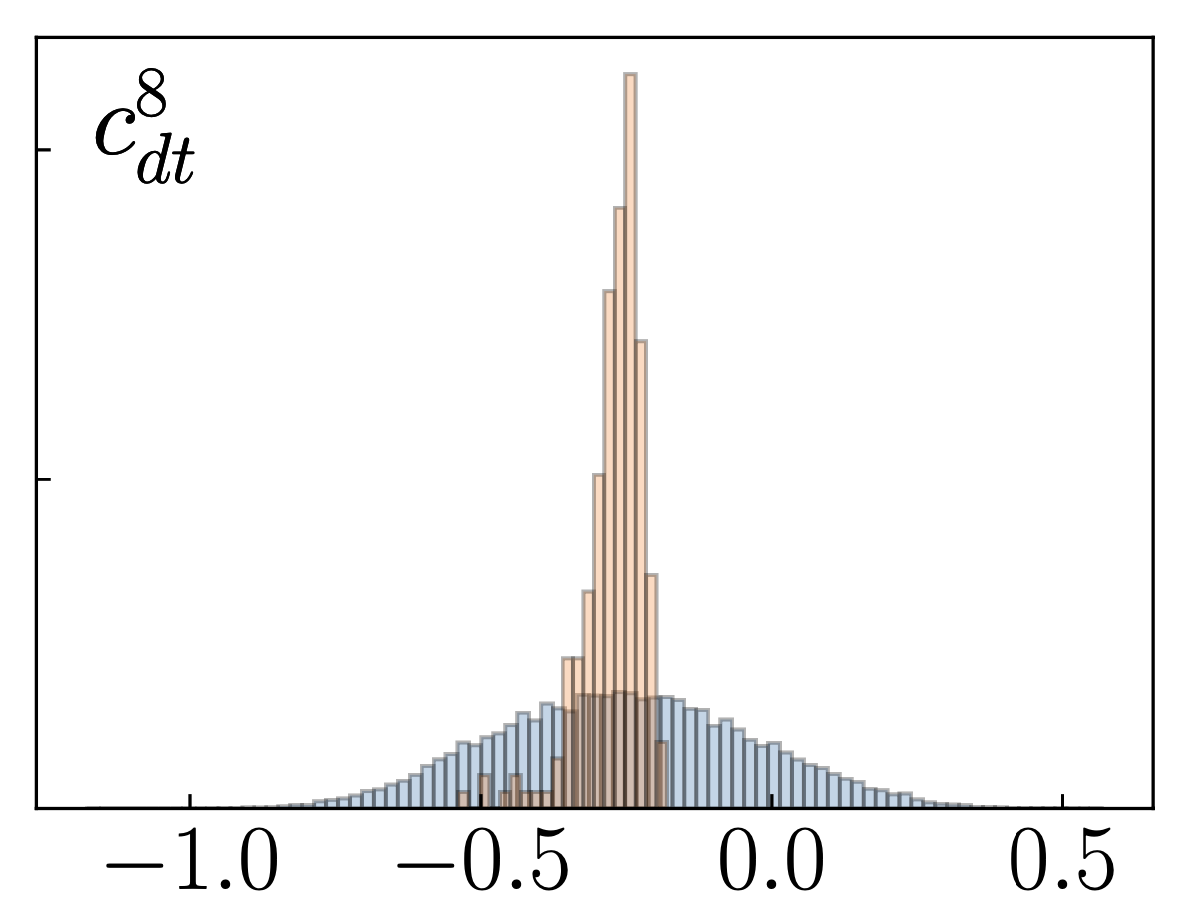}
        \caption{Posterior distributions for a one-parameter fit
          of the four-fermion coefficient $c_{dt}^8$ with the sole experimental
          input being the CMS $m_{t\bar{t}}$ distribution
          displayed in Fig.~\ref{fig:data-theory-CMS}.
          Results are obtained with \smefit and we compare the outcome of Nested Sampling (in blue) with
          that of MCfit (in red)
          for linear (left panel) and quadratic (right panel) EFT fits.
        }
    \label{fig:O8dt-fit}
\end{figure}

In Fig.~\ref{fig:O8dt-fit} we
compare the posterior distributions for a one-parameter fit
of the four-fermion coefficient $c_{dt}^8$ with the sole experimental
input being the CMS $m_{t\bar{t}}$ distribution
displayed in Fig.~\ref{fig:data-theory-CMS}.
Results are obtained with \smefit and we compare the outcome of Nested Sampling (in blue) with
that of MCfit (in red)
for linear and quadratic  EFT fits.
The agreement in the linear fit is lost for its quadratic counterpart,
with the main difference being a sharp peak in the region
$c_{dt}^8 \in (-0.3,-0.2)$ in which the contribution
from the delta function in Eq.~\ref{eq:mc_posterior} is most
marked.

The scenario displayed in Fig.~\ref{fig:O8dt-fit} is chosen to display the maximum effect,
based on a single coefficient with large quadratic EFT corrections, and a dataset
where the SM overshoots the data in the $m_{t\bar{t}}$ region where EFT effects
are the largest.
Within a global fit, these differences are \textit{partially} washed out
(indeed the Bayesian and MCfit posterior distributions mostly agree well for
the quadratic \smefit analysis, as shown in~\cite{Giani:2023gfq}, for the majority
of fitted coefficients).
Nevertheless, at least in its current implementation, Fig.~\ref{fig:O8dt-fit} highlights
that the Monte-Carlo replica method is affected by pitfalls that
prevent its straightforward application to global EFT interpretations of experimental
data which include quadratic corrections.


\bibliographystyle{utphys}
\providecommand{\href}[2]{#2}\begingroup\raggedright\endgroup


\end{document}